\documentclass[twocolumn,tighten]{aastex62}

\newcommand{\SII}{[S\tiny{ }\footnotesize{II}\normalsize{] }}

\newcommand{\MgII}{Mg\tiny{ }\footnotesize{II}\normalsize{ }}
\newcommand{\SiII}{Si\tiny{ }\footnotesize{II}\normalsize{ }}

\newcommand{\kms}{\ifmmode\,{\rm km}\,{\rm s}^{-1}\else km$\,$s$^{-1}$\fi}
\newcommand{\Lya}{\ifmmode\,{\rm Ly}{\rm \alpha}\else Ly$\alpha$\fi}
\newcommand{\Msun}{\mathrm{M}_{\sun}}

\usepackage{lipsum}  
\defcitealias{Dopita16}{D16}

\newcommand{\allcaps}[1]{\verb!#1!}

\newcommand\outflowprofile[2]{{\begin{figure*}
  
  \begin{minipage}[c]{0.55\textwidth}
     \centering
    \includegraphics[width=\textwidth]{fig/#1/#1-outflow-profile.png}
     \centering
    \includegraphics[width=\textwidth]{fig/#1/#1-residual.png}
  \end{minipage}
  \hfill 
  \begin{minipage}[c]{0.44\textwidth}
     \centering
    \includegraphics[width=\textwidth,height=1.25\textwidth]{fig/#1/#1-outflow-lines.png}
  \end{minipage}
  \caption{Same as Figure~17, for #1. 
}
\label{fig:appendix_object_#1}
\end{figure*}}}

\graphicspath{{./}{}}

\shorttitle{The Kinematics of the Interstellar Medium at Cosmic Noon}
\shortauthors{}

\begin{document}
\title{Resolved velocity profiles of galactic winds at Cosmic Noon }

\correspondingauthor{Keerthi Vasan G.C.}
\email{kvch@ucdavis.edu}

\author[0000-0002-2645-679X]{Keerthi Vasan G.C.}
\affiliation{Department of Physics and Astronomy, University of California, Davis, 1 Shields Avenue, Davis, CA 95616, USA}

\author[0000-0001-5860-3419]{Tucker Jones}
\affiliation{Department of Physics and Astronomy, University of California, Davis, 1 Shields Avenue, Davis, CA 95616, USA}

\author{Ryan L. Sanders}
\affiliation{Department of Physics and Astronomy, University of California, Davis, 1 Shields Avenue, Davis, CA 95616, USA}
\affiliation{Hubble Fellow}

\author{Richard S. Ellis}
\affiliation{Department of Physics and Astronomy, University College London, Gower Street, London WC1E 6BT, UK}

\author{Daniel P. Stark}
\affiliation{Steward Observatory, University of Arizona, 933 N Cherry Avenue, Tucson, AZ 85719, USA}

\author[0000-0003-1362-9302]{Glenn G. Kacprzak}
\affiliation{Centre for Astrophysics and Supercomputing, Swinburne University of Technology, Hawthorn, Victoria 3122, Australia}
\affiliation{ARC Centre of Excellence for All Sky Astrophysics in 3 Dimensions (ASTRO 3D), Australia}

\author[0000-0002-2784-564X]{ Tania M. Barone }
\affiliation{Centre for Astrophysics and Supercomputing, Swinburne University of Technology, Hawthorn, Victoria 3122, Australia}
\affiliation{ARC Centre of Excellence for All Sky Astrophysics in 3 Dimensions (ASTRO 3D), Australia}

\author[0000-0001-9208-2143]{Kim-Vy H. Tran}
\affiliation{School of Physics, University of New South Wales, Kensington, Australia}
\affiliation{ARC Centre of Excellence for All Sky Astrophysics in 3 Dimensions (ASTRO 3D), Australia}

\author[0000-0002-3254-9044]{Karl Glazebrook }
\affiliation{Centre for Astrophysics and Supercomputing, Swinburne University of Technology, Hawthorn, Victoria 3122, Australia}
\affiliation{ARC Centre of Excellence for All Sky Astrophysics in 3 Dimensions (ASTRO 3D), Australia}

\author{Colin Jacobs}
\affiliation{Centre for Astrophysics and Supercomputing, Swinburne University of Technology, Hawthorn, Victoria 3122, Australia}
\affiliation{ARC Centre of Excellence for All Sky Astrophysics in 3 Dimensions (ASTRO 3D), Australia}

\begin{abstract}
We study the kinematics of the interstellar medium (ISM) viewed ``down the barrel'' in 20 gravitationally lensed galaxies during Cosmic Noon ($z=1.5 - 3.5$). We use moderate-resolution spectra ($R\sim4000$) from Keck/ESI and Magellan/MagE to spectrally resolve the ISM absorption in these galaxies into $\sim$10 independent elements and use double Gaussian fits to quantify the velocity structure of the gas. We find that the bulk motion of gas in this galaxy sample is outflowing, with average velocity centroid $\left<v_{cent}\right>=-141$~\kms\ ($\pm111$~\kms\ scatter) measured with respect to the systemic redshift. 16 out of the 20 galaxies exhibit a clear positive skewness, with a blueshifted tail extending to $\sim -500$ \kms. We examine scaling relations in outflow velocities with galaxy stellar mass and star formation rate (SFR), finding correlations consistent with a momentum-driven wind scenario. Our measured outflow velocities are also comparable to those reported for FIRE-2 and TNG50 cosmological simulations at similar redshift and galaxy properties. We also consider implications for interpreting results from lower-resolution spectra. We demonstrate that while velocity centroids are accurately recovered, the skewness, velocity width, and probes of high velocity gas (e.g., $v_{95}$) are subject to large scatter and biases at lower resolution. We find that $R\gtrsim1700$ is required for accurate results for the gas kinematics of our sample. This work represents the largest available sample of well-resolved outflow velocity structure at $z>2$, and highlights the need for good spectral resolution to recover accurate properties.

\end{abstract}

% Unified Astronomy Thesaurus concepts: 

\section{Introduction} 

The formation and evolution of galaxies is regulated by feedback from star formation and supermassive black hole growth \citep[e.g.,][]{blackholescaling2003,winds2005-review,quasarfeedback2005,blackholefeedback2012_review,somerville2015}. The energy released by high star formation or black hole accretion rates can drive powerful galactic-scale outflows of gas and dust, limiting future star formation \citep[e.g.,][]{zhang2018-review,theoreticalchallenges-review2017,feedbackAndOutflows2017}. At redshifts $z\simeq2-3$, corresponding to the peak period of cosmic star formation activity \citep[``Cosmic Noon''; e.g.,][]{madau-dickinson-2014-starformation}, virtually {\it all} star-forming galaxies exhibit outflows \citep[e.g.,][]{Frye2002, shapley2003, outflows-highz-sugahara2019}. This is indeed expected based on their high star formation rate (SFR) surface densities \citep{heckman2002,sdss-outflows2016}. 

Outflows in high-redshift galaxies are typically identified by interstellar medium (ISM) features in the rest-frame ultraviolet spectrum. Outflowing gas produces blueshifted absorption, and redshifted emission in \Lya\ and other resonant lines. This signature is observed ubiquitously in $z>2$ star-forming galaxies \citep{deep2outflows, shapley2003,Vanzella2009,  steidel2010, jones2012, du2018}. However, while large samples are available, the spectral resolution $R$ is typically too low to resolve the outflow velocity structure. At $R\sim600$ the full width at half-maximum (FWHM) resolution is $\sim500$~\kms, which is comparable to the maximum observed velocities, whereas in this work we will focus on $R\gtrsim4000$ corresponding to FWHM $\lesssim75$~\kms. Furthermore, many studies rely on stacking analyses which preclude characterizing individual systems. Our current knowledge is thus largely limited to the average velocity centroid, which encompasses both outflows and ambient interstellar material. This leaves key questions unanswered, such as the proportion of gas which is able to escape the galaxy halo (as opposed to low-velocity gas which will remain in the circumgalactic medium (CGM) or recycle back to the galaxy), and the covering fraction of low-ionization gas which regulates the escape of ionizing photons \citep[e.g.,][]{du2018}. Low-resolution data are likewise unable to disentangle outflows from the non-outflowing ISM component.

A promising way forward is to observe bright gravitationally lensed galaxies, which can be magnified by factors of $\sim$10$\times$. Such bright sources enable moderate resolution spectroscopy with good sensitivity on 8--10m telescopes. Early studies of a few individual systems at $z\simeq2-3$ revealed the velocity structure of ISM and outflowing gas spanning $\sim$1000~\kms\ \citep{pettini-cb58-2002, quider2009, quider2010, dessauges-8oclock}. Similarly, \citet{tj2013} and \citet{nicha2016-escape-fraction} used deep spectroscopy of seven strongly lensed $z>4$ galaxies to measure their covering fraction profiles, revealing a considerable diversity among the star-forming population. 

The number of well-characterized strongly lensed systems has grown tremendously over the last decade thanks to all-sky surveys and dedicated lens searches \citep[e.g.,][]{hsc-2018,colin-des-2019,Huang_DESI-decam_2020}.  Previously, \citet{tucker-dustinthewind} conducted a study of 9 bright lensed galaxies from the \verb!CASSOWARY! survey \citep{belokurov2009,danstark-cswa-confirmation}, quantifying their bulk outflow velocities and chemical compositions. This work aims to compile a larger sample of 20 targets observed at moderate spectral resolution ($R\sim 2530-6300$) with the main goal of quantifying the ISM outflow velocity structure in a statistical sense. With these results we seek to aid and improve upon the interpretation of larger samples at lower spectral resolution, by comparing trends in outflow velocities between low and moderate resolution data. Finally, we seek to compare the measured outflow velocities with those obtained in simulations with different feedback prescriptions, and provide a benchmark data sample for future comparison with cosmological simulations.

This paper is organized as follows: Section \ref{sec:sample} describes the lensed galaxy sample and moderate resolution spectroscopy. In Section \ref{sec:vel-structure} we derive velocity profiles of the interstellar and outflowing gas, while Section \ref{sec:ism-features} discusses the kinematic features of the ISM. Section \ref{sec:galaxy-trends} compares the observations of outflow velocities with scaling relations from previous work and simulations. We summarize the main conclusions of this work in Section~\ref{sec:conclusions}. Throughout this paper, we use the AB magnitude system and a $\Lambda$CDM cosmology with $\Omega_{M}=0.3$, $\Omega_{\Lambda}=0.7$ and $\mathrm{H}_0=70$~\kms~Mpc$^{-1}$.

\begin{figure*}[ht!]
\centerline{
\includegraphics[width=\linewidth]{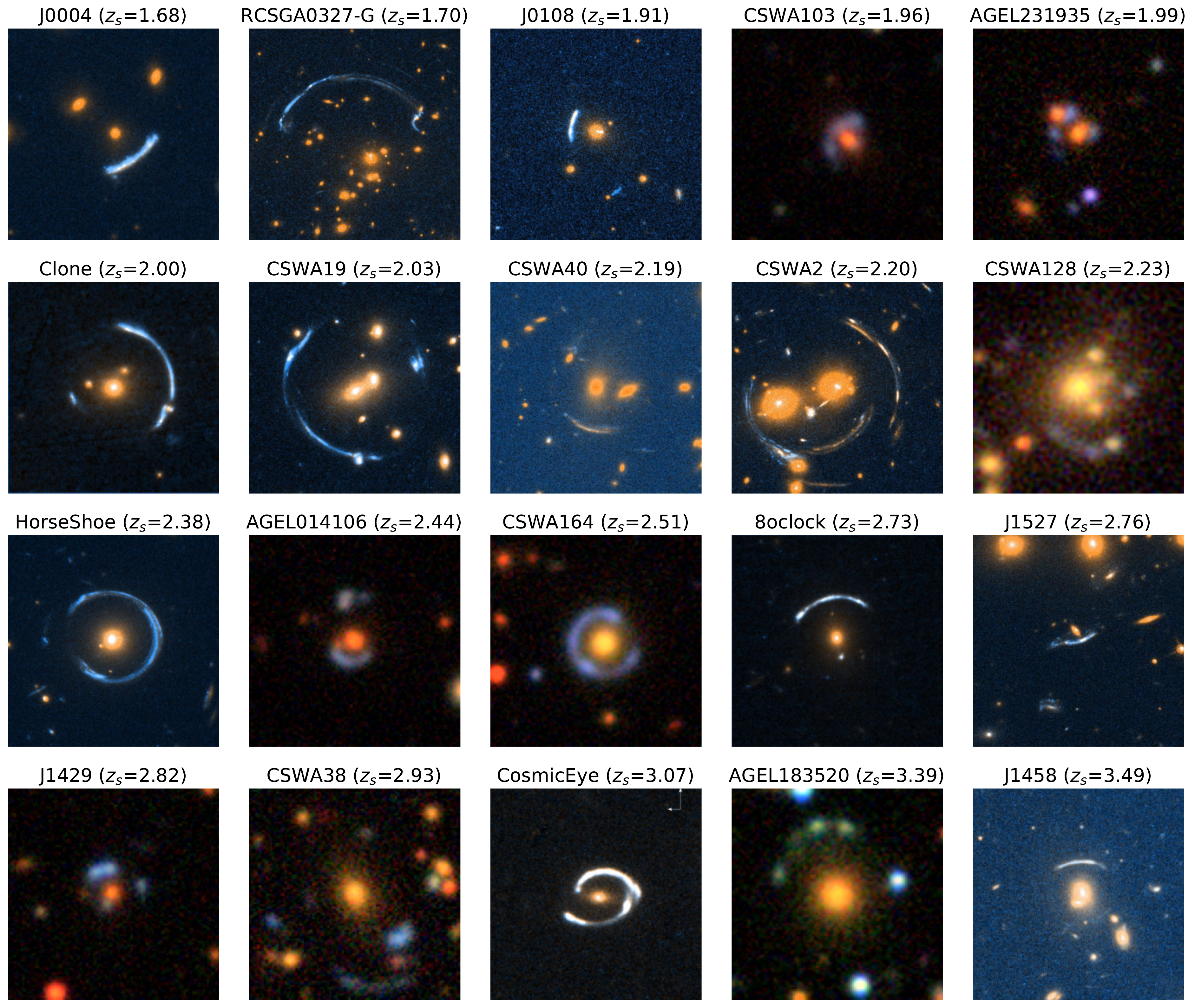}
}
\caption{ 
Color composite images of the gravitationally lensed galaxies used in this paper obtained either from DECaLS, SDSS, or the Hubble Space Telescope archive. The galaxies are arranged in order of increasing outflow velocity parameter ($v_{75,V2}$), from top to bottom. Each image is centered on the deflector(s) contributing to the lensing potential. These lensed galaxies are bright and appear highly magnified on the sky with a mean magnification value of $\mu = 9$. The images are oriented North-up, East-left with the image sizes labeled in arcseconds. The RA, Dec slit position and the position angle (PA) used for observations can be found in Table \ref{tab:redshifts} and the references provided therein. 
\label{fig:image-montage}}

\end{figure*}
\begin{deluxetable*}{|c|c|c|c|c|c|c|c|}[!ht]
    \tablewidth{\textwidth}
    \tablecaption{Table of Galaxy properties.} % Name of table 
    \label{tab:redshifts}
    \tabletypesize{\small}
    % Header %
    \tablehead{ 
    \colhead{Object ID} &  \colhead{RA (slit)} & \colhead{Dec (slit)} & \colhead{PA (slit)} &
    \colhead{$z_s$} &  \colhead{Notes} &\colhead{Survey}  & \colhead{R}
    }
    %data %
    \startdata
J0004             & 00:04:51.685 & $-$01:03:20.86  & parallactic & 1.6812 & stellar absorption & MEGASAURA  & 2750\\
RCSGA0327-G       & 03:27:26.626 & $-$13:26:15.30  & parallactic & 1.70385 & nebular emission & MEGASAURA  & 2830 \\
J0108             & 01:08:42.206 & +06:24:44.41  & parallactic & 1.9099 & stellar absorption & MEGASAURA  & 4380\\
CSWA103           & 01:45:04.38  & $-$04:55:50.8  & 115 & 1.95978 & \ion{C}{3}] & CASSOWARY & 6300\\
AGEL231935+115016 & 23:19:34.66 & +11:50:18.1  & -40 & 1.99256 & ISM absorption & AGEL & 4700\\ % DCLS6967002
Clone             & 12:06:10.65 & +51:44:44.1  & 40 & 2.0026 & stellar absorption & KOA & 4700 \\
CSWA19            & 09:00:02.80 & +22:34:07.1 & 86 & 2.03237 & \ion{C}{3}] & CASSOWARY  & 6300\\
CSWA40            & 09:52:40.29 & +34:34:39.2 & 70 & 2.18938 & stellar absorption & CASSOWARY  & 6300\\
CSWA2             & 10:38:41.88 & +48:49:22.4 & 17 & 2.19677 & \ion{C}{3}] & CASSOWARY  & 6300\\
CSWA128           & 19:58:35.44 & +59:50:52.2 & 60 & 2.22505 & \ion{O}{3}] & CASSOWARY  & 6300\\
HorseShoe         & 11:48:33.264 & +19:29:59.11 & parallactic & 2.3814 & stellar absorption & MEGASAURA  & 3980\\
AGEL014106-171324 & 1:41:06.1273 & $-$17:13:23.545 & 320 & 2.43716 & ISM absorption & AGEL & 4700 \\
CSWA164           & 02:32:49.93 & $-$03:23:25.8 & 158 & 2.51172 & stellar absorption & CASSOWARY & 6300 \\
8oclock           & 00:22:40.36 & +14:31:27.6 & -276 & 2.735 & stellar absorption & KOA & 4700\\
J1527             & 15:27:45.116 & +06:52:19.57  & parallactic & 2.76238 & stellar absorption & MEGASAURA & 2740 \\
J1429             & 14:29:54.857 & +12:02:38.68 & parallactic & 2.8241 & stellar absorption & MEGASAURA  & 3500 \\
CSWA38            & 12:26:51.48 & +21:52:17.9 & 130 & 2.92556 & stellar absorption & CASSOWARY & 6300 \\
CosmicEye         & 21:35:12.7 & $-$01:01:42.9 & parallactic & 3.0734 & stellar absorption & MEGASAURA  & 2530\\
AGEL183520+460627 & 18:35:20.55 & +46:06:35.4 & 16 & 3.38845  & nebular emission & AGEL & 4700 \\ %DESI-278 
J1458             & 14:58:36.143 & $-$00:23:58.17 & parallactic & 3.487 & stellar absorption & MEGASAURA & 4000\\
\enddata
\tablenotetext{}{References for each survey are as follows. MEGASAURA: \citet{rigby-megasura-paper1}, 
CASSOWARY: \citet{tucker-dustinthewind}, 
AGEL: \citet{VyAGEL}, KOA: Keck Observatory Archive. $z_s$ is the source galaxy systemic redshift, and $R$ is the spectral resolution. RA, Dec and PA (position angle) correspond to the slit locations used to observe the galaxies.} The spectral features used to determine $z_s$ are listed under Notes. 
\end{deluxetable*}

\begin{figure*}[ht!]
\centerline{
\includegraphics[width=\linewidth]{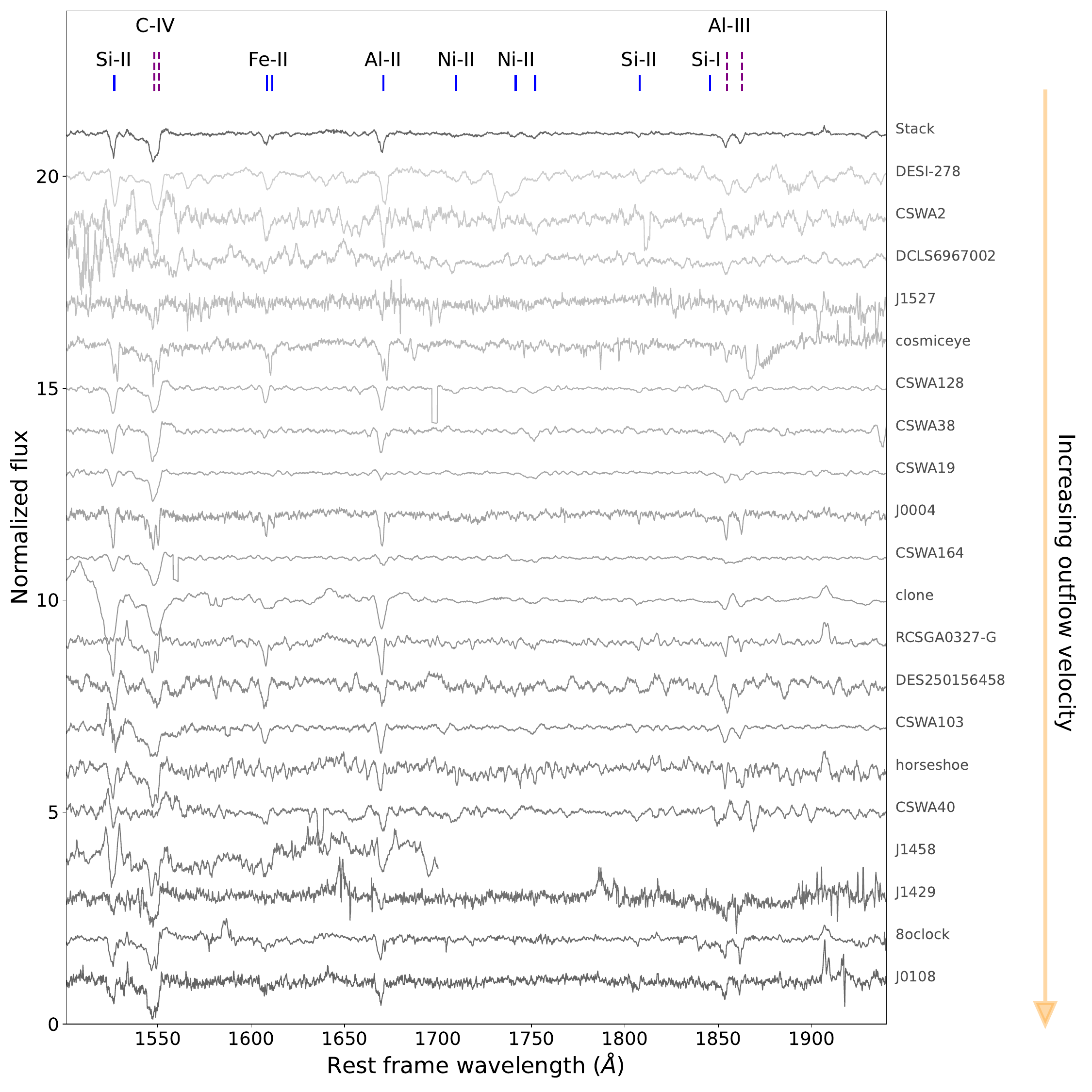}
}
\caption{ Spectra of the 20 lensed galaxies used in this work showing the rest-frame wavelength range 1500--1950 \AA. The spectra are sorted from top to bottom in the increasing value of $v_{75,V2}$ (75\% outflow velocity measured considering only gas with $v<0$; see Table~\ref{tab:velocity-measurements}), and offset for clarity. Low ionization (e.g., \ion{Si}{2}, \ion{Al}{2}) and high ionization (e.g., \ion{C}{4}) lines are marked in blue and purple respectively. The interstellar transitions probe a range of optical depth, from the strong lines such as \ion{Si}{2}~$\lambda$1526 and \ion{Al}{2}~$\lambda$1670 to the weak (optically thin) \ion{Ni}{2} features. A median-stacked spectrum is shown in the top row to demonstrate the various ISM absorption features with higher signal-to-noise.
}\label{fig:all-spectra}
\end{figure*}

\section{Sample and spectroscopic data}\label{sec:sample}

The goals of this work require moderate resolution spectroscopy ($R\gtrsim4000$) in order to sample the ISM absorption profiles with $\sim$10 independent spectral resolution elements. We have compiled a sample from our previous work, other archival data, and new observations from an ongoing survey of bright lensed galaxies discovered in wide area imaging surveys. The full sample used in this work is listed in Table~\ref{tab:redshifts}, and color images of each source are shown in Figure~\ref{fig:image-montage}. Below we describe the spectroscopic data sets.

\begin{enumerate}
    \item \verb!CASSOWARY!: The Cambridge And Sloan Survey Of Wide ARcs in the skY (CASSOWARY, abbreviated CSWA) consists of bright lensed galaxies discovered in Sloan Digital Sky Survey (SDSS) imaging \citep{belokurov2009,danstark-cswa-confirmation}. Followup echellete spectra were taken with ESI \citep{keck-esi} at Keck Observatory using an 0\farcs75 slit width, resulting in $R=6300$ resolution (FWHM = 48 \kms) covering a wavelength range of 3900--11000 \AA. These data are described in \cite{tucker-dustinthewind} including an analysis of the ISM chemical composition. 7 targets from this sample (CSWA2, CSWA19, CSWA38, CSWA40, CSWA103, CSWA128, and CSWA164) have sufficient data quality and coverage of the ISM lines needed for this work. 
    
    \item \verb!MEGASAURA!: The Magellan Evolution of Galaxies Spectroscopic and Ultraviolet Reference Atlas (MEGaSaURA) consists of spectra of lensed galaxies taken with the MagE spectrograph on the Magellan telescopes, extracted over the wavelength range 3200--8280 \AA\ \citep{rigby-megasura-paper1}. 8 targets from this sample (J0004, J0108, J1429, J1458, J1527, CosmicEye, HorseShoe, and RCSGA0327-G) are used in this paper. A range of MagE slit widths were used resulting in spectral resolution ranging from 2530-4400, with an average $R=3300$. 
    
    \item \verb!AGEL!: As part of the ASTRO3D Galaxy Evolution with Lenses (AGEL) project, we have obtained Keck/ESI spectra of bright lensed galaxies discovered from a machine learning search in wide area imaging. The search methodology and a subset of targets are described in \cite{colin-des-2019}. Spectra were taken with a 1\farcs0 slit providing $R=4700$ resolution (FWHM = 64 \kms) covering a wavelength range of 3900-11000 \AA. The observations are described in \citet{VyAGEL}. 3 targets (AGEL231935+115016, AGEL122651+215218, AGEL183520+460627) from the AGEL sample are used in this paper.

    \item KOA: Data for 2 additional bright lensed galaxies (Clone, 8oclock) were obtained from the Keck Observatory Archive (KOA) and reduced using MAKEE written by Tom Barlow\footnote{\url{https://www2.keck.hawaii.edu/inst/esi/makee.html}
    } for inclusion in this analysis. These observations were taken with the same setting as the AGEL sample, using the 1\farcs0 slit. The reduction was performed following the same methods used for the AGEL data \citep{VyAGEL}, with default settings prescribed for ESI. A manual extraction region covering the entire galaxy light was taken to be the continuum, with the rest of the slit considered as the sky to generate the error spectra. Extracted 1D spectra are binned to a common dispersion of 11.5 \kms\ per pixel. 
\end{enumerate}
Our sample is comprised of moderately massive, star-forming main-sequence galaxies (Section~\ref{sec:galaxy-trends}), which show no evidence of AGN in the available spectra. Those with resolved spectroscopic observations exhibit a wide range of kinematic structure \citep[e.g.,][]{stark2008, tucker2013, Wuyts-rcs, boordoloi_rcsga, nicha2016, chisholm2018, BethanJames2018, Shaban_2022} with HorseShoe, Clone and CosmicEye being rotationally supported whereas CSWA2, CSWA19, CSWA38, CSWA128, and RCSGA0327 appear to be mergers/interacting systems. Figure~\ref{fig:all-spectra} plots spectra of the full sample between 1500-1950~\AA, with prominent absorption features labelled.

\subsection{Systemic Redshifts}

Systemic redshifts are needed to characterize ISM kinematics with respect to the stars. Table~\ref{tab:redshifts} lists the redshifts for galaxies in our sample, which span $z = 1.6$--3.5, along with the type of features used for these measurements. In most cases, the systemic redshift is based on stellar photospheric absorption lines with a typical uncertainty of $\leq 30 \kms$. In some cases where suitable stellar features are not reliably measured (e.g., CSWA19), we use nebular \ion{C}{3}] or \ion{O}{3}] emission lines to establish the redshift. Photospheric absorption or nebular emission lines are available for 18 of the targets in our sample. For 2 targets (AGEL231935+115016, AGEL014106-171324) where none of these features are securely measured, we estimate the systemic redshift from the ISM absorption lines themselves in the following way: we find the velocity corresponding to the maximum covering fraction and then apply an offset of $+172\pm19$ \kms\ (i.e.  $z_{sys} = z_{Cf,max}+ 172 \kms/c \pm 19 \kms/c$). This offset value is derived as the median difference and sample standard deviation between the systemic and $z_{Cf,max}$ from the 18 galaxies in the sample with robust systemic redshifts. 
The offset between systemic and ISM absorption velocities in our sample is comparable to measurements from non-lensed galaxies at similar redshifts \citep[e.g.,][]{steidel2010,tj2013}.

 \begin{figure*}[!ht]
  \begin{minipage}[c]{\textwidth}
     \centering
    \includegraphics[width=\linewidth]{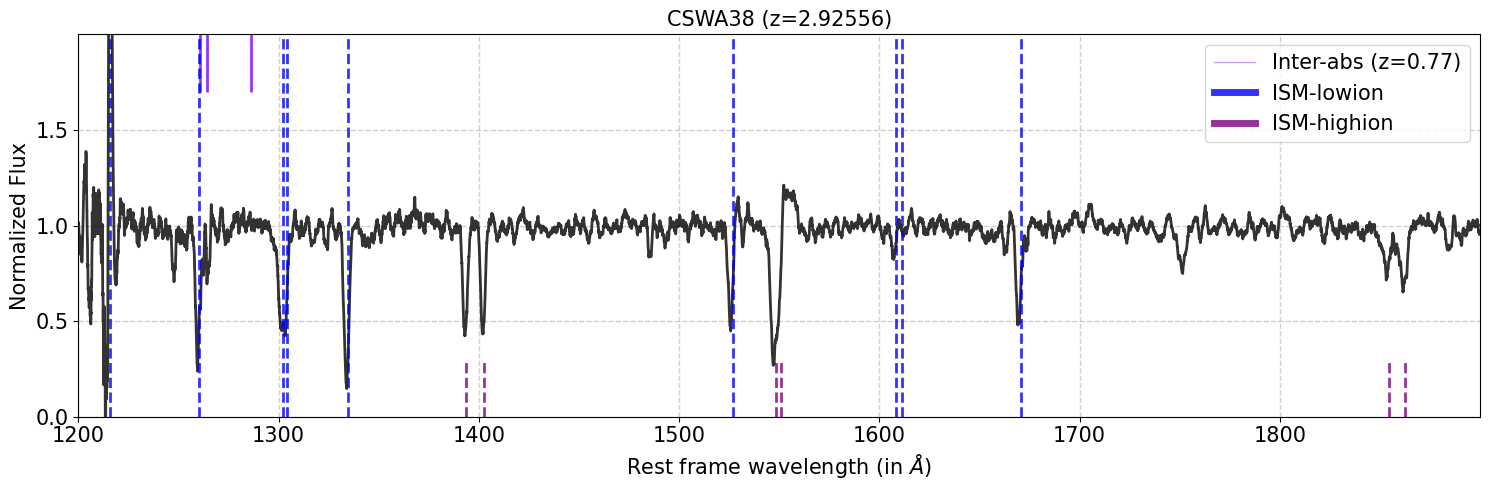}
  \end{minipage}
   \vspace{10pt} 
  
   \begin{minipage}[c]{0.5\textwidth}
     \centering
    \includegraphics[width=\textwidth]{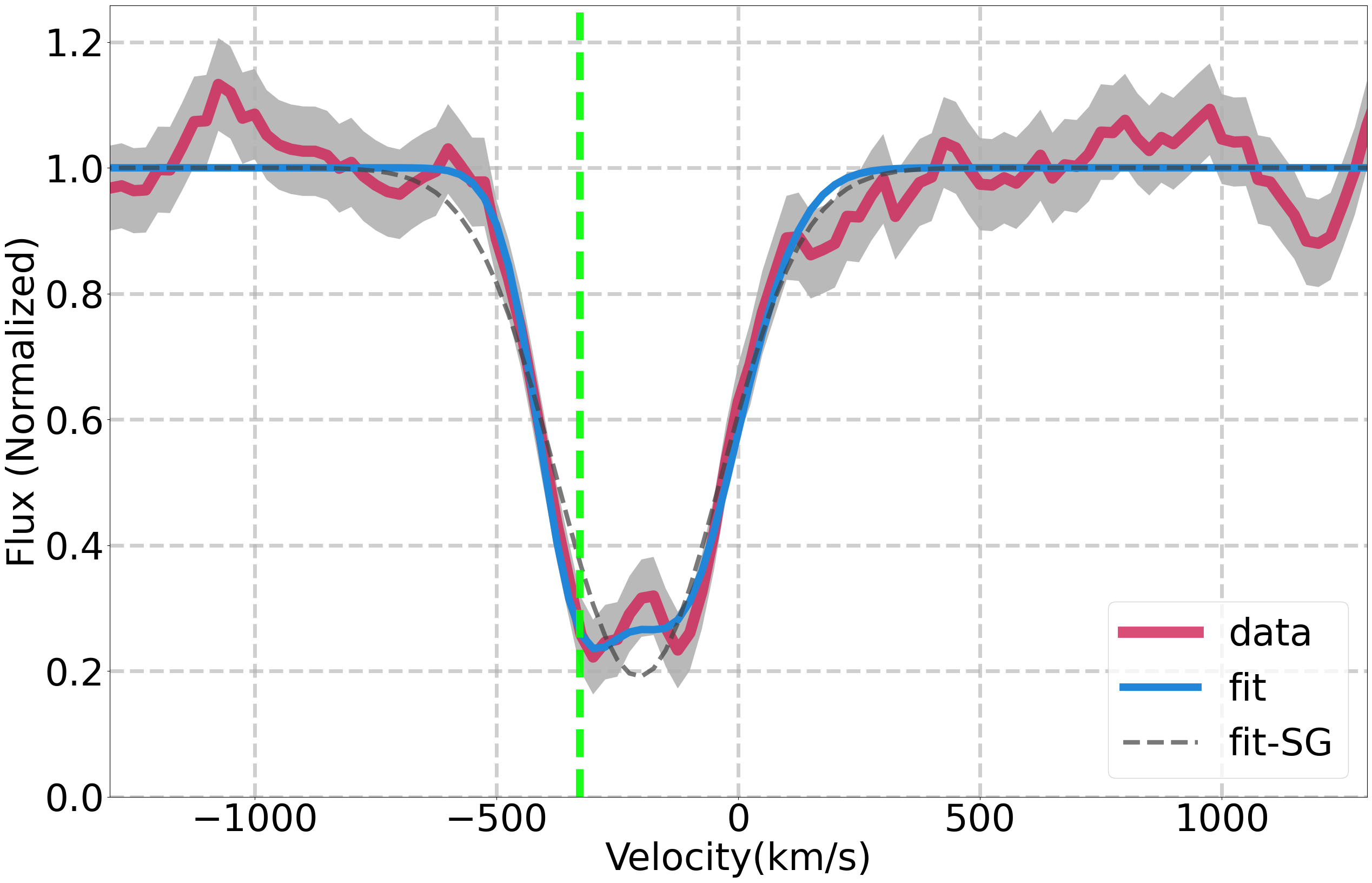}
  \end{minipage}
  \hfill 
  \begin{minipage}[c]{0.45\textwidth}
     \centering
    \includegraphics[width=\textwidth,height=1.2\textwidth]{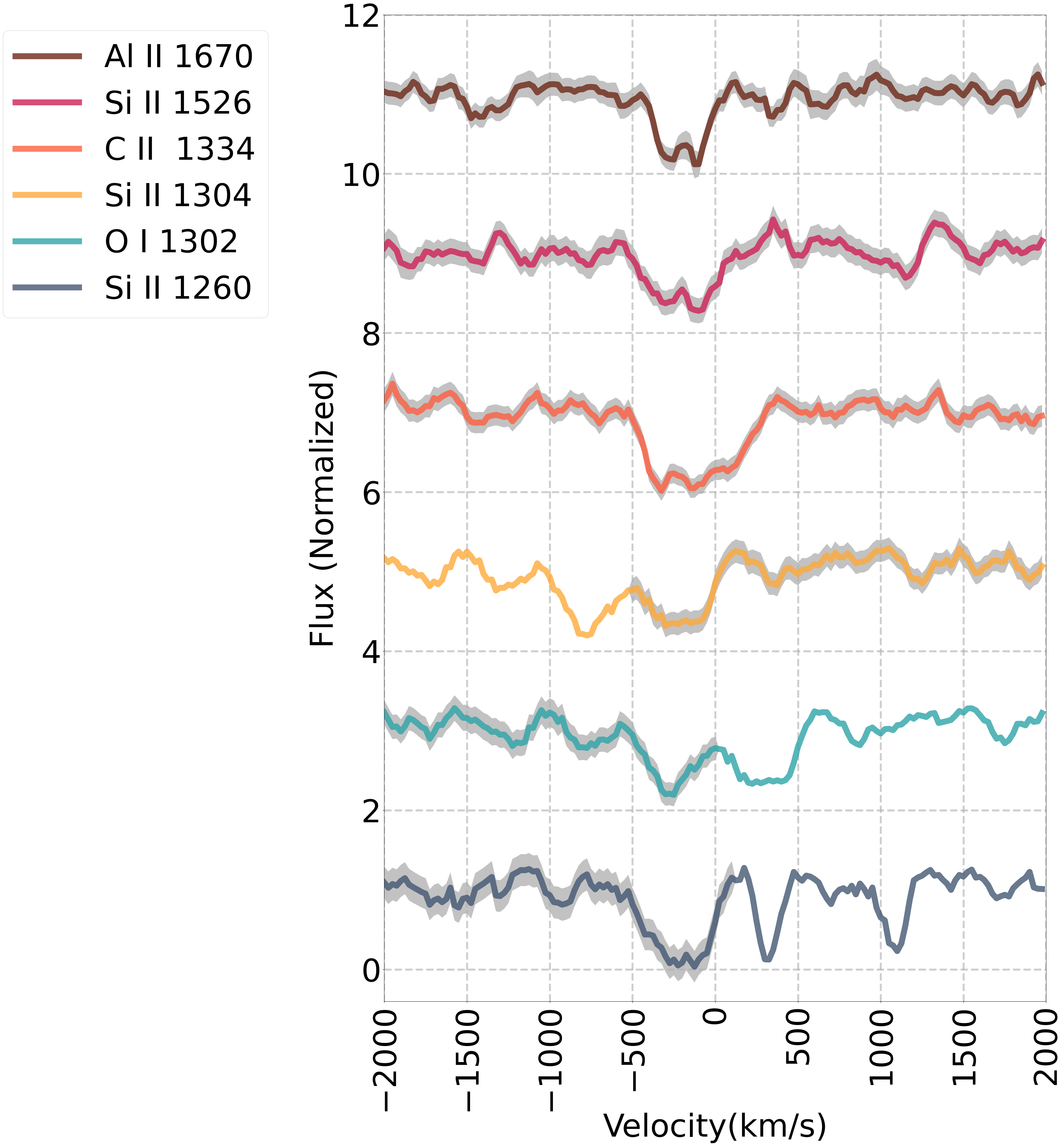}
  \end{minipage}
  \caption{
  \emph{Top:} Example normalized spectrum of one of the lensed galaxies in our sample, CSWA38. Prominent low ionization and high ionization ISM lines from the source galaxy at $z=2.92$ are marked in blue and purple dashed lines. \ion{Mg}{2} absorption from an intervening absorber galaxy at $z=0.77$ \citep{kris2021} is marked in violet. 
  \emph{Bottom Left:}
  Mean absorption profile of gas as a function of velocity obtained from a weighted average of the low-ionization \ion{Si}{2}~$\lambda$1260, \ion{O}{1}~$\lambda$1302, \ion{Si}{2}~$\lambda$1304, \ion{C}{2}~$\lambda$1334, \ion{Si}{2}~$\lambda$1526, and \ion{Al}{2}~$\lambda$1670 lines. The mean absorption profile is related to the covering fraction as 1-$C_f(v)$ (Equation \ref{eq:psi}). The gray shaded regions represent the error spectrum. The blue line is a double Gaussian fit to the data, showing good agreement, while the dashed line is a best-fit single Gaussian (SG) profile. The green vertical line indicates the outflow velocity parameter $v_{75,V2}$ defined as the velocity (in \kms) at 75\% absorption considering only absorption with $v<0$ (see Figure~\ref{fig:velocity-definitions}).
  \emph{Bottom Right:} Spectra of low-ionization lines used to obtain the covering fraction profile. Regions which have no error bars (gray shading) are not included for the weighted average. In these cases the regions are excluded due to absorption from an intervening galaxy (e.g., $>200 \kms$ in \ion{Si}{2}~$\lambda$1260) or the blended nature of the \ion{O}{1}~$\lambda$1302, \ion{Si}{2}~$\lambda$1304 lines. 
  }\label{fig:example-coveringfrac-plot}
 \end{figure*}

\begin{figure*}[ht!]
\centerline{
\includegraphics[width=\linewidth]{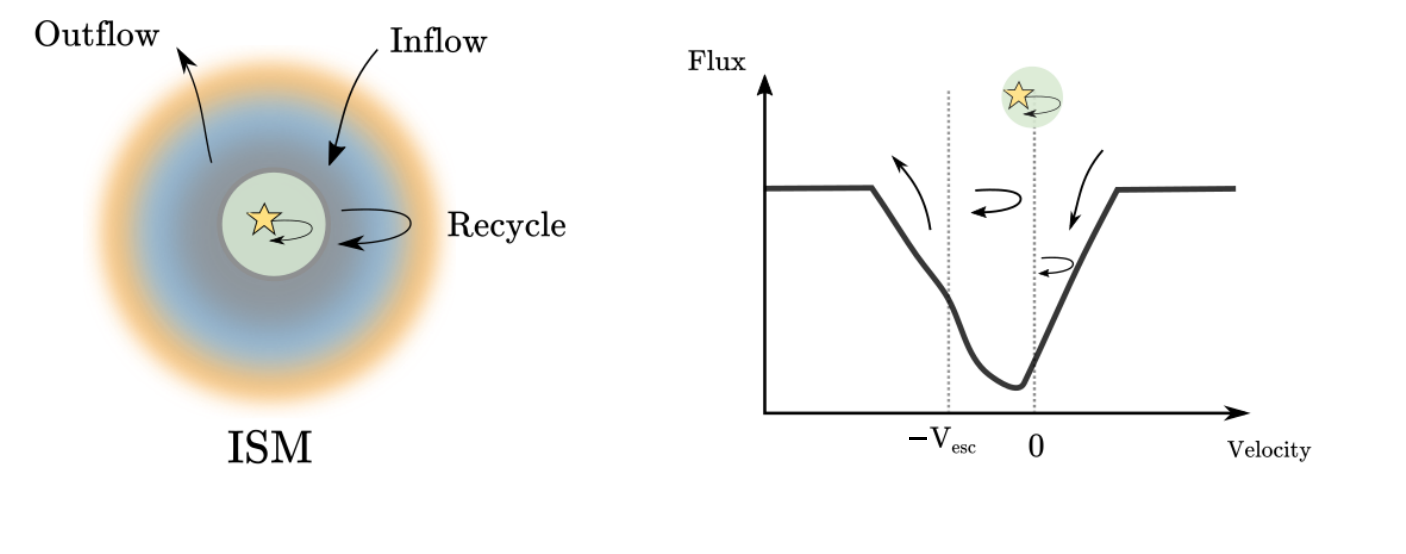}
}
\caption{A guide to interpreting ISM velocity profiles in terms of the baryon cycle, shown as a schematic on the left with corresponding ISM absorption signatures on the right. Outflowing gas has blueshifted absorption (i.e. $v<0$) whereas inflowing gas has redshifted absorption ($v>0$). Recycling gas which arises from the outflowing gas transitioning to inflowing gas has velocities $v \approx 0$. The systemic velocity of the stars and their dispersion is centered at $v=0$ by definition. The region $v<-v_{esc}$ corresponds to gas with velocities greater than the escape velocity of the ISM. 
}\label{fig:baryon-cycle-model}
\end{figure*}

\subsection{Continuum normalization}\label{subsec:cont-norm}

In this work, we are interested in the strength of interstellar absorption relative to the stellar continuum. In order to achieve a constant continuum level around the ISM lines, the spectra from all targets are initially normalized by a running median of 2001 pixels ($\sim$20,000~\kms) which removes any large scale structures in the spectra arising from effects such as dust attenuation, flux calibration, and flat fielding uncertainties (e.g., Figure~\ref{fig:example-coveringfrac-plot} - top). To remove any local scale structures, we consider a region spanning $-$2000 to 2000 \kms\ around the ISM line of interest and divide it by the median value of the local region. We then average the ISM lines (Section~\ref{subsec:lowion-kinematics}) and divide by a third order polynomial fit to the continuum around the absorption profile, to account for any residual structure. This achieves a continuum level close to 1 for the mean absorption profile in all target galaxies (e.g.,  Figure~\ref{fig:example-coveringfrac-plot}). Any absorption can then be interpreted as gas present along the line-of-sight in front of this continuum starlight.

We note that using a third order polynomial normalization increases the $\Delta v_{90}$ line widths by $48 \pm 36$ \kms\ on average compared to using only a median normalization, although there is no effect on the centroid ($-3 \pm 7$ \kms). Additionally, we estimate the typical uncertainty in continuum level using the third order polynomial normalization to be approximately $\sim$1\%, which propagates to a $\sim$3\% average change in the width of ISM absorption as parameterized by $\Delta v_{90}$ (Section~\ref{sec:ism-features}) or similar quantities, while velocity centroids remain consistent within the statistical uncertainties. The effect is such that an underestimated continuum implies an underestimated $\Delta v_{90}$ and absorption equivalent width from best-fit profiles. This uncertainty does not significantly affect the main results and conclusions presented herein.

The normalization procedure described here is relatively insensitive to the ISM line itself. 
In some cases the lensed galaxy spectra are subject to blending with the deflector light due to the nature of the observations, especially with \verb!AGEL! and \verb!MEGASAURA! data. This can affect the relative depth and equivalent width of ISM absorption profiles. However, the kinematic measurements used in this work are robust to blending with other sources, provided they have smooth continuum spectra. Any strong spectral features which interfere with the ISM lines of interest are masked out and not used in our analysis. In some cases, there are strong intervening absorption systems at lower redshift, which are likewise masked and not used in this analysis.

\begin{figure*}[ht!]
\centerline{
\includegraphics[width=\linewidth]{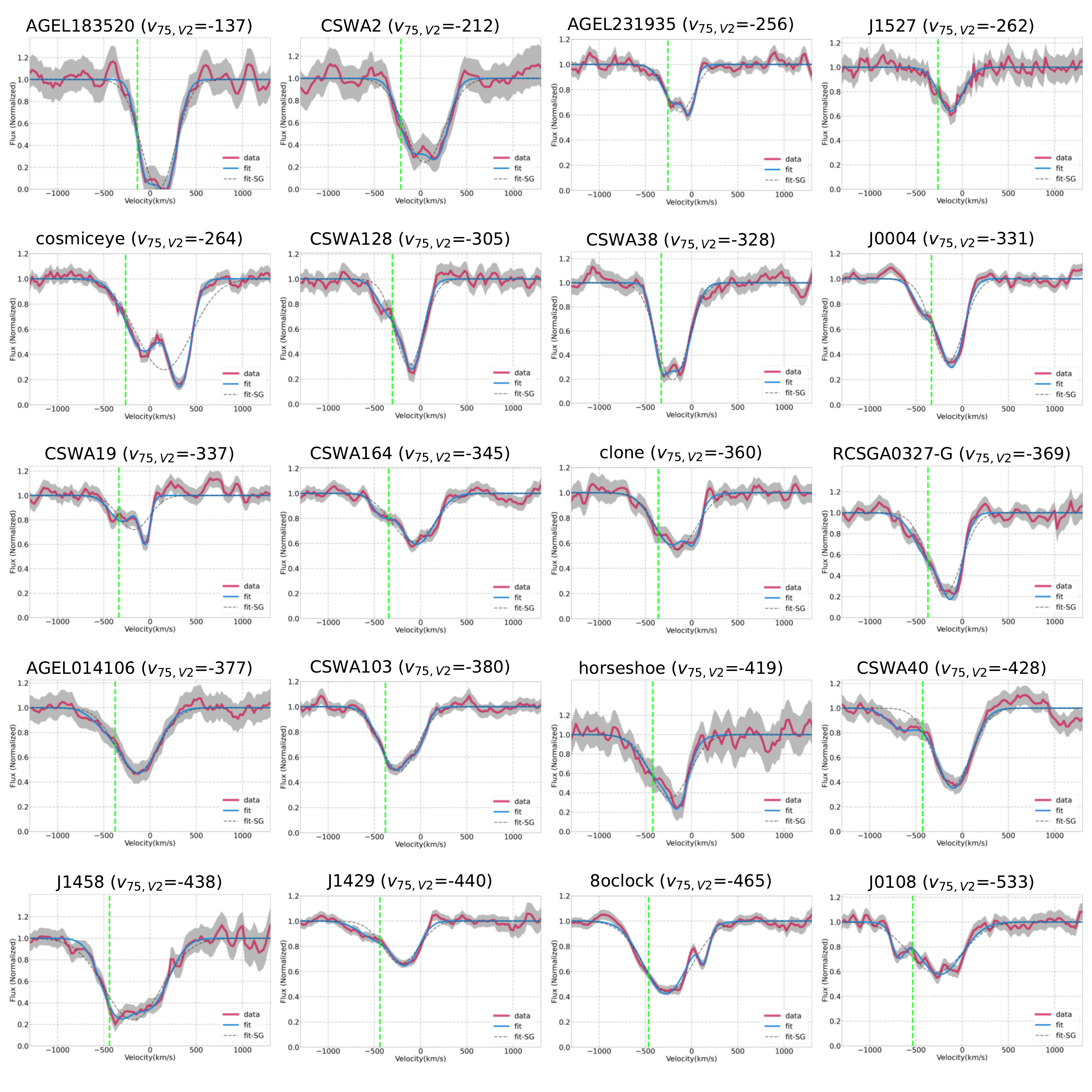}
}
\caption{Plots of ISM absorption profiles for the full sample, sorted by increasing values of the outflow velocity parameter $v_{75,V2}$ (given in \kms). The normalized flux profiles are related to covering fraction as $1-C_f(v)$, and $v_{75,V2}$ is the 75\% percentile of absorption measured by considering gas only at $v<0$, where $v=0$ is the systemic velocity  (see Table.~\ref{tab:velocity-measurements} and Section.~\ref{sec:ism-features} for more details). Denoted in red is the observed velocity profile, gray regions show the $2\sigma$ confidence interval, blue lines are double Gaussian (DG) fits to the data, and dotted lines are single Gaussian (SG) fits. The green vertical lines indicate the measured $v_{75,V2}$ in each case with the value (in \kms) given above each plot.}\label{fig:velocity-profiles}
\end{figure*}

\section{Velocity structure of ISM gas}\label{sec:vel-structure}

Ultraviolet ISM absorption lines probe the velocity structure of gas seen along the line of sight toward (``in front of'') the young stars in a galaxy. Spectrally resolving the absorption velocity profile is a practical and powerful way to probe the baryon cycle, as illustrated schematically in Figure~\ref{fig:baryon-cycle-model}. 
Interstellar gas within the galaxy will absorb at the systemic redshift (i.e., $v=0$) with a velocity range set by the galactic rotation curve and velocity dispersion. 
Inflowing gas gives rise to redshifted absorption (at $v>0$), while outflows result in blueshifted absorption ($v<0$) which may even exceed the escape velocity. 
Recycling gas -- which transitions from outflowing to inflowing at moderately low velocity -- would result in absorption near $v\approx0$. 

In this section we describe our methodology to determine spectrally resolved ISM absorption profiles, in order to characterize the gas kinematics and geometric covering fractions in our sample.
The observed intensity $I$ for an interstellar absorption line is 
\begin{equation}
\frac{I(v)}{I_0} = 1 - \Psi(v), 
\end{equation}
where $I_0$ is the intensity of stellar continuum and $\Psi(v)$ describes the absorption depth as a function of velocity $v$. It is dependent on the covering fraction of gas $C_f(v)$ and the optical depth $\tau$ in the following way:
\begin{equation}
\label{eq:psi}
    \Psi(v) = C_f(v)(1-e^{-\tau}) \approx C_f(v)
\end{equation}
where the latter approximation is valid for the case of optically thick absorption $\tau \gg 1$. In this work, we are interested in studying the ISM gas kinematics by measuring the covering fraction as a function of velocity $C_f(v)$ from galaxy-integrated slit spectra. 
We describe the ISM absorption profiles and velocity structure for lines of different optical depth in Section~\ref{sec:lines-diff-tau}. The profiles are consistent among the stronger transitions indicating $\tau \gtrsim 1$ around the line center, suggesting that they largely trace the covering fraction. For our analysis we use these strongest lines with $\tau \gtrsim 1$, such that Equation~\ref{eq:psi} is a reasonable approximation. We note that if the gas is not optically thick (e.g., as may be the case at higher velocities), then these represent a lower limit on the covering fraction.

\subsection{Kinematics of the low-ionization gas}\label{subsec:lowion-kinematics}

The rest-frame UV spectra used in this work include interstellar absorption from both low- and high-ionization species, as well as stellar features, Ly$\alpha$ in absorption and/or emission, and other features such as nebular and fine structure emission (see Figure~\ref{fig:example-coveringfrac-plot} for an example). We focus on ISM kinematics of the low-ionization phase, from which there are numerous prominent transitions of \ion{Si}{2}, \ion{O}{1}, \ion{C}{2}, \ion{Al}{2} and \ion{Fe}{2}. These metal ion transitions are often optically thick, approximately tracing the gas covering fraction as a function of velocity (Equation~\ref{eq:psi}). This is in contrast to the \ion{H}{1} Ly$\alpha$ profile which is complicated by resonant emission and damping wings. 

For each spectrum we measure the ISM absorption profile $I/I_0$ from an average of the best available strong low-ion metal lines. We select those with good continuum sensitivity which appear to be saturated (based on multiple lines showing similar absorption profiles). Each absorption line is interpolated to a common velocity grid of 25 \kms\ and we take an inverse-variance weighted mean of the median-continuum-normalized flux at each velocity. Those which are affected by features such as strong sky emission, telluric absorption, bad pixels, or intervening absorption systems are excluded from this analysis. For blended transitions (e.g., \ion{O}{1}~$\lambda$1302 and \ion{Si}{2}~$\lambda$1304), only regions of interest corresponding to the transition are taken into account. Specifically, we use typical velocity ranges $v \lesssim 0$~\kms\ and $v \gtrsim -500$~\kms\ for the $\lambda$1302 and $\lambda$1304 transitions, respectively, similar to the approach of \cite{tucker-dustinthewind}.
Other ISM lines are affected to a lesser extent by blending with weak features such as \SII$\lambda$1259 (blended with \SiII$\lambda$1260), \ion{C}{2}*~$\lambda$1335 (affecting \ion{C}{2}~$\lambda$1334), and stellar photospheric features near the \ion{O}{1}~$\lambda$1302 line. These features and their effects on derived ISM absorption profiles are typically not detected in individual galaxy spectra. We therefore do not mask these regions, effectively treating them as part of the stellar continuum (which is generally full of lines with low equivalent width). From analysis of the high-SNR stacked spectrum, we find that these blends can cause an increase in the measured $\Delta v_{90}$ by up to $\lesssim 70$~\kms\ depending on the lines used. This represents a source of systematic uncertainty in the absorption profiles, with magnitude comparable to uncertainty arising from the continuum normalization (Section~\ref{subsec:cont-norm}).

Figure~\ref{fig:example-coveringfrac-plot} illustrates this process for an example galaxy in the sample, with equivalent figures for the full sample displayed in Appendix~\ref{sec:appendix_sample}. We derive the low-ionization ISM covering fraction profiles $C_f(v)$ from these mean absorption profiles for the galaxies in our sample, shown in Figure~\ref{fig:velocity-profiles}.

\subsection{Fitting $C_f(v)$}\label{sec:fitting-cf}

The mean ISM absorption profiles shown in Figure~\ref{fig:velocity-profiles} encode the key observational results of this paper. From these profiles we can examine the typical outflow velocities, the maximum velocities with substantial gas covering fractions, and diversity within the sample, among other properties. 
For analysis purposes, it is useful to have an analytic form which captures the velocity structure of ISM absorption profiles. For quasar sightlines a Voigt profile is appropriate to describe distinct absorption components, but this is not suitable for galaxy spectra whose profiles represent a large number of interstellar clouds. 

Although we adopt the weighted mean profile measurements shown in Figure~\ref{fig:velocity-profiles} as the ground truth, we also fit two analytic functions to each profile. The first is a single Gaussian (hereafter SG) function of the form: 
\begin{equation}
C_f(v) = A_{sg} \exp[(v-v_{sg})^2/2 \sigma_{sg}^2] 
\end{equation}
where the subscripts indicate a single Gaussian (sg). This does not capture the clear asymmetries seen in most of the sample (Figure~\ref{fig:velocity-profiles}). It is nonetheless instructive since this fit captures the information equivalent of a low-resolution ($R\sim300-1000$) spectrum, in which the absorption would be only marginally resolved. 
The second profile is a double Gaussian (hereafter DG) function of the following form:
\begin{equation}
C_f(v) = A_0 \exp[(v-v_0)^2/2 \sigma_0^2] + A_1 \exp[(v-v_1)^2/2 \sigma_1^2] 
\end{equation}\label{eq:cfv}
where ($A_{0}$, $v_{0}$, $\sigma_{0}$, $A_{1}$, $v_{1}$, $\sigma_{1}$) are the parameters to be fit. We adopt a convention that $v_0 < v_1$. The DG is relatively simple but versatile. We find that it yields a reasonable fit to the velocity substructure detected in our sample. The median residuals of the best fit DG model measured between the velocity range $v_{99}$ and $v_{01}$ are $\sim0.03$ whereas for the SG they are $\sim5\times$ higher. We therefore make use of the DG fits to derive kinematic properties such as the velocity centroid and width (Section~\ref{sec:kinematics-observed}). The SG fits are used mainly as an emulator of lower spectral resolution data.

To quantify the uncertainty in each parameter, we fit each weighted mean absorption profile with 250 realizations of the Basin-Hopping stochastic algorithm \citep{basinhopping}. For each realization we add random $1\sigma$ noise to $C_f(v)$ based on the error spectrum. The velocity centroids ($v_{sg},v_0,v_1$) are allowed to vary from $-700$ to 500 \kms, dispersions ($\sigma_{sg},\sigma,\sigma_1$) from 50 to 700 \kms, and absorption depth ($A_{sg},A_0,A_1$) from 0.1 to 1. For each realization, all parameters are initialized to random values within the above ranges. These bounds are chosen based on the observed covering fraction profiles such that they sample the entire parameter space. We place an additional constraint $-800~\mbox{km~s}^{-1}<(v_0-v_1)<0~\mbox{km~s}^{-1}$ when fitting the double Gaussian. This ensures that the same component ($v_0$) always captures the blueward absorption, which we will generally attribute to outflowing gas. We note that this approach is somewhat more general than that of e.g. \citet{boordoloi_rcsga} in which one component's centroid is fixed to represent the systemic component; here we do not require any component to exactly trace the systemic velocity.

\begin{figure*}[!htb]

   \begin{minipage}[c]{0.49\textwidth}
     \centering
    \includegraphics[width=\textwidth]{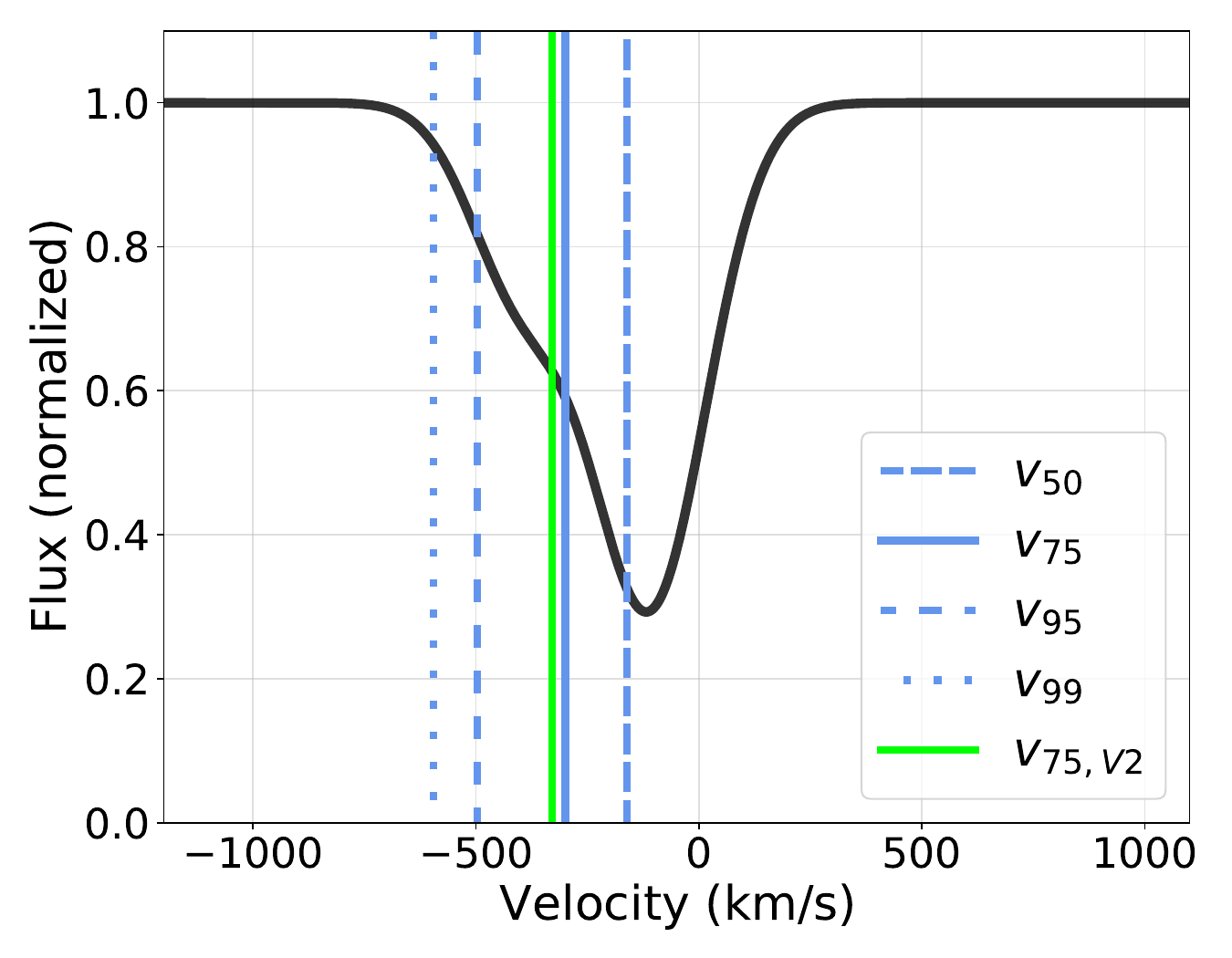}
  \end{minipage}
  \hfill 
  \begin{minipage}[c]{0.49\textwidth}
     \centering
    \includegraphics[width=\textwidth]{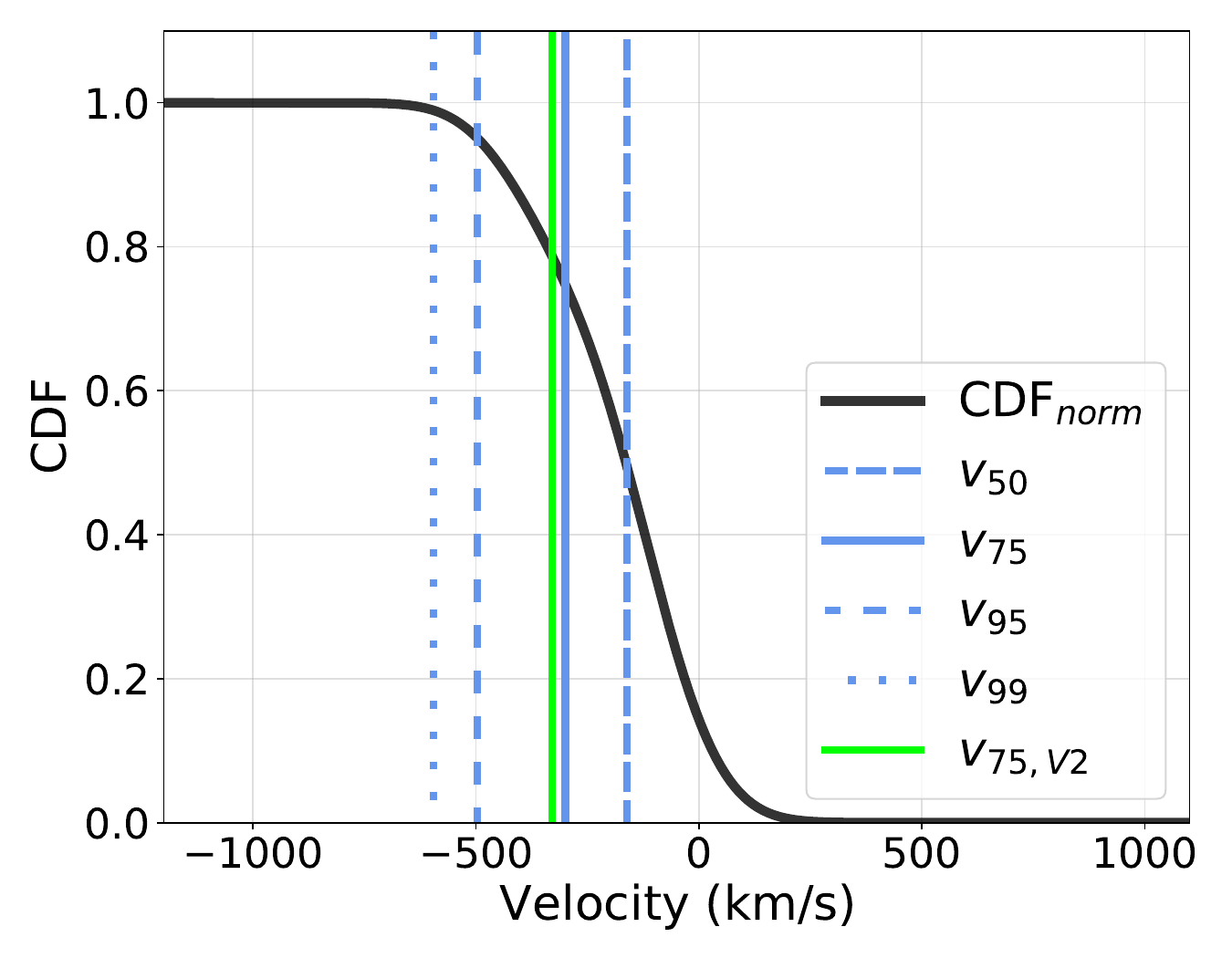}
  \end{minipage}

    \caption{Illustration of the velocity measurements used in this work.
    \emph{Left:} An example normalized flux profile analogous to those in Figure~\ref{fig:velocity-profiles}. \emph{Right:} The normalized Cumulative Distribution Function (CDF) of the absorption profile shown in the left panel. This is essentially the CDF of the covering fraction $C_f(v)$ (Equation~\ref{eq:cfv}).
    Both panels show velocities $v_{50}, v_{75}, v_{95}$, and $v_{99}$ which are defined as the 50\%, 75\%, 95\% and 99\% percentiles of absorption (as seen from the CDF). A subscript of V2 (e.g., $v_{75,V2}$ shown in green) indicates that only $v<0$ was considered; essentially the CDF is normalized to zero at $v=0$ for this case.  Table~\ref{tab:velocity-measurements} lists the entire set of quantities used.}
    \label{fig:velocity-definitions}

\end{figure*}
\begin{deluxetable*}{|C|C|c|}[!htb]
    \tablecaption{Definitions of velocity measurements presented in this paper. We note that $v_{50}$ is the centroid velocity and is used interchangeably with $v_{cent}$.  \label{tab:velocity-measurements} }
    \tablewidth{0.5\textwidth}
    \tabletypesize{\footnotesize}
    % Header %
    \tablehead{ 
     \colhead{Parameter-DG} &  \colhead{Parameter-SG} &\colhead{Description}  
    }
    %data %
    \startdata
    v_{05} &  v_{05,SG} & Velocity at 5\% absorption \\
    v_{50} = v_{cent} &  v_{50,SG} = v_{cent,SG} & Velocity at 50\% absorption \\
    v_{90} &  v_{90,SG} & Velocity at 90\% absorption \\
    v_{95} &  v_{95,SG} & Velocity at 95\% absorption \\
    v_{99} &  v_{99,SG} & Velocity at 99\% absorption \\
    \Delta v_{90} & \Delta v_{90,SG} & $v_{95} - v_{05}$ \\ \tableline 
    v_{05,V2} &  v_{05,SG,V2} & These quantities are \\
    v_{50,V2} &  v_{50,SG,V2} & calculated  in the same   \\
    v_{90,V2} &  v_{90,SG,V2} & way as described   \\
    v_{95,V2} &  v_{95,SG,V2} & above but considering \\
    v_{99,V2} &  v_{99,SG,V2} & only absorption  with $v<0$ 
    \enddata
\end{deluxetable*}

The median fit values obtained at the end of 250 realizations are used to estimate all velocity measurements used in this paper. The $1\sigma$ standard deviation of each quantity is calculated as $\sigma \approx \text{MAD}/0.675$ where MAD = Median Absolute Deviation. Unlike the mean and standard deviation which are easily affected by spurious outliers, the median value and MAD offer better quantifiable values to describe the fits. The resulting best-fit profiles are plotted in Figure~\ref{fig:velocity-profiles} along with the data and observational uncertainties. Tables~\ref{tab:single-gaussian-fit} and \ref{tab:double-gaussian-fit} in Appendix~\ref{sec:appendix_sample} list fit parameters obtained for each of the targets along with the derived uncertainties. 

In all targets we find that the DG fits are able to capture the broad asymmetric wings which are ubiquitously present in the absorption profiles. In addition, they also accommodate complex absorption profiles such as the Cosmic Eye which includes a strong redshifted component. The performance of the SG fits on the other hand varies heavily depending on the asymmetry of the profile. In some cases (e.g., CSWA103), they provide reasonably good fits whereas in more asymmetric cases (e.g., CSWA128) there are large residuals, especially at high velocities. Encouragingly, the residuals obtained for the DG fits are consistently centered around 0 with the standard deviation being generally compatible with the signal-to-noise of each spectrum, indicating a reasonable fit to the data.

\section{Kinematic Features of the Gaseous ISM at Cosmic Noon}\label{sec:ism-features}

Having obtained a covering fraction profile including parametric fits for each galaxy, we now explore kinematic properties of the sample. We measure various standard quantities to facilitate comparison of these moderate resolution down-the-barrel results with other probes (including quasars, low resolution galaxy spectra, and theoretical simulations). 
To best compare with the literature we adopt two parallel lines of analysis: (a) considering the entire covering fraction profile; and (b) considering only the absorption at $v<0$ (i.e., blueshifted) which we denote with a $_{V2}$ subscript. The latter is useful in comparison with theoretical studies which consider only outflowing gas. However we note that the $v<0$ absorption still includes approximately half of the systemic component. 
 
For both analyses we measure the velocity corresponding to the percentiles $5\%$, $75\%$, $95\%$, $99\%$, and $50\%$ of absorption (denoted as $v_{05}$, $v_{90}$, etc.), larger percentiles being more blueshifted. Here $v_{50} = v_{cent}$ is the velocity centroid ($v_{50}$ and $v_{cent}$ are interchangeable). We also measure the velocity width $\Delta v_{90} = v_{05} - v_{95}$, spanning the 5--95 percentile of absorption.
Table~\ref{tab:velocity-measurements} lists all quantities used in our analysis, and we illustrate some of these for an example velocity profile in Figure~\ref{fig:velocity-definitions}. All quantities are calculated for both the SG and DG fits, with results given in Table~\ref{tab:derived-velocity-vals}.
These quantities have been found useful to describe the kinematics in observational and simulation studies in the literature, and we adopt the same conventions for ease of comparison.

\subsection{Bulk outflow motion of ISM gas }\label{sec:kinematics-observed}

A visual inspection of the global covering fraction profiles obtained in Section~\ref{sec:vel-structure} and Figure~\ref{fig:velocity-profiles} indicates that the bulk motion of the gas in the ISM is outflowing, with blushifted velocity centroids ($v_{cent} < 0$). Quantifying the kinematic properties is an important step towards understanding the feedback processes which drive these outflows and impact the host galaxy evolution. In this section we probe the outflow velocity structure quantitatively in terms of $v_{cent}$, $v_{75}$, and $v_{95}$.

\begin{figure}[!ht]

    \centerline{
        \includegraphics[width=0.97\columnwidth]{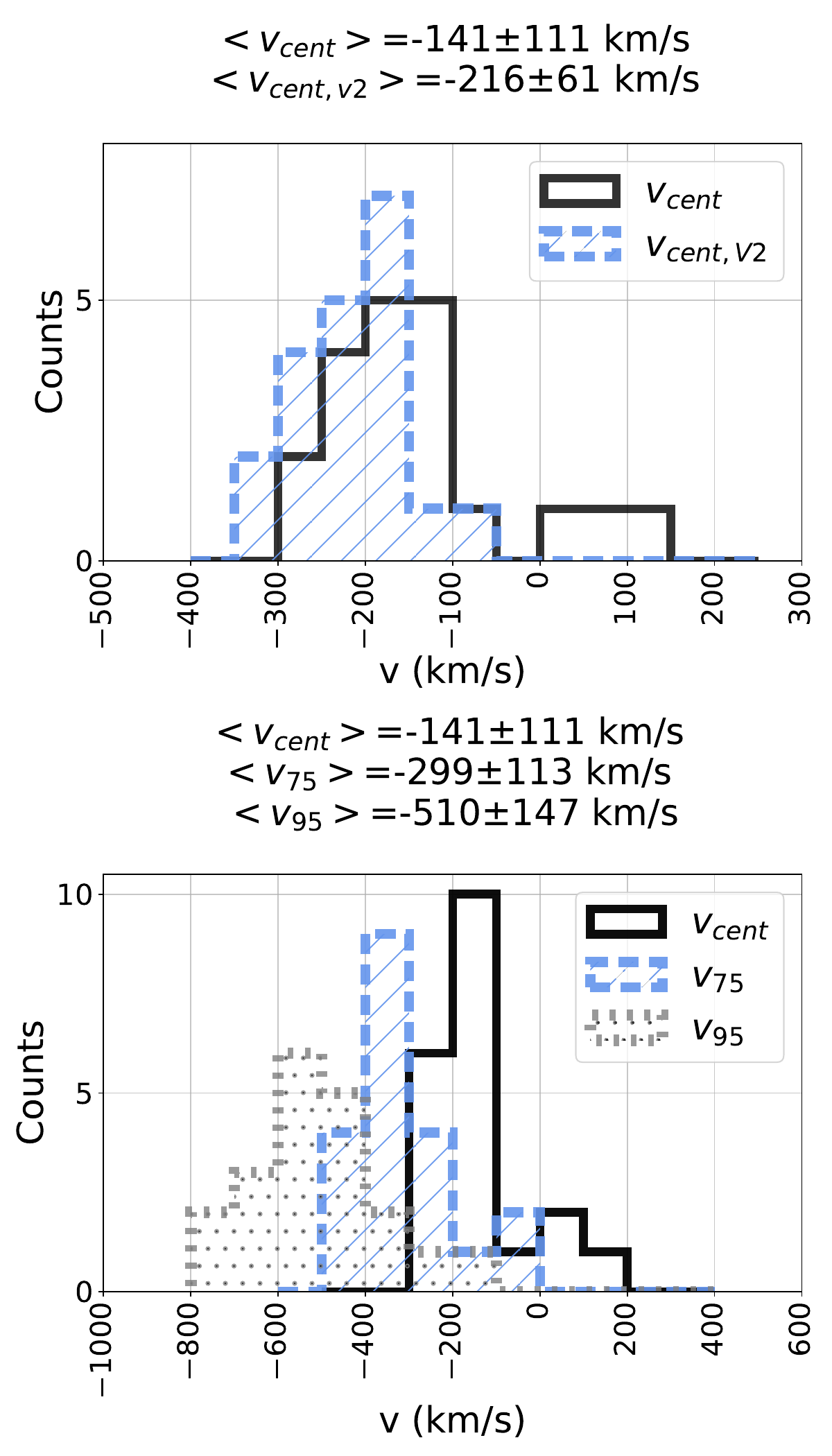}
        }
    \caption{\emph{Top:} Histogram of the absorption velocity centroid $v_{cent}$ for the sample measured by considering (i) the entire profile (solid), and (ii) only $v<0$ \kms\ (dashed). 
    \emph{Bottom:} Histogram of $v_{cent}$, $v_{75}$, and $v_{95}$ values for all targets in the sample. 
    The mean and sample standard deviation of each quantity are given above the plots. 
    The aggregate sample shows typical centroids blueshifted by $\sim$150~\kms\ relative to the systemic velocity, with significant absorption extending to outflow velocities of 300--500~\kms\ or more. 
    }
    \label{fig:centroid-histogram}

\end{figure}

The centroid $v_{cent}$ gives a measure of typical outflow velocities, which can readily be compared with other samples. The median absorption centroid and its sample standard deviation for the galaxies in the sample is $-141 \pm 111$ \kms\ for the full profiles, and $-216 \pm 61$ \kms\ if we consider only the velocities $v<0$. This latter number is a lower limit to the purely outflowing gas component (as opposed to the total including systemic interstellar absorption). Figure~\ref{fig:centroid-histogram} illustrates the histogram obtained for both these metrics (as listed in Table~\ref{tab:derived-velocity-vals}). These values are similar to measurements from larger samples of $z\simeq2$--3 galaxies at lower spectral resolution (e.g., $-168 \pm 16$~\kms\ from \citealt{steidel2010} compared with our sample median $-141\pm25$~\kms). The covering fraction at $v_{cent,V2}$ ranges from 18-95\% for galaxies in this sample with a median of 50\%, suggesting a patchy covering fraction of the outflowing gas with substantial variations within the sample. 

The $v_{75}$ and $v_{95}$ values probe the high-velocity blueshifted tail of outflowing gas. The distributions of these values for the lensed sample are also shown in Figure~\ref{fig:centroid-histogram} (lower panel). Compared to the centroid velocity ($v_{cent} = v_{50}$), we find that the median $v_{75} \approx 2 \times v_{cent}$ and $v_{95} \gtrsim 3 \times v_{cent}$. Thus we see clear signatures of outflows at $>$3 times the centroid velocity, with absolute $v_{95}$ typically extending beyond 450~\kms, although the covering fraction is smaller at larger absolute velocity.

\subsection{Quantifying the asymmetry in absorption}\label{sec:skewness}
\begin{figure}[!htb]

    \centerline{
        \includegraphics[width=\columnwidth]{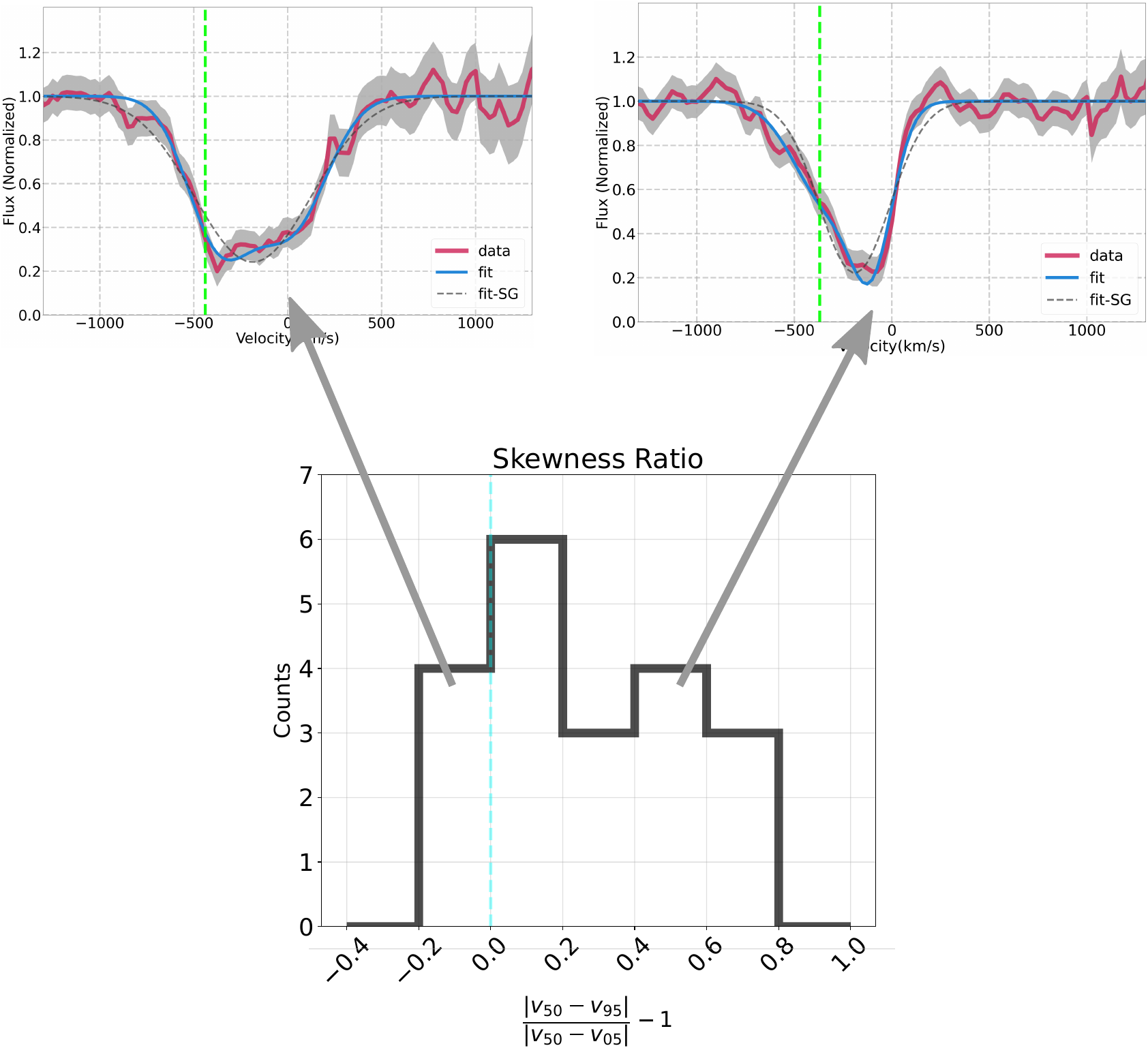}
        }
    \caption{Histogram of skewness ratios. A value of 0 (denoted by cyan dashed line) indicates no skewness between the blue and redward absorptions, i.e., absorption which is symmetric about the centroid. 
    A skewness value $>0$ indicates that the slope of the blueward absorption is shallower than the redward absorption, while values $<0$ correspond to shallower redward slopes. Examples of positively and negatively skewed profiles are shown at the top. 
    Nearly all galaxies in the sample show positive skewness, indicating asymmetric profiles with a broad tail of blueshifted absorption from outflowing gas. }
    \label{fig:skewness}

\end{figure}

Another significant visual feature of the covering fraction profiles is the asymmetry. The quantities $|v_{50}-v_{05}|$ and $|v_{50}-v_{95}|$ trace the extent of the gas present redward and blueward of the bulk outflowing gas velocity. The median $|v_{50}-v_{05}|$ and $|v_{50}-v_{95}|$ measured with the DG are 292 \kms\ ($\approx 1.7 \times |v_{50}|$) and 
357 \kms\ ($\approx 2.1 \times |v_{50}|$) respectively, indicating a clear skewness on average with a shallower slope for the blueshifted velocity range. In comparison, a SG fit gives 340 \kms\ ($\approx 2 \times |v_{50}|$) for the same quantities, which are identical by symmetry of the single Gaussian. Figure \ref{fig:skewness} plots a histogram of the skewness ratio defined as 
\begin{equation}
    \text{Skewness Ratio} = \frac{|v_{50}-v_{95}|}{|v_{50}-v_{05}|} - 1 
\end{equation}
where a positive Skewness Ratio indicates that the blue wing is more extended than the red wing.

16 out of the 20 galaxies in our sample have positive skewness (i.e. Skewness Ratio $>0$). Looking at the covering fraction profiles of galaxies which have Skewness ratio $<0$ (e.g., J1458), one can clearly see that they have an inverted skewed profile wherein the redshifted side has a shallower slope (e.g., Figure \ref{fig:skewness}) which can give rise to a negative skewness (i.e. $|v_{50}-v_{95}| < |v_{05}-v_{50}|$). The origin of this skewness in the profile is an interesting but challenging question which we do not tackle in this paper, but in a simplistic sense, the different skewness ratios could be interpreted as the response of the ISM gas to a galactic wind captured either at different points in time or viewing angles. A key point is that such details about the kinematic structure are not captured by the SG fits (nor by low resolution spectra). This illustrates the need for good spectral resolution to reveal the complex velocity structure of outflowing gas.

\subsection{Width of absorption using $\Delta v_{90}$}
The $\Delta v_{90}$ diagnostic is commonly used in the literature for quasar absorption systems, and is usually defined as the velocity range spanning 5\% to 95\% of the
total column density. However there are some key differences between quasar probes and our measurements.
First, quasars probe the full line-of-sight through a halo
(distances $-\infty$ to $\infty$) whereas our “down-the-barrel”
galaxy spectra sample only half the halo (0 to $\infty$). Our
spectra do not probe the redshifted outflowing gas on the
far side of the galaxy, causing $\Delta v_{90}$ to be smaller than for
a background quasar at impact parameter $b = 0$. Sec
ond, quasars probe a narrow “pencil beam” area which
is prone to stochastic sampling of absorbing gas clouds
\citep[e.g.,][]{marraQuasarsSimsComparision}, whereas our galaxy spectra encompass a much larger cross-sectional area of several
kpc$^2$. Thus we may expect our galaxy spectra to be more representative of the gas covering fraction. Third, the absorption profiles from Section~\ref{sec:vel-structure} are constructed from the strongest ISM lines, which are more sensitive to gas covering fraction as opposed to column density. 
In summary, the $\Delta v_{90}$ values for our sample represent approximately the velocity width of covering fraction profiles through half of the host galaxy halos.

\subsubsection{Kinematics at different optical depths($\tau$)}\label{sec:lines-diff-tau}

\begin{figure}[!ht]
    \centering
    \centerline{
        \includegraphics[width=0.6\columnwidth]{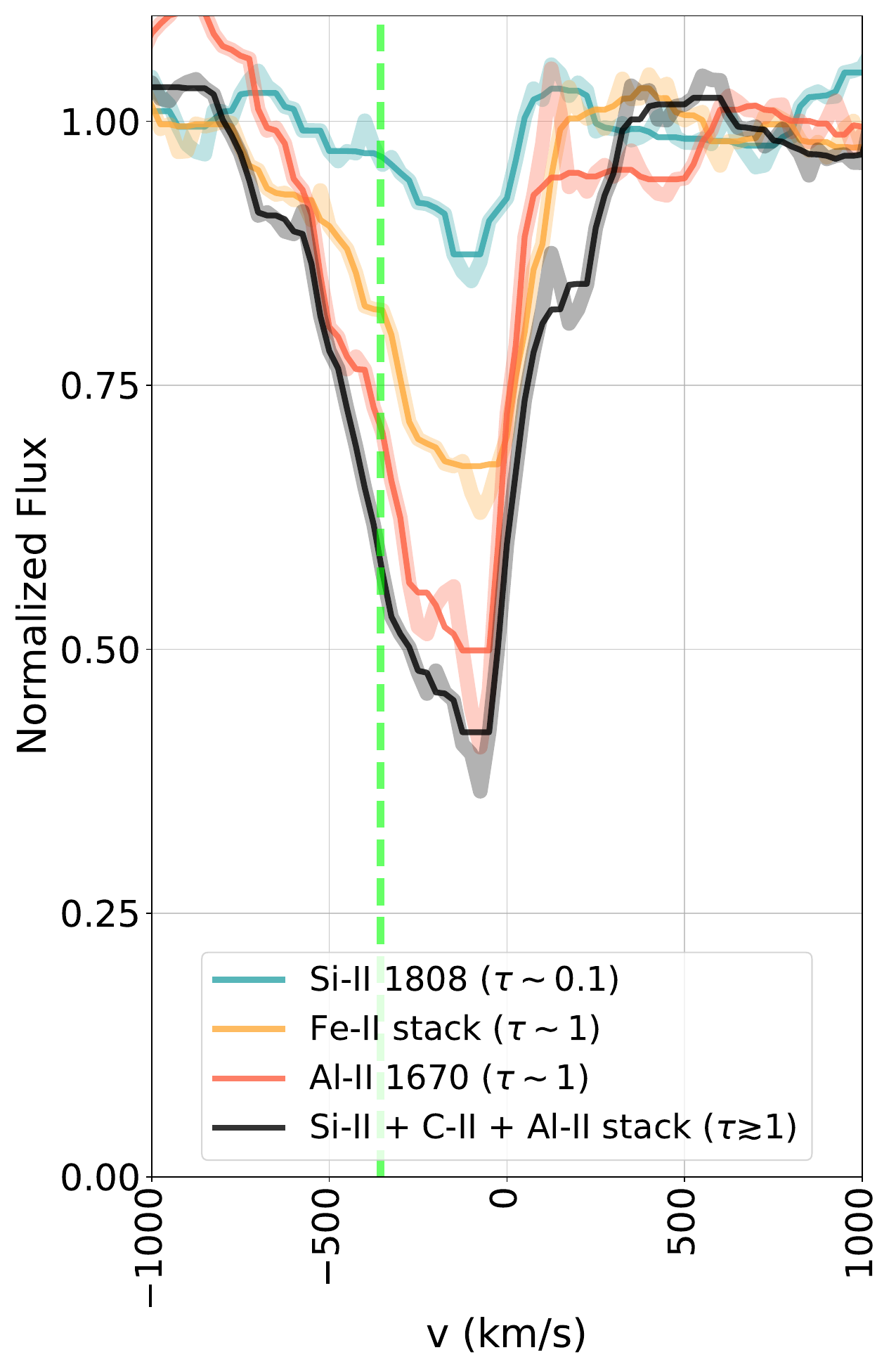}
        }
    \centerline{
        \includegraphics[width=0.85\columnwidth]{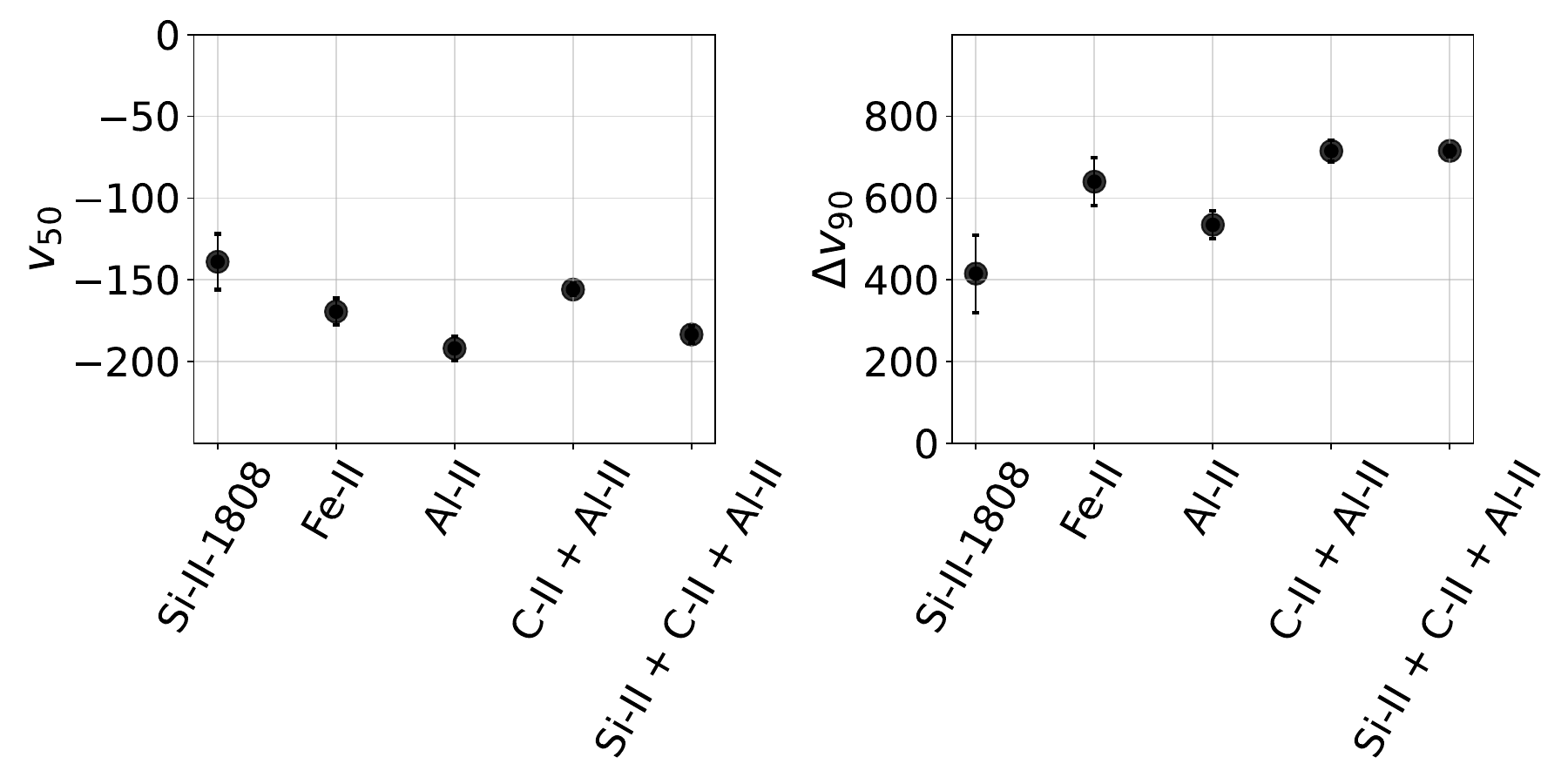}
        }    
        \caption{\emph{Top:} Median stacked absorption profiles for lines with different optical depths ($\tau$) : \ion{Si}{2}~$\lambda1808$ (green), \ion{Fe}{2}~$\lambda$1608 (orange), and \ion{Al}{2}~$\lambda$1670 (red).  A joint stack of the strongest low ion transitions (\ion{Si}{2}, \ion{C}{2}, \ion{Al}{2}) is shown in black. Bold lines are running medians of each profile whereas the lighter shade is the full-resolution data. \ion{Si}{2}~$\lambda1808$ probes optically thin gas ($\tau \ll 1$), while \ion{Fe}{2} and \ion{Al}{2} have apparent optical depths of order unity ($\tau \sim 1$) and the strong transitions have $\tau \gtrsim 1$. The green vertical line is the median $v_{75,V2} = -327$~\kms\ measured from this work shown as a reference point. Notably the profiles of different $\tau$ all show similar mean blueshifted velocities, indicating that velocity centroids are robust to the choice of absorption lines.  \emph{Bottom:} $v_{50}$ (centroid) and $\Delta v_{90}$ measurements for the four profiles shown above, with $\tau$ increasing toward the right. We determine $v_{50}$ and $\Delta v_{90}$ from a DG fit to the absorption profiles. The main results are unchanged if SG fits are used. $v_{cent}$ is approximately constant across the range of optical depths, as can be seen visually in the top panel. In contrast, $\Delta v_{90}$ is smaller for the optically thin lines which trace the bulk of the total column density, although the difference is comparable to the uncertainty. This indicates that high-velocity absorption seen in the strongest transitions is likely a small fraction of the total outflow mass. \label{fig:stacking-lowions}}
\end{figure}

To assess how well the absorption profiles from strong ISM lines trace the column density, we compare them with weaker ISM absorption lines whose apparent optical depth is $\tau \sim 0.1$--1. The low ion velocity profiles are typically constructed from the strongest ISM transitions, 
%\ion{Si}{2}, \ion{Al}{2}, \ion{Fe}{2} and/or \ion{C}{2} 
with $\tau \gtrsim 1$ (see Table~\ref{tab:Lines-used}). We compare these with the \ion{Al}{2}~$\lambda$1670 and \ion{Fe}{2}~$\lambda$1608 lines which are often unsaturated ($\tau \sim 1$), as well as the optically thin ($\tau \ll 1$) transition \ion{Si}{2}~$\lambda1808$. Median velocity profiles for each of these lines are obtained by stacking the spectra from all objects in the sample with the relevant wavelength coverage. Figure~\ref{fig:stacking-lowions} shows a plot of the median stacked profiles for these different ISM absorption features as function of optical depth.

Visually inspecting the profiles reveals a remarkable similarity in the kinematics probed by the different transitions, despite the varying optical depths. 
Assuming that the stack of strong lines traces the covering fraction at $\tau \gg 1$ (Equation~\ref{eq:psi}), the maximum absorption depth suggests $\tau\simeq1.5$ for \ion{Al}{2}~$\lambda$1670, $\tau\simeq0.6$ for \ion{Fe}{2} , and $\tau\simeq0.1$ for \ion{Si}{2}~$\lambda1808$ (supporting an optically thin interpretation). 
The \ion{Si}{2}~$\lambda1808$ profile exhibits blueshifted absorption consistent with the stronger features, although at lower signal-to-noise ratio. We perform a DG fit to the strongest ISM absorption line profiles and \ion{Si}{2}~$\lambda1808$ line to derive the velocity centroid ($v_{cent}$) and $\Delta v_{90}$ values, as described above. Figure~\ref{fig:stacking-lowions} plots these quantities (lower panels). We find the velocity centroid is $\approx -160$~\kms\ for all transitions. The $\Delta v_{90}$ for the stronger low-ion transitions is $\sim 630$ \kms, including for \ion{Fe}{2} which has apparent $\tau<1$, whereas for \ion{Si}{2}~$\lambda1808$ it is $\sim 400\pm100$ \kms. Visually, this difference in $\Delta v_{90}$ between the optically thin and thick lines likely arises from the higher outflow velocity regions, which may be affected by lower $\tau$ in addition to reduced signal-to-noise. Nonetheless the line widths are broadly similar across a range of optical depth, indicating that we can use the $\Delta v_{90}$ measurements obtained from strong low ion transitions to compare with measurements based on optical depth from quasar sightlines, with the caveat that values based on optical depth may be lower by $\sim 250$ \kms.

% \ion{Si}{2}~$\lambda1808$ which has lower signal-to-noise}, and a DG fit to the other profiles.
\subsubsection{Comparison to quasar sightlines}\label{subsec:quasar-sightlines}
\begin{figure}[!htb]
    \centerline{
        \includegraphics[width=\columnwidth]{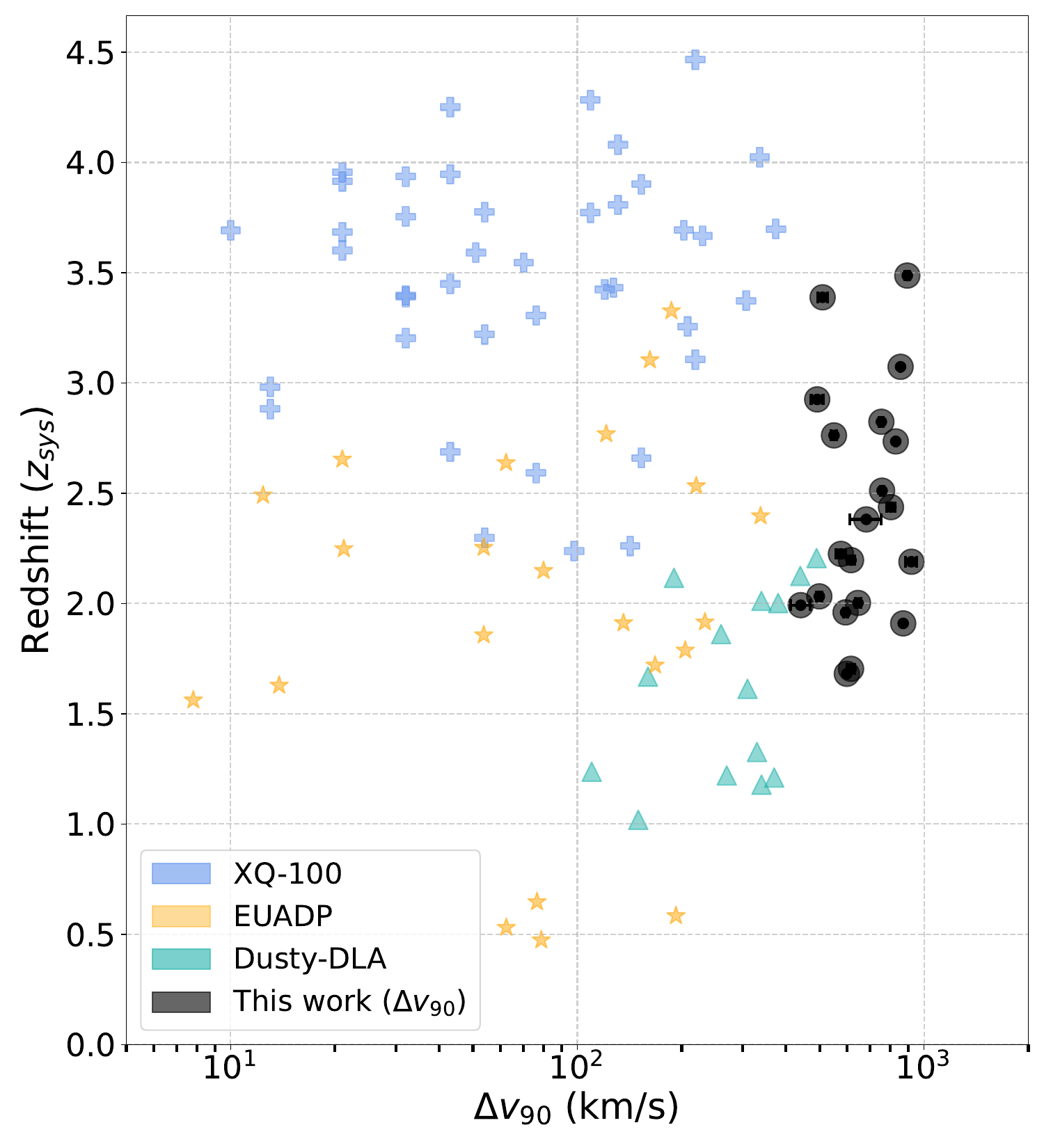}
        }
    \caption{Comparison of $\Delta v_{90}$ measured in this work with those obtained from surveys of background quasar sightlines: XQ-100 \citep{xq100}, EUADP \citep{eudap-sub-damped-lya} and Dusty-DLAs \citep{dustydlas}. Compared to quasar absorber samples, this work probes gas at very low impact parameters ($b=0$) and larger cross-sectional area, while sampling only the foreground region ($R=0\rightarrow\infty$ c.f. background quasars which probe $R=-\infty\rightarrow\infty$). Despite probing only half of the halo, the typical $\Delta v_{90} \simeq 600$~\kms\ for this work is considerably larger than for quasar absorber systems. Only the most extreme quasar systems have comparable $\Delta v_{90}$ values. This suggests that the high-velocity gas which is ubiquitous in our down-the-barrel galaxy sightlines is located at small impact parameters which are extremely rare in quasar samples. DLAs at similar redshifts with known hosts have been found to probe impact parameters $b\lesssim25$ kpc and $\Delta v_{90} \lesssim 350$ \kms\ \citep[Figure~\ref{fig:quasars-impactparameter};][]{DLAhosts2013}.
    } \label{fig:v90-comparision-othersurveys}

\end{figure}

In this subsection, we compare the width of absorption measured using $\Delta v_{90}$ and Equivalent Width (EW) as we step away from ``down-the-barrel'' observations to pencil beam quasar sightlines probing larger impact parameters. Figure \ref{fig:v90-comparision-othersurveys} plots the $\Delta v_{90}$ measurements as a function of redshift ($z$). 
These values are compared with various quasar surveys: XQ-100 \citep{xq100}, EUADP \citep{eudap-sub-damped-lya}, and ``Dusty DLAs'' with $2175$~\AA\ dust attenuation bumps \citep{dustydlas}. The galaxies from this work have $\Delta v_{90}$ values ranging between 440 and 920 \kms\ with a median of 630 \kms. These galaxy values are $\sim 6$ times greater than those observed in the quasar absorption samples, falling near and beyond the largest values seen toward quasars. However, we caution that there are two main caveats in this comparison: (1) $\Delta v_{90}$ for the low ions is likely overestimated by $\sim 250$ \kms\ compared to the optically thin lines (Section~\ref{sec:lines-diff-tau}), and does not separate the systemic interstellar gas from outflowing and inflowing components. (2) $\Delta v_{90}$ for the galaxies probes only one side of the galaxy (along our line-of-sight), such that it is smaller than would be observed toward a background source which would capture the highly redshifted outflowing gas on the far side of the galaxy. Despite these caveats, whose effects are in opposite directions, it is clear that the galaxy absorption profiles span velocity ranges comparable to the largest seen in quasar absorber systems at similar redshifts. 

\begin{figure}[!htb]

       \centerline{ \includegraphics[width=0.78\columnwidth]{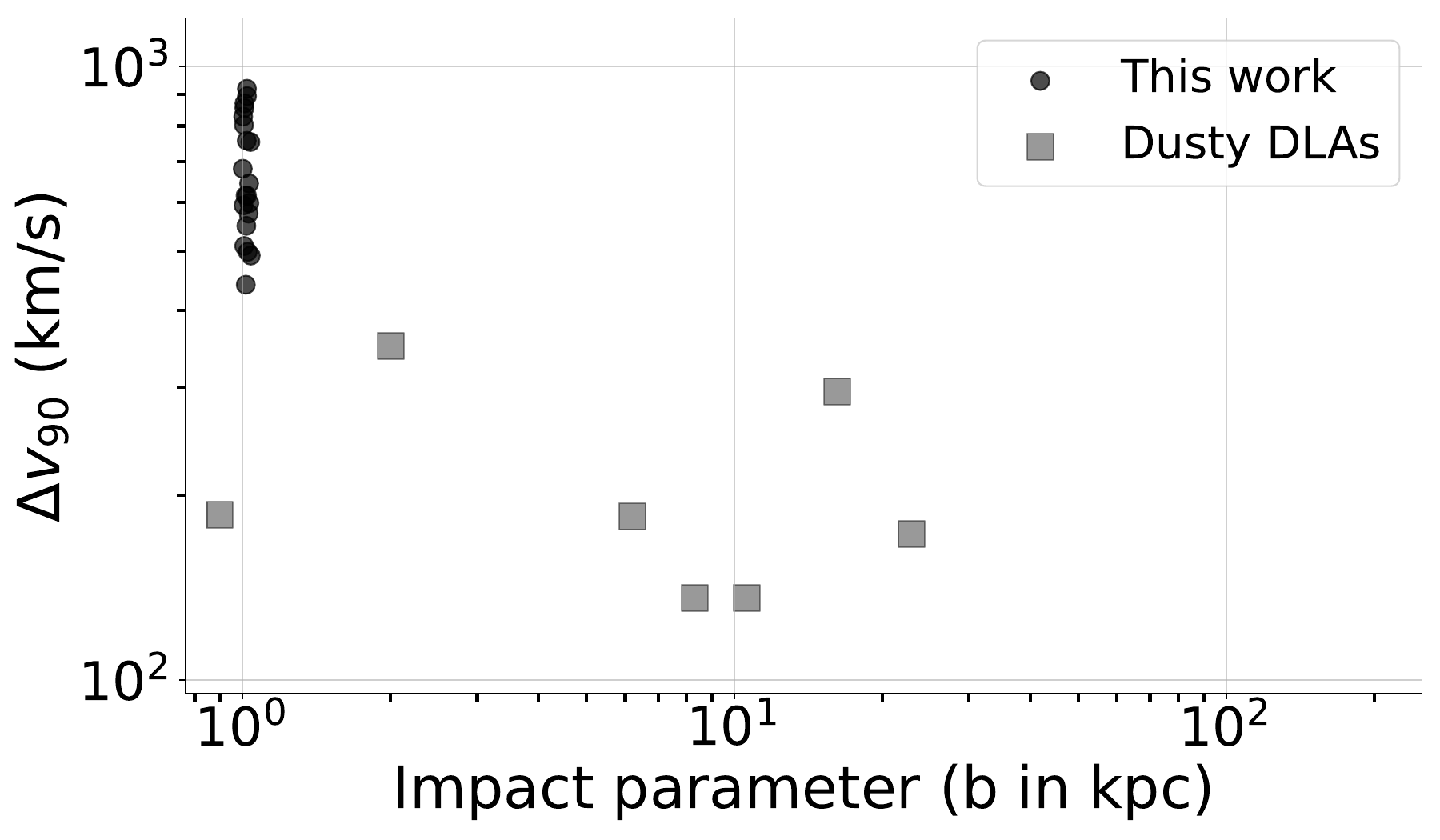}
        }
     \centerline{
        \includegraphics[width=0.93\columnwidth]{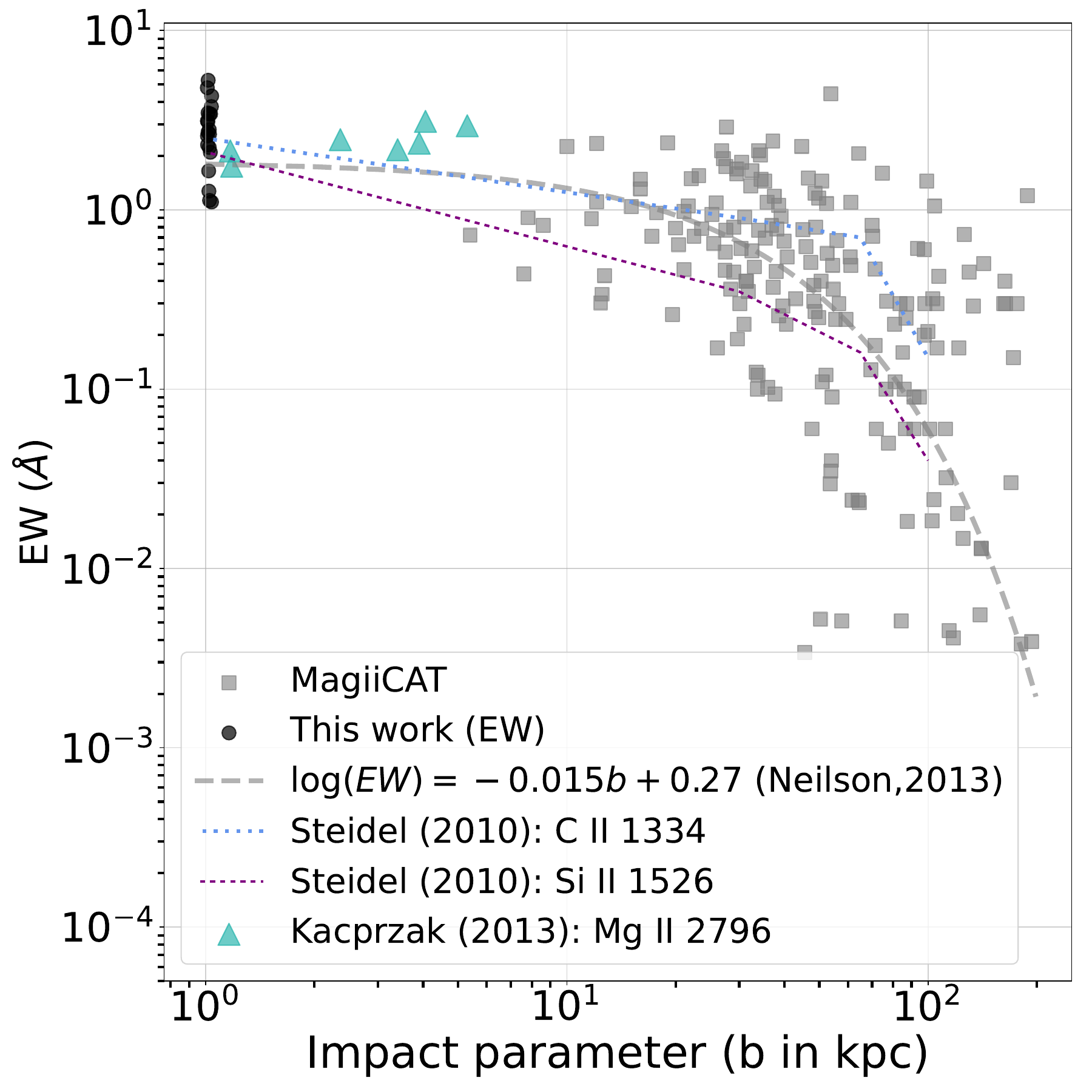}
        }       
        
    \caption{{\emph Top: } Comparison of $\Delta v_{90}$ as a function of impact parameter b for our sample (at b~$=0$ but plotted here at b~$=1$ kpc for clarity), and for DLAs at $z\sim2$ probed by quasar sightlines \citep{DLAhosts2013}. Even at impact parameters $\lesssim 25$ kpc, these DLAs have smaller $\Delta v_{90}$ values than those obtained from ``down-the-barrel'' galaxy observations. {\emph Bottom:} Comparison of absorption equivalent width (EW, in \AA) as a function of impact parameter (b) for our sample and the MAGIICAT survey \citep{NielsonMagiicatFitVals} of \MgII\ absorbers in background quasar sightlines. The average EW clearly drops off by orders of magnitude as we move to higher impact parameters. The dashed line denotes the best fit to the MAGIICAT sample extrapolated to lower impact parameters, and it appears to be consistent with the EWs obtained from this work. We note that the large EW for our sample at small impact parameters is primarily due to the large velocity widths from outflows, whereas large EW for MAGIICAT galaxies might also arise from narrower absorption (smaller $\Delta v_{90}$) with higher covering fractions. EW values obtained from quasar sightlines at low impact parameters \citep[$b\lesssim5$kpc;][]{glennLowImpact2013} and from galaxy-galaxy pairs at $z\sim2$ from \citet{steidel2010} are shown in green points and dotted lines respectively, also showing a similar trend.}
    \label{fig:quasars-impactparameter}
\end{figure}

The large $\Delta v_{90}$ values in our sample are likely driven by gas at smaller impact parameters than probed towards quasars. For most quasar absorbers, the host galaxy position and hence impact parameter is unfortunately unknown. This is particularly true for galaxies at higher redshifts since they are fainter and harder to identify from available imaging. Figure \ref{fig:quasars-impactparameter} (top panel) plots the $\Delta v_{90}$ measurements from this work alongside those obtained from quasar sightlines of a DLA sample with known host location \citep{DLAhosts2013} as a function of impact parameter ($b$). As shown in the figure, these DLAs have lower $\Delta v_{90}$ measurements even at modest impact parameters $b\sim10$ kpc. 


Dedicated surveys such as MAGIICAT \citep{magiicat} and MEGAFLOW \citep{megaflow} have studied quasar absorption associated with known host galaxies. This provides information on trends with impact parameter, although the hosts are at lower redshifts than our sample. 
MAGIICAT galaxies have measurements of equivalent width (EW) of \ion{Mg}{2}, a low-ion species with which we can directly compare. For a subset of our sample which has spectral coverage and good SNR for both \ion{Mg}{2} and shorter-wavelength low ionization lines, we find that the \ion{Mg}{2} profile closely traces the ISM absorption line profiles used in this work, including in the high-velocity wings. Therefore, we convert the low ion covering fraction to an expected EW of \ion{Mg}{2}~$\lambda$2796 as follows:

\begin{equation}
    EW (\text{in \AA}) = \frac{W_{vel}}{c} \lambda_{2796}
\end{equation}
where 
\begin{equation}
    W_{vel} = \sum C_f(v) \Delta v
\end{equation}
and $\lambda_{2796}$ corresponds to the rest frame wavelength of \ion{Mg}{2}. 
This EW estimate assumes $\tau\gtrsim1$ for \ion{Mg}{2}~$\lambda$2796 absorption in our sample, which we expect based on the observed low ions. Here $W_{vel}$ is effectively an equivalent width in units of velocity, calculated by summing the covering fraction profiles.

Figure \ref{fig:quasars-impactparameter} (bottom panel) compares the EW obtained for our sample, MAGIICAT (with typical $z\sim0.4$), and stacks of $z\sim2$ galaxy-galaxy pairs from \citet{steidel2010}. The galaxy-galaxy pairs are a useful comparison since they probe the cross-sectional area of a background galaxy, similar to down-the-barrel spectra. The galaxies from \citet{steidel2010} have similar stellar mass and SFR as our sample, and those from MAGIICAT have similar stellar mass \citep{magiicatVirialmass}.
Our $z\sim2$ sample at $b\simeq0$ kpc spans EW~$=1$--5~\AA, whereas MAGIICAT probes larger impact parameters and has a median EW of 0.43~\AA. The width of absorption drops by orders of magnitude as $b$ increases away from the galaxy. Extrapolating the trend line obtained from \citet{NielsonMagiicatFitVals} for MAGIICAT ($EW \propto 10^{-0.015 b}$) to lower impact parameters provides a good match to our sample average. This result is complimented by the galaxy-galaxy pairs which also show low ion EW decreasing similarly at higher impact parameters.
We can see that the galaxy-pair trend line obtained for \ion{C}{2} 1334 is a better match to the MAGIICAT sample, whereas \ion{Si}{2} 1526 falls below this trend line. This may be due to lower optical depth of \ion{Si}{2} 1526 compared to both \ion{C}{2} 1334 and \ion{Mg}{2} 2796. 

Looking at the maximum EW obtained in the quasars, one can find some values which seem to have comparable EW to the high-z sample. These may be associated with orientation effects where the quasar sightline probes near the minor axis where we would expect outflows, or if the sightline incidentally passes through a fast moving cloud of gas. 
The background galaxy samples are likely to show smaller scatter because of the greater cross-sectional area probed. 
For our sample, the effective area varies for each source and is typically of order half the total cross-section of the galaxy (based on spectroscopic slit placement), or several square kpc (and $\gtrsim 1$~kpc$^2$ in all cases). Arc tomography studies probing $\sim$kpc$^2$ regions have indeed found smaller scatter than observed toward quasar sightlines \citep{kris2021,lopez-nature}. 

In summary, we find that the absorption width (EW and $\Delta v_{90}$) values obtained in this work at $b\approx0$ are higher than those typically observed in quasar sightlines at larger impact parameter. However, extrapolating the trend in quasar absorption ($EW \propto 10^{-0.015 b}$) to lower impact parameters offers reasonable agreement. We find similar agreement with measurements from galaxy-galaxy pairs at $z\sim2$, suggesting a smooth decrease in absorption equivalent width with impact parameter with little redshift dependence. This indicates that the large EW and $\Delta v_{90}$ in our down-the-barrel spectra arises from gas at small distances ($\lesssim10$~kpc) from the host galaxy. Spatially resolved emission line studies mapping \ion{Fe}{2}* and \ion{Mg}{2} in one of the lensed galaxies in this work \citep[RCSGA0327-G;][]{Shaban_2022} and other star-forming galaxies \citep[e.g.,][]{Finley_2017,Burchett_2021} find similar spatial extent, further supporting a relatively small distance for the gas associated with down-the-barrel absorption.

\begin{figure}[!ht]

    \centerline{
        \includegraphics[width=0.75\columnwidth]{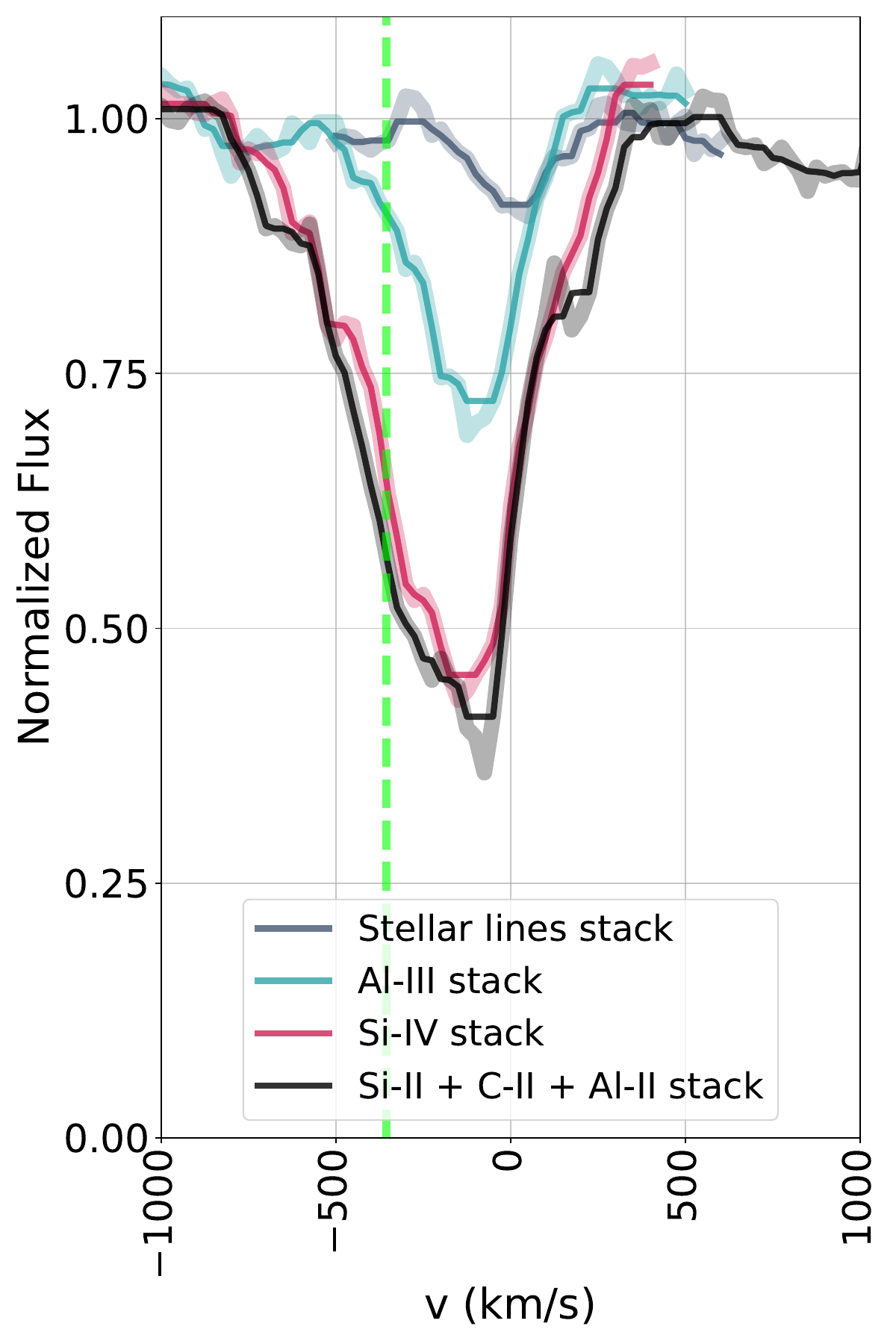}
        }

    \caption{Median absorption profiles of gaseous species with different ionization potentials, compared to the stellar kinematics. We show stacks of stellar photospheric lines in dark blue, \ion{Al}{3}~$\lambda\lambda$1854,1862 in cyan, a stack of strong low ions (\ion{Si}{2}, \ion{C}{2} and \ion{Al}{2}) in black, and \ion{Si}{4}~$\lambda\lambda$1393,1402 in red. The green line shows the median $v_{75,V2}=-327$~\kms\ value measured from this work, for reference. 
    Stellar absorption kinematics show a median $v=0$ as expected, while the ISM profiles are clearly blueshifted due to prominent outflows (associated with the baryon cycle schematic illustrated in Figure~\ref{fig:baryon-cycle-model}). 
    The low ions, \ion{Al}{3} and \ion{Si}{4} all show nearly identical kinematics suggesting that these phases are co-spatial and powered by the same outflow mechanism. We note that the \ion{Al}{3} transitions appear to have moderate optical depth $\tau \lesssim 1$, while other interstellar absorption profiles appear to be optically thick and thus trace the gas covering fraction. 
    }
    \label{fig:stacking-allions}

\end{figure}

\begin{figure*}[!htb]
    \centerline{
        \includegraphics[width=0.335\linewidth]{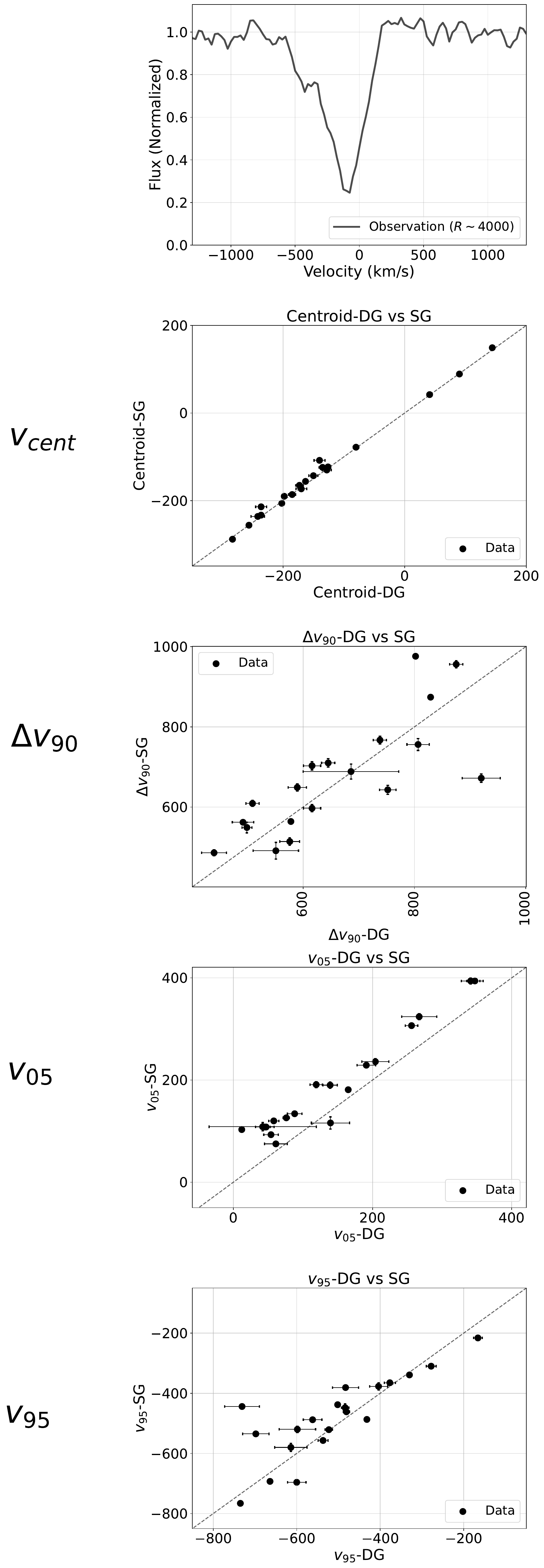}
        \includegraphics[width=0.245\linewidth]{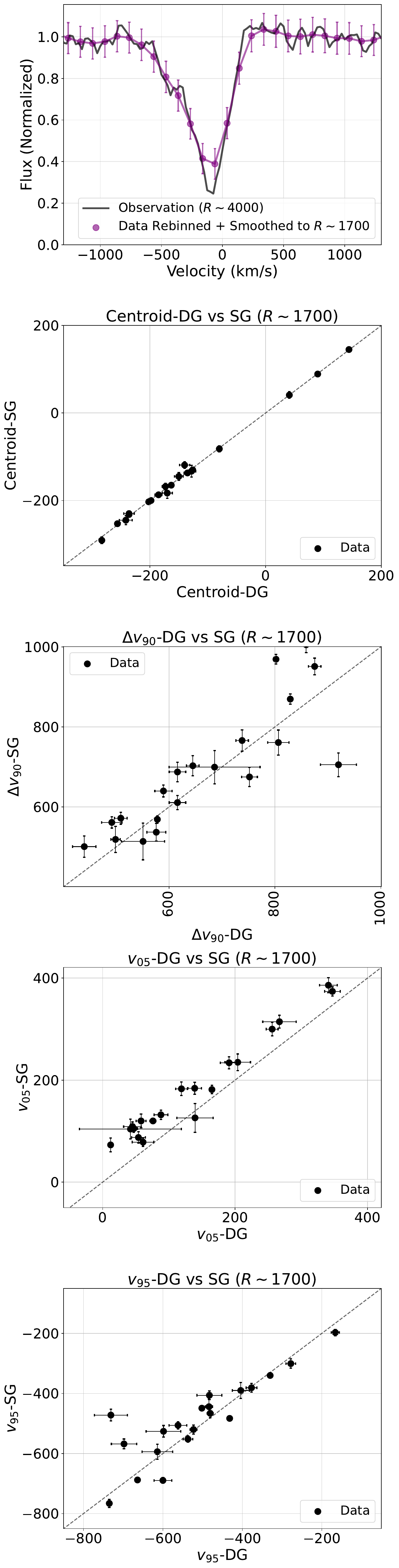}
        \includegraphics[width=0.245\linewidth]{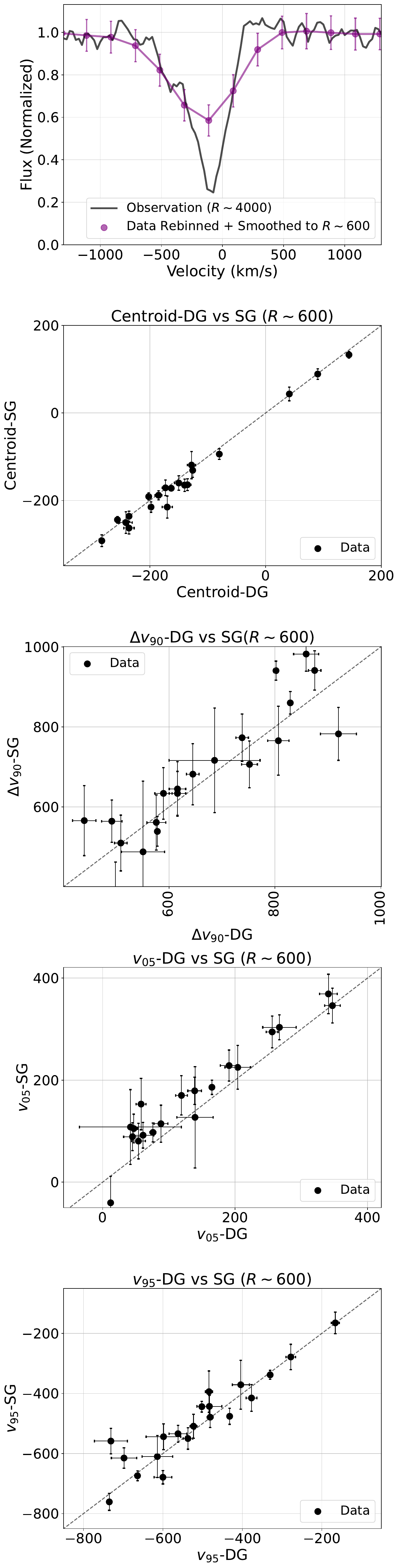}
        }
    \caption{\emph{Top panels}: Example of an absorption profile used for the SG fitting at different spectral resolution. The gray line denotes the profile in its native resolution, whereas purple lines show the profile after smoothing and rebinning.
    \emph{Lower panels}: $v_{cent}$, $\Delta v_{90}$, $v_{05}$ and  $v_{95}$ values obtained from double Gaussian (DG) fits compared with single Gaussian (SG) fits to the covering fraction profiles, at (i) the native resolution ($R\gtrsim4000$; \emph{Left}), (ii) $R\sim1700$  (\emph{Middle}) and (iii) $R\sim600$  (\emph{Right}). We describe the method used to transform our observed data to lower spectral resolution in Section~\ref{subsubsec:smoothed-rebinned-data-comparision}. The SG fits are representative of the information content for lower spectral resolution data, whereas DG fits accurately capture the full velocity structure resolved in our sample. The black dashed line in each panel represents one-to-one correspondence between the two fits. The centroid velocity measurements of the SG fits agree well with those obtained from the DG, whereas the $\Delta v_{90}$, $v_{05}$ and  $v_{95}$ measurements show larger scatter around the average linear trend. Some metrics are clearly biased in the SG fits, most clearly seen for $v_{05}$ where the SG value is systematically larger. This scatter and bias is explained by asymmetry in the observed profiles, with most galaxies having a skewness ratio $>0$ (see Section~\ref{sec:skewness} for discussion), which the SG is unable to capture. Specifically, galaxies with higher skewness ratios are systematically more biased (see  Figure~\ref{fig:resolution-skewness-dependence}). }
    \label{fig:derived-fits}      
\end{figure*}
\begin{figure*}[!htb]
    \centerline{
        \includegraphics[width=0.85\linewidth]{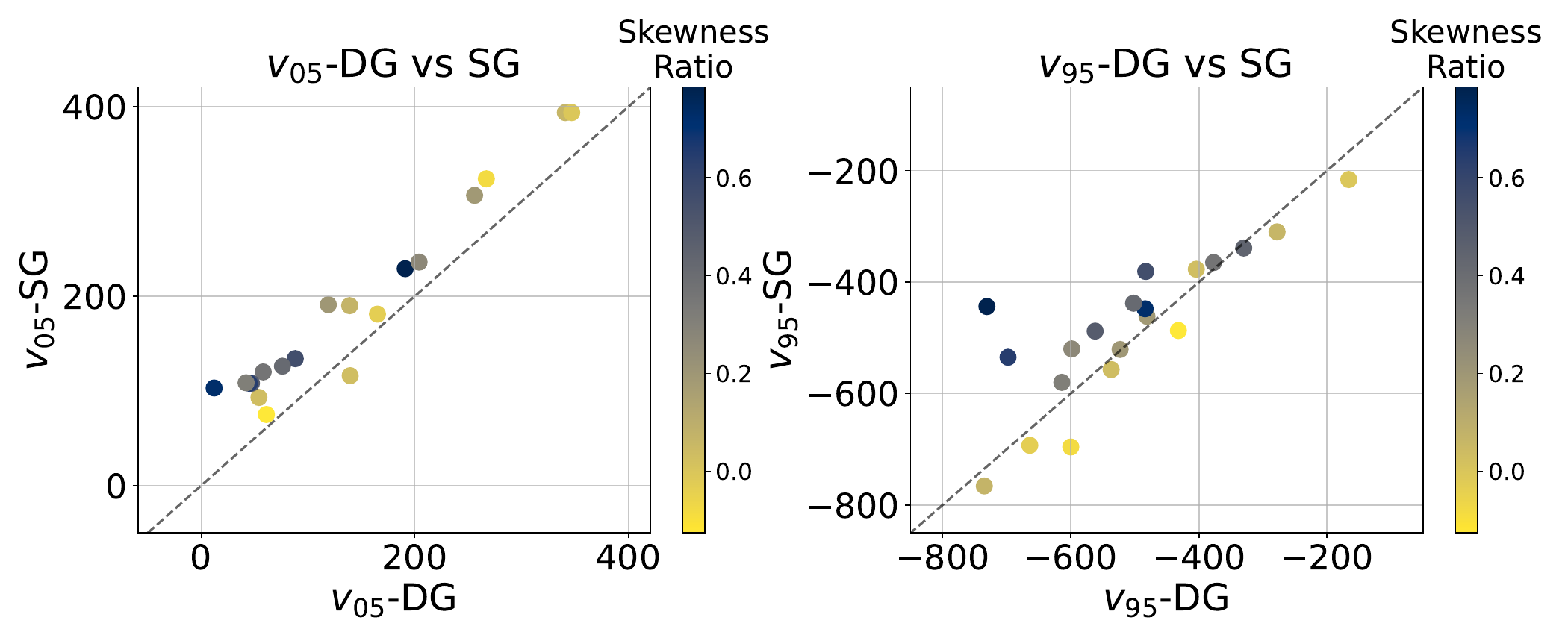}
        }
     \centerline{
        \includegraphics[width=0.6\linewidth]{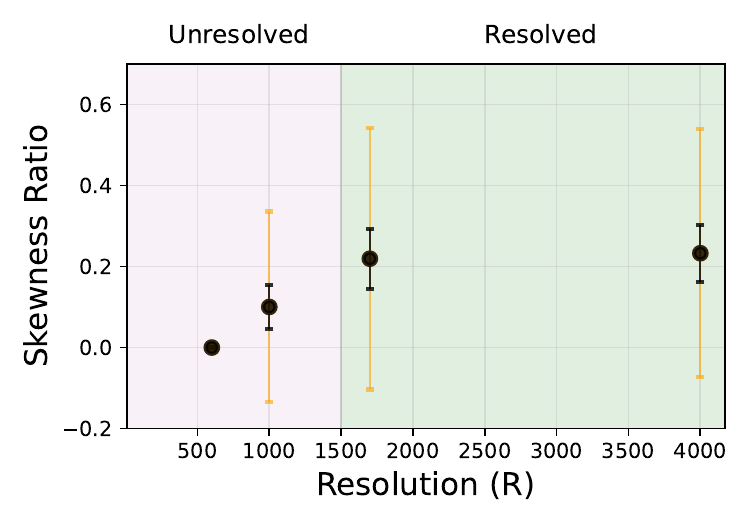}
        }       
    \caption{
    \emph{Top:} $v_{05}$ and $v_{95}$ values obtained from double Gaussian (DG) fits compared with single Gaussian (SG) fits to the covering fraction profiles at $R\gtrsim4000$, color coded by Skewness Ratio of the absorption profile (Section~\ref{sec:skewness}). $v_{05}$ measurements tracing the redshifted velocities show a clear bias, whereas $v_{95}$ which probes the high velocity outflowing gas has a lower bias but a larger scatter of 85~\kms\ (see Table~\ref{tab:low-res-comparision}). Absorption profiles which have higher Skewness Ratio values (e.g., J0004, RCSGA0327-G, CSWA128) show a clear bias in both  $v_{05}$ and $v_{95}$ measurements. This suggests that velocity metrics other than the centroid ($v_{50}$) are not captured by symmetric fitting profiles, and thus are largely unreliable at low spectral resolution. 
    \emph{Bottom:} Average Skewness Ratio of absorption profiles in the sample, fit with a Double Gaussian (DG) after smoothing to different spectral resolution ($R$). 
    The sample standard deviation and uncertainty in the mean are denoted in black and orange error bars respectively. 
    At $R<600$, the profiles uniformly appear symmetric (Skewness Ratio~$=0$) even when fit with a DG. 
    We find that $R\gtrsim1700$ is essential to recover the shape of the velocity profiles (e.g., skewness) and reduce biases that might be introduced due to lower resolution. This region is labeled as 'Resolved' indicating where the intrinsic profile skewness in our sample is recovered with a DG fit. This threshold corresponds to a FWHM spectral resolution of $4\times$ smaller than the $\Delta v_{90}$ line width, to adequately resolve the asymmetry.}
    \label{fig:resolution-skewness-dependence}      
\end{figure*}

\subsection{Kinematics at intermediate and high ionization states}\label{sec:high-ions-comparision}

The warm ISM and CGM gas with $T\sim 10^4$ K is multiphase, with contributions from \ion{H}{1}, \ion{H}{2}, and a range of metal species. In previous sections we have focused on the low-ionization metal species which are thought to predominantly trace \ion{H}{1}. Here we briefly examine species of different ionization potential in order to assess whether the low-ion results are applicable to other phases. 

We construct median stacks of \ion{Al}{3}~$\lambda\lambda$1854, 1862, and \ion{Si}{4}~$\lambda\lambda$1393, 1402 absorption lines with the same methodology as in Section~\ref{sec:lines-diff-tau}. These span ionization potentials from 1--3.3 Rydberg. Figure~\ref{fig:stacking-allions} compares the stacked velocity profiles of these species along with the stacked low ions used in previous sections. An equivalent stack of stellar photospheric lines (\ion{Si}{3}~$\lambda$1294, \ion{Si}{3}~$\lambda$1417, \ion{S}{5}~$\lambda$1501, and \ion{N}{4}~$\lambda$1718) is also plotted to show the stellar velocity range, which likely reflects that of the systemic (as opposed to outflowing) gas. We confirm that the stellar absorption is symmetric about $v=0$ as expected. The kinematic structure of \ion{Si}{4} and \ion{Al}{3} is similar to the low ion stack, suggesting that these species exist co-spatially. We note that \ion{Al}{3} is unsaturated with $\tau \lesssim 1$, as the $\lambda$1854 line is clearly stronger than $\lambda$1862, whereas the low ions appear optically thick.

In summary, all of these ions -- which are typically associated with $\sim$10$^4$~K gas -- exhibit similar kinematics. 
\citet{chisholm2018} have also analyzed \ion{O}{6} for one of the lensed galaxies in this sample (CSWA38), and find that this hotter \ion{O}{6} phase is likely also co-spatial with the low ions, although with a different column density profile. We conclude from the similar absorption profiles that the various ions associated with $\sim$10$^4$~K gas are likely co-spatial, tracing the same outflows.

\subsection{Implications for low spectral resolution surveys}\label{subsec:low-res-comparision}

In this section we assess the extent to which lower-resolution spectra can accurately capture the kinematics of outflowing gas. 
There have been several large surveys of galaxies at $z\gtrsim2$ which have characterized ISM absorption at lower spectral resolution $R\lesssim1000$ \citep[e.g.,][]{shapley2003,Vanzella2009,steidel2010,mosdef-lris-2022}. In order to examine which kinematic properties can be reliably obtained with such data, we consider two scenarios below. First, we examine single Gaussian fits to the absorption profiles, which represents an idealized case. We then perform an equivalent analysis after smoothing and rebinning the data to mimic lower resolution surveys, with potentially detrimental effects from blending of adjacent spectral features. In practice, such blending may also affect the continuum normalization which would result in larger biases (i.e., worse performance) than the idealized case we consider herein

We note that the data used as the basis of comparison in this section has finite resolution $R\sim4000$. Given the line widths, correcting for the instrument line spread function (LSF) has a small effect: $\Delta v_{90}$ decreases by 8 \kms\ on average and $v_{50}$ remains unchanged. As this is a small difference relative to the uncertainties, we report measurements directly from the $R\sim4000$ spectra without correcting for the LSF. The true intrinsic line widths are thus $\sim$1\% smaller than these reported values.

\subsubsection{Single Gaussian fit to $R\sim4000$ data}\label{subsubsec:R4000-data}

Single Gaussian (SG) fits to the ISM absorption profiles are described in Section~\ref{sec:fitting-cf} along with the resulting velocity metrics. 
Figure~\ref{fig:derived-fits} (left panel) compares the  quantities $\Delta v_{90}$, $v_{cent}$, $v_{05}$, and $v_{95}$ obtained from the SG and Double Gaussian (DG) fits, both at the native $R\sim4000$ spectral resolution. The mean offset and scatter between SG and DG fits for each metric are listed in Table~\ref{tab:low-res-comparision}. 
Velocity centroids show excellent agreement, with a mean offset $\left< v_{cent,DG} - v_{cent,SG} \right>$ of only $-4\pm1$~\kms\ and sample standard deviation of 10~\kms.

The limitations of SG fits (and of low resolution spectra) are nonetheless apparent in higher-order velocity measurements. The $v_{05}$ velocity in Figure~\ref{fig:derived-fits} shows a clear bias (mean offset of $-52 \pm 4$~\kms) and substantial scatter (indicating error for individual objects) with SG fits. This bias is also evident in $v_{95}$ and $\Delta v_{90}$  which have a smaller mean offset but larger scatter. The bias is a consequence of the intrinsic asymmetry in observed line profiles which is not captured by a SG fit; a single Gaussian profile cannot recover the skewness (Section~\ref{sec:skewness}). 
Consequently we also find that absorption profiles with higher Skewness Ratios have larger biases in SG fits (Figure~\ref{fig:resolution-skewness-dependence}, \emph{Top}). 

These results demonstrate that asymmetric fitting profiles are essential to recover the covering fraction, skewness, and higher-order velocity measurements of absorption profiles. While a symmetric SG profile is able to recover accurate velocity centroids, the quantities describing both the blue- and red-shifted velocity extremes (such as $v_{95}$ and $v_{05}$) are subject to large scatter and systematic biases. Consequently the spectral resolution must be sufficiently high to distinguish the asymmetric profile shapes.

\begin{deluxetable*}{|C|CC|CC|CC|CC|}
    \tablecaption{Performance of a single Gaussian (SG) fit to absorption profiles at different spectral resolutions compared to a double Gaussian (DG) fit at higher $R\sim4000$. The mean offset (e.g., measured as $\left< v_{cent,DG} - v_{cent,SG} \right>$ for $v_{cent}$) and the sample standard deviation ($\sigma$; e.g., of $v_{cent,DG} - v_{cent,SG}$) are given for the velocity metrics $\Delta v_{90}$, $v_{cent}$, $v_{05}$, and $v_{95}$. 
    This sample $\sigma$ represents the typical error introduced by not resolving deviations from a symmetric Gaussian profile, which varies for individual objects depending on their actual profile shape, such as skewness (Figure~\ref{fig:resolution-skewness-dependence}).
    The skewness ratio (discussed in Section~\ref{sec:skewness}) as measured by a single Gaussian fit is 0 by symmetry, whereas it varies between -0.2 to 1.0 for the double Gaussian.   }
    \tablewidth{0.5\textwidth}
    \tabletypesize{\small}
    % Header %
    \tablehead{ \colhead{Quantity measured} & \colhead{$R\sim4000$} & \colhead{} & \colhead{$R\sim1700$}  & \colhead{}& \colhead{$R\sim1000$} & \colhead{} & \colhead{$R\sim600$} & \colhead{}\\ \hline
    \colhead{} & \colhead{Mean offset} & \colhead{Sample $\sigma$} & \colhead{Mean offset}  & \colhead{Sample $\sigma$} & \colhead{Mean offset} & \colhead{Sample $\sigma$} & \colhead{Mean offset} & \colhead{Sample $\sigma$}\\ \hline
    \colhead{} & \colhead{\kms} & \colhead{\kms} & \colhead{\kms}  & \colhead{\kms} & \colhead{\kms} & \colhead{\kms} & \colhead{\kms} & \colhead{\kms}
    }
    \startdata
    v_{cent}       & -4\pm1  & 10  & 0\pm2   & 11  & 5\pm2 & 13 & 8\pm4 & 21 \\
    \Delta v_{90}  & -20\pm6 & 97  & -21\pm8 & 88 & -16\pm10 & 84 & -14\pm17 & 104 \\
    v_{05}         & -52\pm4 & 40  & -48\pm5 & 39 & -40\pm6 & 41 & -34\pm10 & 55  \\
    v_{95}         & -28\pm5 & 85  & -22\pm6 & 77 & -20\pm6 & 69 & -15\pm10 & 70
    \enddata
    \tablenotetext{}{}\label{tab:low-res-comparision}
\end{deluxetable*}

\subsubsection{Profile fits at lower spectral resolution}\label{subsubsec:smoothed-rebinned-data-comparision}

We now consider the quantitative effects of fitting to data of lower spectral resolution, where the intrinsic asymmetry of absorption profiles is less apparent.
We smooth the absorption profiles to a spectral resolution of $R\simeq1700$, 1000, and 600 via convolution with a Gaussian kernel (of $\sigma_{smooth}=75$, 125, and 200~\kms\ respectively). The smoothed spectra are also rebinned to $\sigma_{smooth}$ per spectral pixel. The set of $R$ is chosen to span an illustrative range, with the lowest resolution being comparable to large $z\gtrsim2$ galaxy samples observed with Keck/LRIS and VLT/FORS2.

Figure~\ref{fig:derived-fits} (top row) shows the rebinned and smoothed absorption profile of an example target spanning the range of resolutions considered here. We fit the smoothed and rebinned data with a SG and DG profile following the same methods as for $R\sim4000$ (Section~\ref{sec:fitting-cf}). Parameters from the SG and DG fits are then corrected for the effect of smoothing (i.e., deconvolved from the smoothing kernel). Mathematically this can be expressed as:

\begin{equation}
    v_{deconv} = v_{fit} 
\end{equation}

\begin{equation}
    \sigma_{deconv} = s \times \sigma_{fit}
\end{equation}

\begin{equation}
    A_{deconv} = \frac{A_{fit}}{s} 
\end{equation}

\begin{equation}
   \text{where}\  s=\sqrt{1-\left(\frac{\sigma_{smooth}}{\sigma_{fit}}\right)^2}, 
\end{equation}
with $\sigma_{fit}$ being the best-fit Gaussian velocity dispersion to the smoothed profile. The value of $\sigma_{deconv}$ can then be compared directly to the velocity dispersion obtained at higher resolution. The median scaling factor to correct for instrument resolution is $s=0.92$ at $R\sim1700$ and $s=0.64$ at $R\sim600$. In other words, the intrinsic line profiles are broadened by a factor $\frac{1}{s} \simeq 1.6$ at $R\sim600$, which can be reasonably corrected in most cases. Velocity metrics are measured from this scaled velocity profile using the same methods described earlier (Figure~\ref{fig:velocity-definitions}). We note that at $R\sim600$, the intrinsic $\Delta v_{90}$ absorption profile widths of our targets are sampled with only $\lesssim$1.5 independent FWHM spectral elements, while objects in our sample with the smallest widths (e.g., J1527) are effectively unresolved.
The middle and right columns in Figure~\ref{fig:derived-fits} compare the $v_{cent}$, $\Delta v_{90}$, $v_{05}$, and $v_{95}$ values obtained from the SG fits at different spectral resolution to the DG fits (at $R\sim4000$). Table~\ref{tab:low-res-comparision} summarizes the mean offset and sample standard deviation for each quantity at different $R$.

We find that the results of SG fits are generally unaffected by degraded spectral resolution, agreeing within $1\sigma$ of the SG fits to $R\sim4000$ data (Section~\ref{subsubsec:R4000-data}). This is expected since the spectral resolution has been corrected using precise knowledge of the smoothing kernel. 
We thus obtain approximately the same systematic bias and scatter in SG fits to lower-$R$ data as for the case of $R\sim4000$.

The performance of SG fits discussed here should be taken as an optimal scenario given the good signal-to-noise ratio (SNR) of the lensed galaxy sample. Typical survey data will have larger statistical uncertainty. SNR is not necessarily a limiting factor however, as line width metrics are limited by the intrinsic scatter found between DG and SG fits (e.g., $\sim$100~\kms\ scatter for $\Delta v_{90}$ seen in Figure~\ref{fig:derived-fits} and Table~\ref{tab:low-res-comparision}). 

Our analysis has demonstrated that quantifying the asymmetric structure of absorption profiles is necessary for accurately measuring quantities such as the maximum outflow velocity (e.g., $v_{95}$). This in turn is crucial for establishing galaxy scaling relations with outflow velocity, and comparing to feedback models (as we discuss in Section~\ref{sec:galaxy-trends}). We now quantify the resolution needed to recover the full asymmetric covering fraction profile structure of our sample. We make use of the skewness ratio defined in Section~\ref{sec:skewness} as a reliable measure of this asymmetry. A skewness ratio of $0$ indicates a symmetric profile, whereas most of the galaxies in our sample (80\%) have skewness ratios $>0$. Figure~\ref{fig:resolution-skewness-dependence} (\emph{bottom}) plots the skewness ratio obtained by a DG fit to the rebinned and smoothed absorption profile at different resolution ($R$).  We find that at $R\sim1700$, the shape of the profile is largely recovered: the mean and sample standard deviation in skewness ratio is $0.22\pm0.32$ compared to $0.23\pm0.30$ for the $R\sim4000$ data. We also find that at $R\sim1700$, other velocity metrics ($\Delta v_{90}, v_{95}, v_{05}$) have a mean offset of $\sim0$~\kms\ and a modest sample scatter of $\sim 50$~\kms\ cf. $R \sim 4000$ measurements. 
However, at $R\sim600$, the skewness ratios are uniformly near zero indicating that the diversity and asymmetry of absorption profile shapes is not recovered for any of our targets at such low resolution. 
At the intermediate $R\sim 1000$, the average recovered skewness ratio is approximately half that of the high-resolution data. 
Individual galaxies with narrower profiles will have worse results at degraded resolution. In other words, the $R$ required to distinguish asymmetric structure depends on the profile width. In this case the threshold $R\gtrsim1700$ corresponds to sampling the average $\Delta v_{90}$ with $\simeq$4 independent FWHM resolution elements. For samples with different gas kinematics, the required resolution should scale as the inverse of the profile width (e.g., $R \propto 1/\Delta v_{90}$).

To summarize, lower resolution data are sufficient to recover $v_{cent}$, while higher resolution $R\gtrsim1700$ is required to recover the full asymmetric covering fraction profile structure and outflow velocity metrics for our sample (e.g., $v_{95}, \Delta v_{90}$). The threshold $R$ required for reliable results will vary with the intrinsic profile width, which effectively corresponds to the gas outflow velocity. 
This analysis also demonstrates that the well-resolved profile shapes of our sample (Figure~\ref{fig:velocity-profiles}) can provide guidance for trade studies of spectral resolution and SNR for future surveys, 
which may be optimized for different scientific goals.

\section{Trends with Galaxy properties}\label{sec:galaxy-trends}

In order to understand the feedback effects of galactic outflows, we seek to compare outflow properties with galaxy demographics such as stellar mass and star formation rate (SFR). We necessarily restrict this analysis to the subset of the lensed sample with suitable ancillary data. In particular, for accurate stellar population properties, we require photometry at infrared observed wavelengths, as well as a lens model to correct for magnification by the foreground deflector galaxy. 

\begin{deluxetable}{|c|C|C|C|C|}
    \tablecaption{Stellar mass, SFR (in units of $\Msun~\mathrm{yr}^{-1}$), and magnification values for the targets presented in this paper. All values have been scaled to a \cite{Chabrier2003} IMF.}  
    \tablewidth{0.4\textwidth}
    \tabletypesize{\small}
    % Header %
    \tablehead{ \colhead{Objid} & \colhead{$\log\left(\frac{M_*}{\Msun}\right)$} & \colhead{SFR} & \colhead{$\mu$}    & \colhead{Ref}  
    }
    %data %
    \startdata
CSWA2                       & 9.1^{+0.3}_{-0.3}       & 32^{+23}_{-13}   & 8.4 & A0\\ 
RCSGA0327-G                 & 9.80_{-0.05}^{+0.05}    & 40_{-10}^{+10}   & 17.2 \pm 1.4 & A5,A6 \\CSWA38                      & 9.8_{-0.2}^{+0.2}           & 10_{-0.2}^{+0.2}        & 7.5 \pm 1.5 & A7 \\
8oclock                     & 9.90^{+0.12}_{-0.13}    & 162^{+124}_{-95} & 5 \pm 1 & A4\\    
Horseshoe                   & 9.9^{+0.2}_{-0.3}       & 210^{167}_{-167} & 10.3 \pm 5.0 & A1\\    
J1527                       & 9.9^{+0.3}_{-0.4}       & 116^{+86}_{-60}  & 15 & A0\\
CSWA128                     & 9.9^{+0.1}_{-0.1}       & 11.69^{+2}_{-1}      & 10 & A0\\
Clone                       & 10.1_{-0.2}^{+0.2}      & 68_{-44}^{+24}   & 13.1 \pm 0.7 & A1\\
CSWA103                     & 10.4^{+0.1}_{-0.2}      & 23^{+18}_{-7}    & 4.7 & A0\\
CSWA19                      &  10.5^{+0.1}_{-0.1}     & 27^{+10}_{-5}    & 6.5 & A0\\    
CosmicEye                   & 10.76^{+0.07}_{-0.08}   & 37.6^{+4.3}_{-4.3}          & 3.69 \pm 0.12 & A2\\
CSWA40                      & 10.8^{+0.2}_{-0.2}      & 169^{+146}_{-66} & 3.2 & A0\\
J1429                       & - & 90  & 8.8 & A3\\
\enddata
\tablenotetext{}{
References are as follows.
A0: \citet{ramesh2023}
A1: \citet{tj2013}
A2: \citet{Johan2011}
% A3: \citet{nicha2016}
A3: \citet{J1429_SFR}
A4: \citet{dessauges-8oclock}
A5: \citet{rcsga0327-wuyts}
A6: \citet{wuyts2010}
A7: \citet{Salimano2022_CSWA38}
} \label{tab:stelmass-sfr}
\end{deluxetable}

\begin{figure*}[!htb]

      \centerline{
        \includegraphics[width=0.5\linewidth]{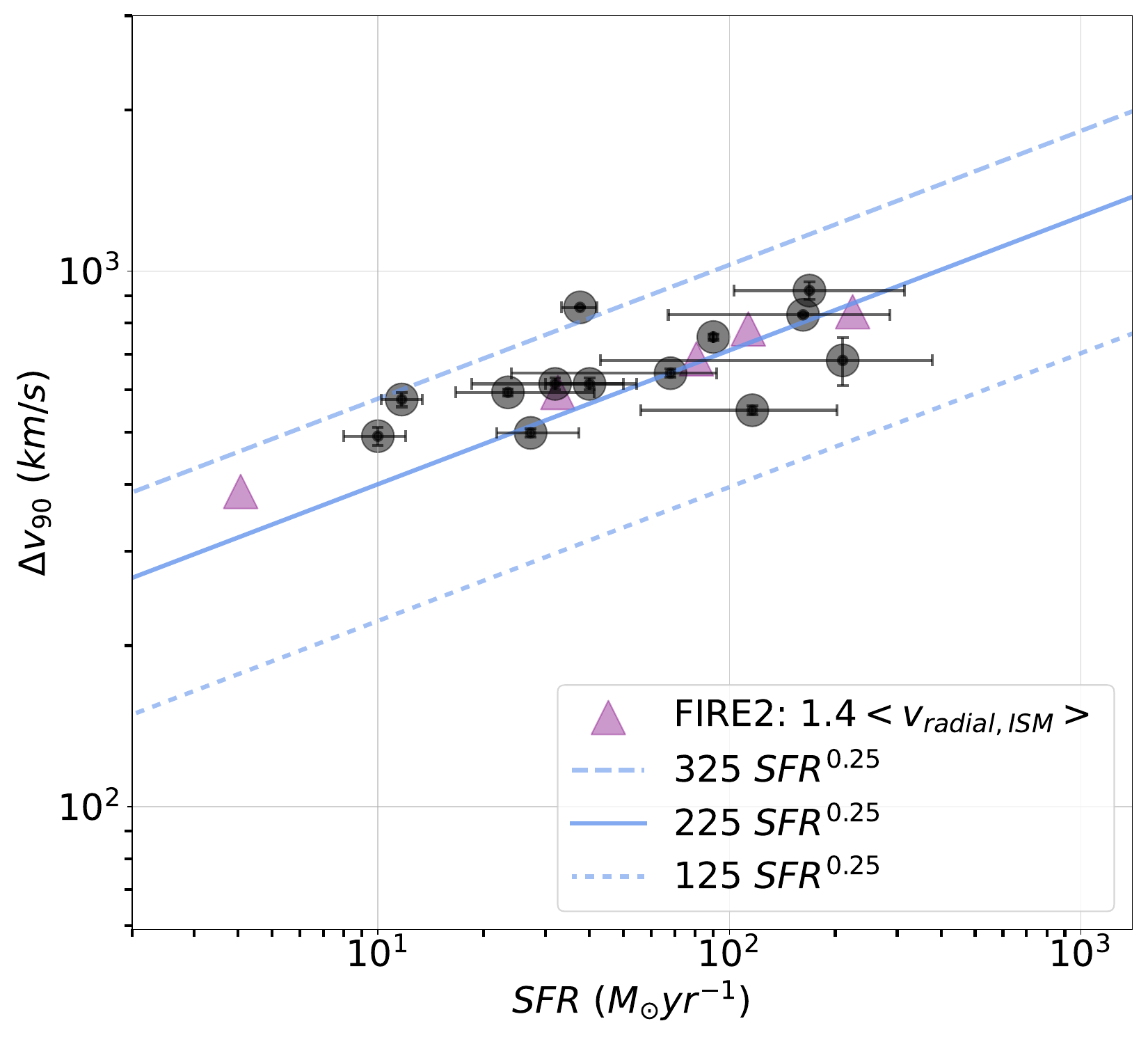}
        } 

      \centerline{
        \includegraphics[width=0.85\linewidth]{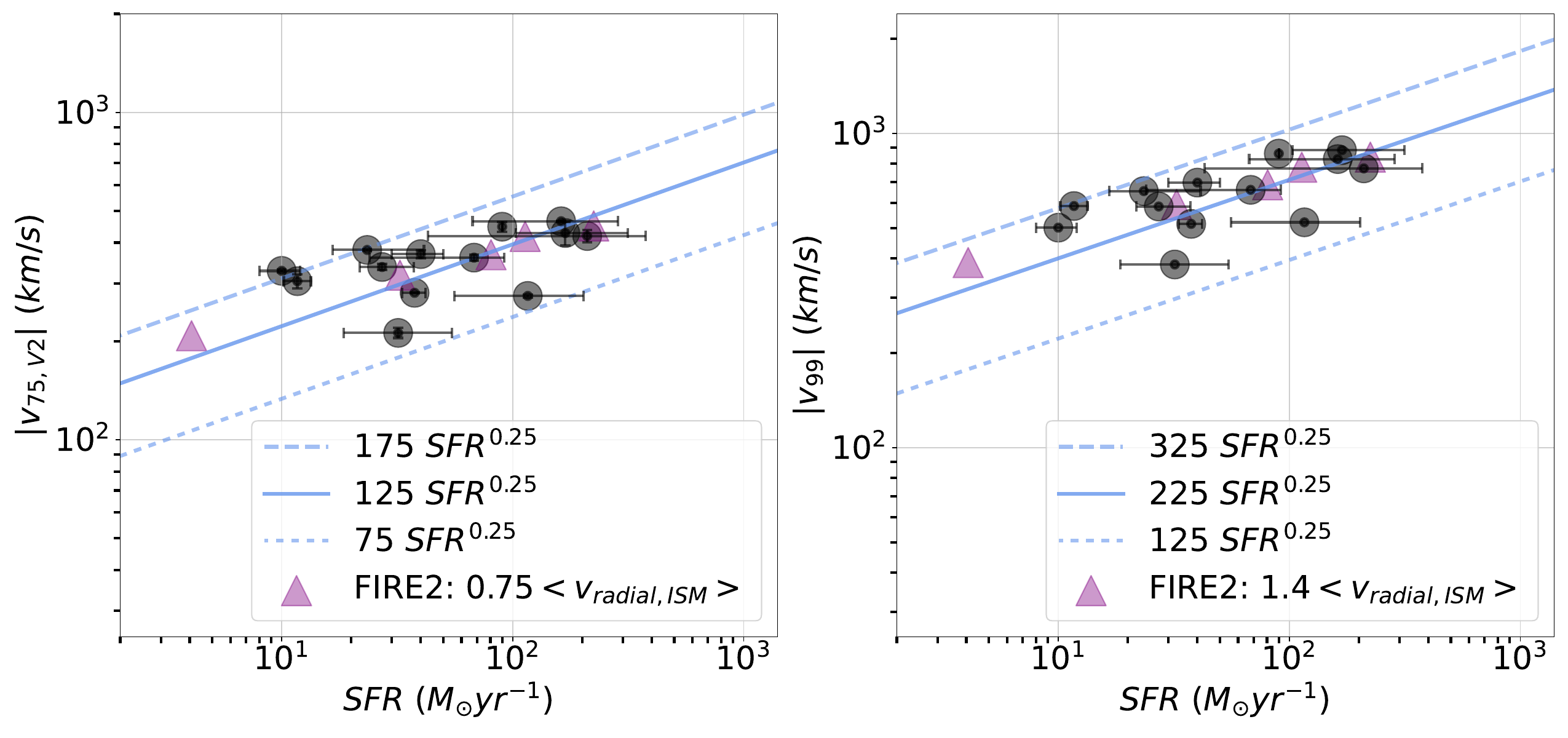}
        }    
    \caption{ Plot of $|v_{75,V2}|$, $|v_{99}|$ and $\Delta v_{90}$ versus SFR for the galaxies in our sample which have reliable SFR measurements in the literature. There is a clear correlation in that galaxies which have high SFR values also tend to have higher outflow velocities.
    Trend lines of the form $v \propto (SFR)^{0.25}$ corresponding to a momentum-driven wind scenario are shown in the figure for reference. We can see that $v \approx 225 (SFR)^{0.25}$~\kms\ is able to capture the $|v_{99}|$ and $\Delta v_{90}$ velocity measurements from this work, while $125 (SFR)^{0.25}$~\kms\ better describes $|v_{75,V2}|$. This supports a positive correlation of outflow velocity and SFR which is close to the expected scaling relation for momentum-driven outflows. We note that this result holds for various ways of defining the outflow velocity as shown in each panel. Mass weighted radial velocity values measured in the ISM of FIRE-2 simulated galaxies at similar redshifts are denoted in purple, and are scaled by a constant factor to match the approximate average trend of the lensed galaxies. 
       }
    \label{fig:sfr-velocity}    
\end{figure*}

Out of the 20 targets in our sample, 12 have reliable stellar mass and SFR measurements (and one more has SFR only). 
Masses and SFRs for six CASSOWARY targets are reported by \citet{ramesh2023}, while measurements for other sources are compiled from the literature. Table~\ref{tab:stelmass-sfr} lists the adopted stellar mass, SFR, and lensing magnification ($\mu$) values along with the original references. 
All stellar population parameters are scaled to the \cite{Chabrier2003} IMF where necessary. The stellar masses span $\log{\mathrm{M_{*}/\Msun}} = 9.1-10.8$ and the SFRs range from 10--210 $\Msun~\mathrm{yr}^{-1}$, which are typical of moderately massive star forming main-sequence galaxies at these redshifts \citep[e.g.,][]{SpeagleSFMainSequence}. 

One caveat in comparing the galaxy properties to the outflow properties is that the inferred SFR and stellar mass are global galaxy properties, whereas outflows may vary across different star-forming clumps \citep[e.g.,][]{Bordoloi2014}. 
In the following section, we assume that the global galaxy averaged outflow properties are sufficiently captured by our slit spectra, which probe several square kpc in the source plane.

\subsection{Galaxies with high SFR also have high outflow velocities and absorption widths}\label{subsec:high-sfr-high-outflow}

One of the key trends we want to explore is whether higher outflow velocities -- traced by $v_{75,V2}$ and $\Delta v_{90}$ values for example -- correlate with higher SFR in the host galaxies. Such a correlation may be naturally expected since the outflows are driven by energy and momentum released by star formation. We consider a simple power law response of the following form: $v=v_0 SFR^{\alpha}$ and $\Delta v = \Delta v_0 SFR^{\alpha}$. Physically, the value for $\alpha$ is determined by the mode of feedback. For example, 
\citet{murray2005} find that in galactic winds primarily driven through momentum injection from supernovae, the luminosity ($L$) scales with the galaxy velocity dispersion ($\sigma$) as $L \propto \sigma^4$ whereas an energy-driven wind would follow $L \propto \sigma^5$.
They also find that for starburst galaxies at high-z, momentum driven winds are more favorable, as have other studies \citep[e.g.,][]{Dave2011}. Therefore, taking the SFR to be a tracer of luminosity and the outflow velocity to roughly scale linearly with the galaxy dispersion \citep[e.g.,][]{sdss-outflows2016}, we might expect the kinematics of the outflowing gas also to scale as $SFR^{0.25}$ (i.e. $v \propto SFR^{0.25}$ and $\Delta v \propto SFR^{0.25}$). 

Figure~\ref{fig:sfr-velocity} plots the SFR versus the outflow velocity ($v_{99}$, $v_{75,V2}$, and $\Delta v_{90}$ metrics) along with a power law scaling relation motivated by the momentum-driven wind scenario. As we can see from the figure, a power law fit is a reasonably good description of our measurements. Notably, we find a similar power law correlation for all three outflow velocity metrics, which primarily differ in normalization as expected. We overlay scaled mass-weighted radial outflow velocities (e.g., $v = 1.4\left<v_{rad,ISM}\right>$) obtained in FIRE-2 simulated galaxies at $z=2-4$ (discussed further in Section~\ref{sec:comparision-theory}), which also show reasonable agreement with this power law correlation. We find that among the different metrics tested, $\Delta v_{90}$ correlates well with $SFR$, having a Spearman correlation coefficient of 0.7 and p-value $=0.007$. The best fit power law between $\Delta v_{90}$ and $SFR$ for our sample is given by $(389 \pm 77) \, SFR ^{0.13 \pm 0.05}$. The slope of the best-fit relation is somewhat shallower than expected for momentum-driven winds at $\sim2\sigma$ significance, although this power-law slope is consistent with the relation between SFR and velocity FWHM (with $\alpha=0.19$) found at low redshift by \citet{ClassyXu2022} for their sample which probes a larger dynamic range in SFR. We note that \citet{ClassyXu2022} also find a marginally steeper slope ($\alpha=0.22$) in SFR versus outflow velocity, and find good overall agreement with a momentum driven wind scenario.

\begin{figure*}[!ht]
    \centerline{
        \includegraphics[width=\linewidth]{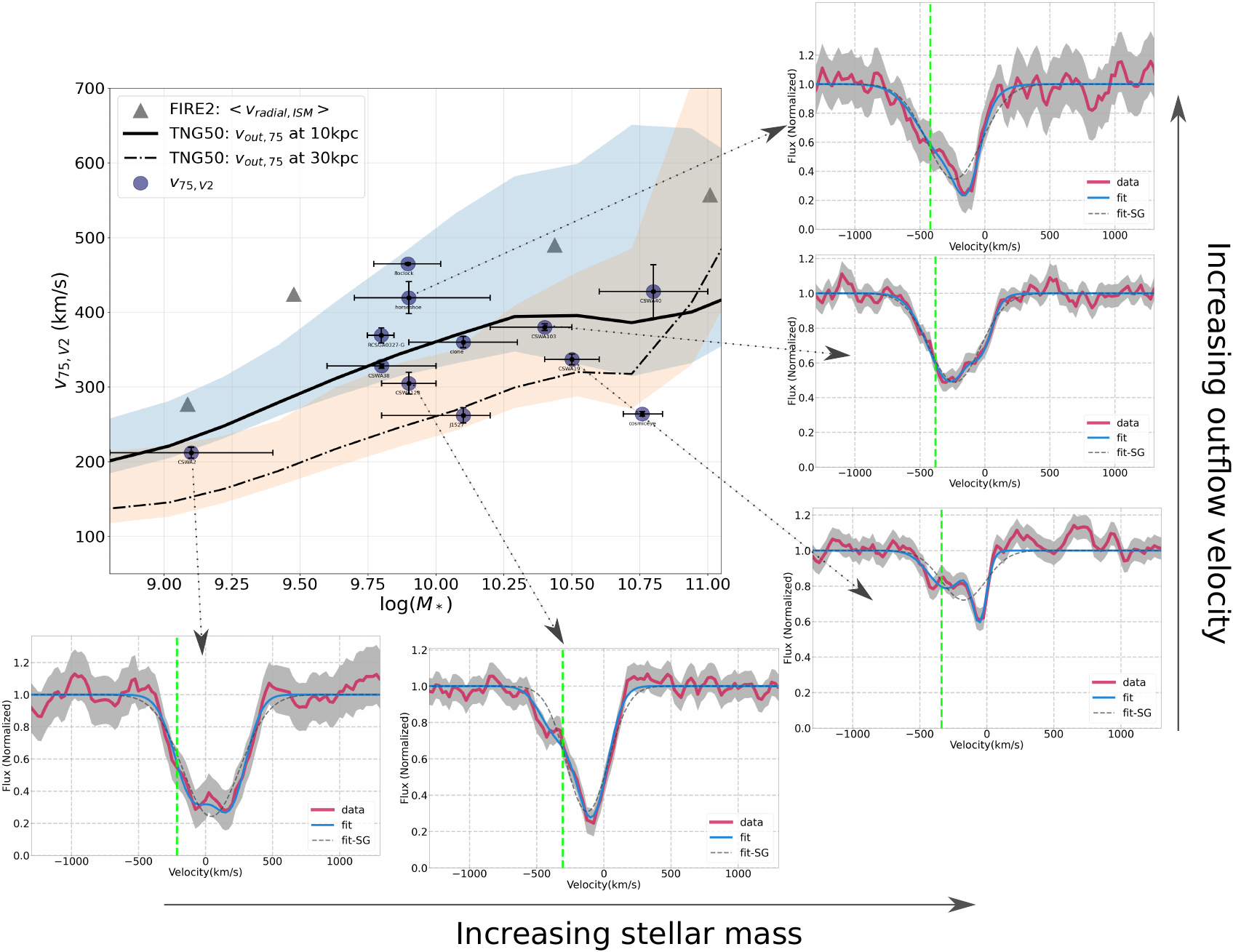}
        }
    \caption{{\it Top left:} Comparison of measured outflow velocities for the lensed galaxies ($v_{75,V2}$: black points with error bars) with 75th percentile outflow velocities in the TNG50 simulation and mass-weighted radial velocity from FIRE-2. The solid and dotted black trend lines from TNG50 are the median values at $r=10$ and 30~kpc, while shaded regions show the scatter (16--84 percentile range). The observations and TNG50 simulations appear generally comparable in terms of the overall trends and scatter (see Section~\ref{subsec:high-sfr-high-outflow}). 
    {\it Surrounding panels:} Velocity profiles for several of the lensed galaxies are shown to illustrate the velocity structure across the range of properties spanned by the sample (see Figure~\ref{fig:velocity-profiles} for the full sample), with $v_{75,V2}$ values denoted by vertical green dashed lines. 
    A key issue for comparison is that the observations include absorption from interstellar (systemic) and recycling gas, whereas TNG50 and other simulations can isolate purely outflowing material. We also note that the TNG50 results include highly ionized gas, although our analysis shows no significant difference in the velocity structure of low and intermediate ion species (see Section~\ref{sec:high-ions-comparision}). The FIRE-2 radial outflow velocities are comparable to those obtained from the 75th percentile outflow velocities from TNG50 at 10kpc.\label{fig:tng50-comparision}    }  
\end{figure*}

Various studies of galaxy outflow velocities and their scaling relations, spanning a wide range of redshift and galaxy properties, have found that the power law coefficient $\alpha$ (where $v \propto SFR^{\alpha}$) ranges from $\alpha \in(0.03-0.35)$ \citep[e.g.,][]{martin2005,rupke2005,deep2outflows,steidel2010,martin2012,erb2012,Bordoloi2014,chisholm2015,Chisholm2016,heckman2016,sugahara2017}. A key challenge is that different studies employ inhomogeneous data and analysis techniques,  including the definition of outflow velocity (such as the maximum outflow velocity $v_{max}$, ISM velocity centroid $v_{IS}$, and others). Our analysis suggests that $\Delta v_{90}$ is potentially of broad use for comparison, as it is more readily measured than quantities such as $v_{99}$, while it correlates well with other metrics and is reasonably robust to spectral resolution effects (Section~\ref{subsec:low-res-comparision}).

\subsection{Comparison to cosmological simulations}\label{sec:comparision-theory}

In this section, we focus on comparing our observational results with predictions for outflow properties obtained from recent cosmological simulations which incorporate stellar feedback. Such comparisons are a valuable test of feedback models used in these simulations, and we also highlight pathways which would be beneficial for future investigation. 
In particular, we compare observations with results from two sets of simulations: TNG50 \citep{tng50} and FIRE-2 \citep{pandya2021}. These were chosen due to the availability of suitable outflow velocity metrics. 

One challenge in comparing with simulations is that the radial distribution of gas responsible for the absorption in our sample is unknown. The observational data probe the total projected velocity of gas along the line-of-sight only on the near-side of a target galaxy. This does not necessarily correspond with the metrics used in theoretical analysis or reported by simulations, where full 3-D spatial and velocity information is available. 
For example, TNG50 and FIRE-2 are able to examine gas outflow velocities as a function of radius from the host galaxy. 
To provide better context for comparison, we thus first consider the likely radial distribution and dynamical timescale of absorbing gas in the observed galaxy sample. 
Following \citet{tucker-dustinthewind}, we expect that the majority of absorption occurs within at most a few tens of kpc from the host galaxy. If we assume that outflowing gas starts at radius $r=0$ and is driven at constant velocity, then its radial distance after a time $t$ is 
\begin{equation}
    r = \frac{v}{-150~\mathrm{\kms}} \frac{t}{100~\mathrm{Myr}} \times 15~\mathrm{kpc}. 
\end{equation}
Given typical velocities $\sim$150~\kms\ and $r\lesssim 50$~kpc (and quite possibly much smaller $r$) deduced from comparison to quasar sightlines, this implies the gas seen in absorption was launched $\lesssim 300$ Myr ago. 

In Figure~\ref{fig:tng50-comparision} we compare our measured outflow velocities with TNG50 simulated galaxies at $z=2$ and FIRE-2 galaxies in the $z=2-4$ bin, as a function of stellar mass. Specifically, we compare $v_{75,V2}$ values from this work with the 75 percentile mass-weighted velocities at different radii in TNG50 simulations. We expect these to be comparable, although they are not strictly identical measures. 
The observations are well bounded by TNG50 values for $r\lesssim30$~kpc as shown in Figure~\ref{fig:tng50-comparision}, indicating reasonable agreement between the data and simulations. FIRE-2 galaxies, on the other hand, have measurements of mass-weighted radial velocity ($\left< v_{radial,ISM} \right>$) for gas in the radius range $r=0.1-0.2r_{vir}$ with typical $r_{vir} \gtrsim 150$ kpc. Encouragingly, these values are also comparable to those seen in observations and those from TNG50 at $r=10$~kpc. Therefore, the feedback prescriptions used in the TNG50 and FIRE-2 simulations yield outflow velocities comparable to those seen in observations. We discuss prospects for future work in this direction in Section~\ref{subsec:spatial-dist}

\subsubsection{Enrichment of the CGM/IGM via outflows}

A key question for galaxy formation is the amount of outflowing material which is able to escape a galaxy's gravitational potential, as opposed to remaining in the CGM and potentially recycling back into the galaxy, and how this varies with galaxy mass. To address this, we compare our measured outflow velocity profiles with estimated escape velocities of the sample. 

The escape velocity $v_{esc}$ is related to the rotational velocity of a galaxy ($v_{rot,max}$) and the virial radius ($r_{vir}$). In the case of an isolated galaxy with a truncated isothermal sphere mass distribution, the relation is 
\begin{equation}
    v_{esc}(r)= v_{rot,max} \sqrt{1 + \log\left(\frac{r_{vir}}{r}\right)}
\label{eq:vesc}
\end{equation}
for gas at radius $r$ \citep{winds_ara}. We estimate rotation velocities $v_{rot,max} \simeq 150-200$~\kms\ for the lensed sample based on the width of stellar photospheric features (Figure~\ref{fig:stacking-allions}; Section~\ref{sec:high-ions-comparision}), which is also supported by rotation curves of galaxies with similar redshift and stellar mass \citep[e.g.,][]{wisnioski2015_kmos3d, forsterschreiber2018_sins}. Assuming $r\sim0.1-0.2 \, r_{vir}$, the escape velocity for these galaxies is 200-300~\kms\ from Equation~\ref{eq:vesc}. The mean 75\% outflow velocity seen in our lens sample is  $|v_{75}| \sim300$~\kms\ (Section~\ref{sec:kinematics-observed}), suggesting that $\sim25\%$ of the gas absorption profile has sufficient velocity to escape into the IGM. However, this simple analysis does not account for the interaction of outflows with the ambient CGM and the role of environment, such that the actual amount of gas exceeding the escape velocity may be smaller.

\begin{figure}[!htb]
    \centerline{
        \includegraphics[width=\linewidth]{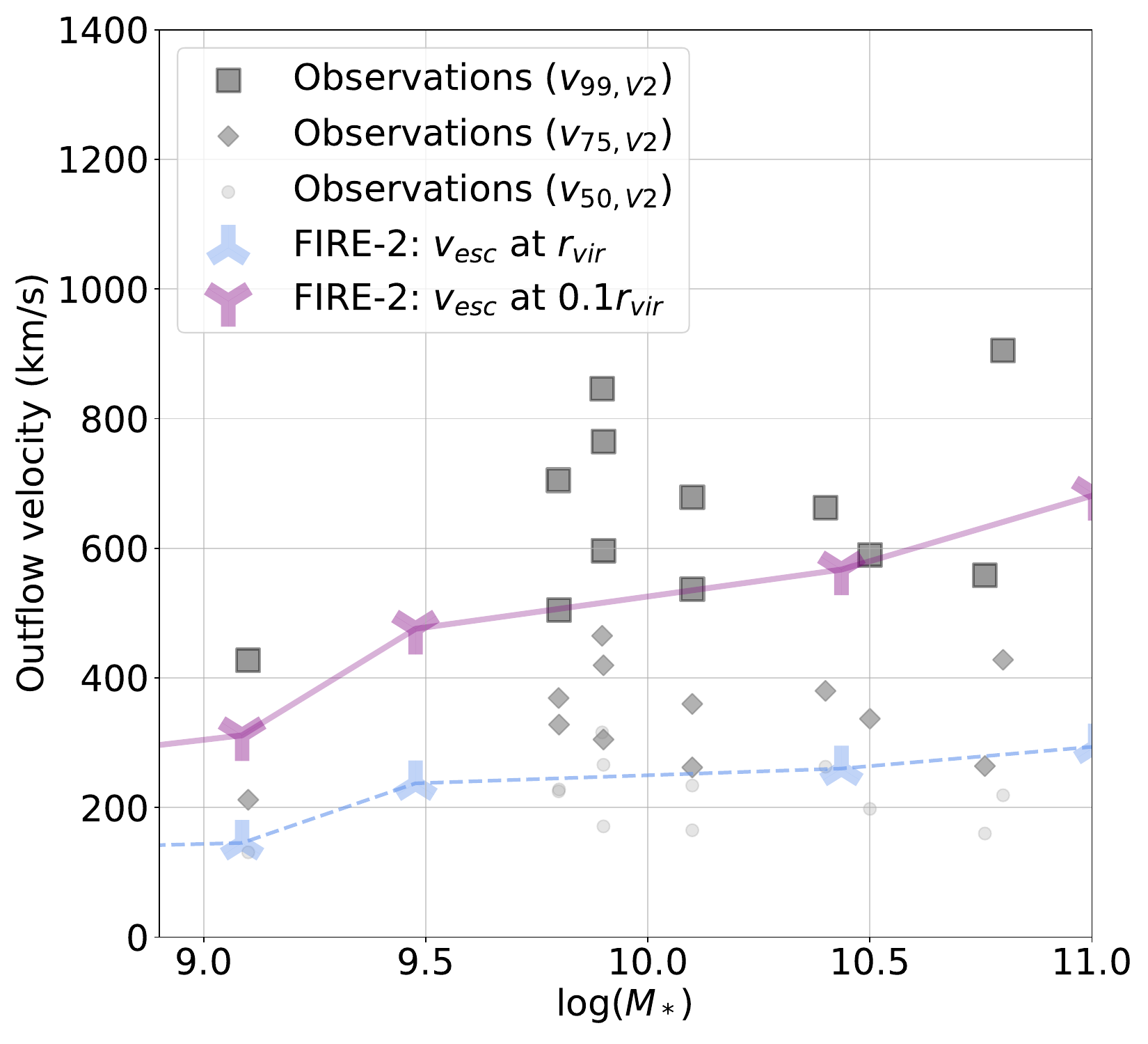}
        }
    \caption{Comparison of the escape velocity $v_{esc}$ obtained in FIRE-2 simulations with outflow velocities measured from the lensed sample. The $v_{99}$ metric corresponds approximately to the largest velocity at which outflowing gas is detected in absorption. The simulations show a trend of $v_{\tt{esc}}$ increasing with mass, and decreasing with radius, as expected. 
    The measured $v_{99,V2}$ values are comparable or larger than $v_{esc}$ even at small radii ($\sim0.1 r_{vir}$), such that the highest velocity gas is able to escape the galaxies' gravitational potential. However $v_{50,V2}$ is typically below the escape velocity even at the virial radius, indicating that the majority of outflowing gas will remain gravitationally bound. }
 
    \label{fig:fire2-tng50-comparision}      
\end{figure}

Figure~\ref{fig:fire2-tng50-comparision} shows the escape velocity of gas at $0.1r_{vir}$ and at the halo radius ($r_{vir}$) obtained in the FIRE-2 simulations, compared to outflow velocities measured for the lensed sample (specifically $v_{99,V2}$, $v_{75,V2}$, and $v_{50,V2}$ corresponding to the 99, 75, and 50 percentiles of absorption blueward of systemic velocity). The $v_{99,V2}$  and $v_{75,V2}$ metrics trace the faster moving outflowing gas seen in absorption, whereas $v_{50,V2}$ traces the bulk motion of gas (Section~\ref{sec:kinematics-observed}). From the figure, it is clear that the $v_{99,V2}$ values are higher than those needed to escape the gravitational potential of the simulated galaxies and their halos, whereas the gas at $v_{75,V2}$ velocities would be able to escape only if the absorbing gas is located at large radii ($>0.1 r_{vir}$). On the other hand, the mean outflow velocity centroid for the sample is $|v_{cent,V2}| = 188$~\kms\ which is below the escape velocity even at $r_{vir}$.

Based on this analysis, the majority of the $T\sim10^4$~K outflowing gas, although moving at over a hundred \kms, appears to be bound within the halo and/or ISM of the galaxy (i.e. it is recycling gas; Figure~\ref{fig:baryon-cycle-model}). The fastest moving gas seen in absorption ($v>v_{75,V2}$) is capable of escaping into the CGM/IGM, enriching it with heavy metals, but is subject to deceleration from interactions with gas and dust along its path. This is consistent with results from \citet{rudie2019}, who find that 70\% of the galaxies with detected metal absorption in the CGM also have unbounded metal-enriched gas capable of escaping the halo.

\subsubsection{Spatial distribution of the ISM gas}\label{subsec:spatial-dist}

Finally, we return our attention to the spatial distribution of the ISM gas around a galaxy. This is an essential quantity for determining outflow rates, mass loading factors, and whether outflowing gas will become unbound and escape into the IGM. However it is challenging to determine, as the observed absorption profiles do not directly depend on galactocentric radius. 
As discussed in Section \ref{subsec:quasar-sightlines}, we can place constraints on the radius of outflowing gas seen in absorption based on comparison with background sightline samples at different impact parameters. The large absorption velocities and equivalent widths seen in our sample indicate the bulk of outflowing gas is at relatively small radius (conservatively within a few tens of kpc). Here we briefly consider prospects for future work. 

Considering the encouraging comparison with simulations, a promising approach is to compare measured outflow velocity profiles with ``mock spectra'' generated from simulations where the spatial distribution of gas is known. This could be useful to assess the likely radial distribution of gas seen in absorption, and as a further test of feedback prescriptions used in simulations. Simulations can also be used to disentangle the outflowing, systemic, and recycling gas components and assess their relative contributions to the total absorption profile. 
Tools such as TRIDENT \citep{trident-hummels-2017} and FOGGIE \citep{FOGGIE} are promising for such analyses. However, a challenge for such work is to self-consistently model the incident spectra and ionization state of the gas; in this case the host galaxy stellar emission may dominate over the extragalactic UV background. 
z
Finally, the technique of arc tomography (in which lensed arcs are used to spatially map CGM gas of lower-$z$ galaxies in absorption) has recently proven to be highly effective \citep[e.g.,][]{lopez2018,lopez2020,kris2021}. While current studies are limited to $z<1$, expanding to higher redshifts with multiple-arc systems is a promising future avenue. Strong lensing galaxy clusters such as the Hubble Frontier Fields \citep[e.g.,][]{Mahler2018} may prove valuable for such analyses.

\section{Summary and Conclusions}
\label{sec:conclusions}

In this paper, we have used moderate resolution spectra ($R\gtrsim4000$) to characterize the ISM and outflowing gas in a sample of 20 strongly lensed galaxies at $z=1.5-3.5$ observed ``down-the-barrel.'' We construct the covering fraction profile ($C_f$) of absorbing gas, and measure various metrics of the gas kinematics. In this work, we examine the outflow velocities (parameterized by $v_{50}$, $v_{75}$, etc.), width of absorption ($\Delta v_{90}$), skewness of absorption profiles, and optical depth ($\tau$) of absorbing gas. We also explore the relations between outflowing gas kinematics and the host galaxy properties (e.g., $M_*$ and SFR), and compare them with those obtained in cosmological simulations. We demonstrate the importance of having good spectral resolution in studies of outflowing gas by considering which of our results can be accurately recovered from lower resolution spectra ($R \lesssim 1000$), and which results would be biased. Below we summarize the main properties of the absorbing gas kinematics found from this work: 

\begin{enumerate}
    \item 
    The low ionization gas is characterized by a diverse range of covering fraction profiles (Figure~\ref{fig:velocity-profiles}; Sections~\ref{sec:sample}, ~\ref{sec:vel-structure}). The profiles are asymmetric, typically with a steep ingress at redshifted velocities and a shallow egress at blueshifted (outflowing) velocities. 80\% of the sample exhibits this skewness toward blueshifted velocities (Figure~\ref{fig:skewness}). 
    A double Gaussian fit is sufficient to capture the structure of ISM absorption kinematics as measured at $R\simeq4000$ and SNR~$\simeq10$ for the full sample. 
    
    \item
    We observe ubiquitous outflows with a typical median velocity of $v_{50} \simeq -150$~\kms, with the extent of detected absorption reaching $3\times$ this median value in most cases ($\sim-500$~\kms; Section~\ref{sec:ism-features}). The typical width of absorption profiles is $\Delta v_{90} \simeq 600$~\kms, which is around 6 times larger than in typical DLA systems at similar redshifts probed by quasar spectra. Given the large absorption widths, it is likely that our down-the-barrel spectra are predominantly probing gas close to the center of the host galaxies (within a few tens of kpc or $\sim$10\% of the virial radius), whereas quasar absorption systems typically sample larger impact parameters. 
    We note that our $\Delta v_{90}$ values are measured for strong transitions which probe the gas covering fraction. Stacks of optically thin transitions suggest that the column density profile width is likely smaller ($\Delta v_{90} \sim 400$~\kms; Figure~\ref{fig:stacking-lowions}), although still very large compared to quasar DLA systems.

    \item
    The lensed sample spans more than an order of magnitude in stellar mass and SFR, allowing us to examine scaling relations with outflow properties along the star forming main sequence at these redshifts (Section~\ref{sec:galaxy-trends}). We observe a positive correlation of outflow velocities and absorption widths ($\Delta v_{90}$) with both SFR and stellar mass, although the correlations are of modest significance within this sample.
    Among the metrics tested, $\Delta v_{90}$ correlates well with $SFR$ with a Spearman coefficient of 0.7 at 2.7$\sigma$ significance (p-value = $0.007$).
    We compare these measured trends in outflow velocity with the TNG50 and FIRE-2 cosmological simulations, and find reasonable agreement, which is encouraging for future work using simulations to help interpret outflow properties. 
    The observed scaling relations are consistent at the $2\sigma$ level with expectations for momentum-driven outflows.

    \item 
    To assess which kinematic properties can be recovered from low-resolution spectra, we compare results from the well-resolved velocity profiles with quantities derived from a single Gaussian fit (Figure~\ref{fig:derived-fits}; Section~\ref{subsec:low-res-comparision}), both at $R\sim4000$ and at degraded resolution (down to $R\sim600$). A single Gaussian is appropriate for the information content of marginally-resolved spectra, and applying such fits at different $R$ allow us to assess possible biases. We find that for single Gaussian fits, velocity centroids are largely reliable, having a mean difference $\left< v_{cent,DG} - v_{cent,SG} \right> = 8\pm4$~\kms\ and a scatter of only $\pm 21~\kms$ ($1\sigma$) at $R\sim600$. Centroid measurements are nonetheless more precise and have lower scatter with increasing spectral resolution (Table~\ref{tab:low-res-comparision}). Velocity widths such as $\Delta v_{90}$ are affected by large scatter with single Gaussian fits and require caution to avoid bias. Velocity metrics which are sensitive to the asymmetry in the absorption profile, such as $v_{95}$ or other indicators of ``maximum'' outflow velocity, show a large scatter and clear bias even at $R\sim4000$ when fit with a symmetric Gaussian profile, illustrating that such metrics are only reliable when the resolution and measurement method is sufficient to capture asymmetric structure. We find that $R\gtrsim1700$ is needed to adequately capture the shape (e.g., skewness) of the absorption profiles in our sample. This corresponds to a FHWM resolution element $\lesssim \frac{\Delta v_{90}}{4}$. These results highlight the important role that spectral resolution plays in inferring key outflow properties. 

\end{enumerate}

This work represents the largest sample to date of well-resolved velocity profiles of gas outflows driven by star forming galaxies at cosmic noon ($z\sim2$--3). We have robustly characterized the typical outflow kinematics and diversity among the galaxy population, with $\sim$10 independent resolution elements across the velocity profiles. While such analysis is currently practical only for galaxies which are highly magnified by gravitational lensing, this sample provides context for interpreting outflow properties from far larger existing samples of high-redshift galaxies with lower spectral resolution. For example, our findings that the $v_{50}$ and $\Delta v_{90}$ metrics can be robustly recovered at low spectral resolution validate their use to characterize outflow scaling relations across larger samples and broader dynamic range than in this work. Moreover, these results can inform the optimal spectral resolution to be used for $z>2$ galaxy surveys with upcoming 30-meter class extremely large telescopes (ELTs). 

A promising avenue for future work is to explore spatially resolved outflow structure, along with the local conditions which launch strong galactic winds. 
As an immediate next step, some targets from this work are being followed up using the Keck Cosmic Web Imager to spatially map these ISM lines. Some will also be part of the galaxy evolution Key Science Program with KAPA \citep[Keck All-sky Precision Adaptive Optics;][]{wizinowich2020} which will provide kinematic maps of the nebular emission at $\sim$100 parsec resolution, providing a detailed view of the star formation morphology and ionized gas kinematics. Combining spatially resolved galaxy structure with spatially+spectrally resolved outflow properties will provide greater insight into the physical process responsible for the feedback which regulates galaxy formation.

\acknowledgements{
This research has made use of the Keck Observatory Archive (KOA), which is operated by the W. M. Keck Observatory and the NASA Exoplanet Science Institute (NExScI), under contract with the National Aeronautics and Space Administration.
Some of the data presented herein were obtained at the W. M. Keck Observatory, which is operated as a scientific partnership among the California Institute of Technology, the University of California and the National Aeronautics and Space Administration. The Observatory was made possible by the generous financial support of the W. M. Keck Foundation. We thank Jane Rigby and the MEGaSaURA team for for making their spectra publicly available.
TJ and KVGC gratefully acknowledge support from the Gordon and Betty Moore Foundation through Grant GBMF8549, from the National Science Foundation through grant AST-2108515, and from a Dean’s Faculty Fellowship. Support for this work was provided by NASA through the NASA Hubble Fellowship grant HST-HF2-51469.001-A awarded by the Space Telescope Science Institute, which is operated by the Association of Universities for Research in Astronomy, Incorporated, under NASA contract NAS5- 26555. RSE acknowledges funding from the European Research Council 
under the European Union Horizon 2020 research and innovation program 
(grant agreement No. 669253). This research was supported by the Australian Research Council Centre of Excellence for All Sky Astrophysics in 3 Dimensions (ASTRO 3D), through project number CE170100013. }The authors wish to recognize and acknowledge the very significant cultural role and reverence that the summit of Maunakea has always had within the indigenous Hawaiian community.  We are most fortunate to have the opportunity to conduct observations from this mountain.

\bibliography{ism_kinematics}

\appendix

\section{Absorption profiles and best-fit parameters for the lensed sample}
\label{sec:appendix_sample}

In this appendix we provide further information on the construction and fitting of absorption line velocity profiles for all galaxies in the lensed sample. 
Figures~\ref{fig:appendix_object_CSWA38} to \ref{fig:appendix_object_clone} 
show the equivalent of the bottom panels of Figure~\ref{fig:example-coveringfrac-plot} for each object. Profiles of all individual transitions used in the stack are shown for each object, and a list of the lines used is given in Table~\ref{tab:Lines-used}. Additionally, a histogram of the residuals obtained from fitting the absorption profile with a single Gaussian (SG) and double Gaussian (DG) profile is included in the bottom left panel of each figure. The DG fits have generally smaller residuals than SG fits in the region of ISM absorption, and in both cases the residuals are found to be centered around 0. 
Best fit parameters are given in Tables~\ref{tab:single-gaussian-fit} and \ref{tab:double-gaussian-fit} for the SG and DG fits, respectively, along with their uncertainties. The derived velocity metrics used in this work ($v_{cent} \equiv v_{50}$, $\Delta v_{90}$, $v_{75}$, etc.) are listed in Table~\ref{tab:derived-velocity-vals}.

\newpage

\outflowprofile{CSWA38}{Low ionization}

\outflowprofile{CSWA128}{Low ionization}

\outflowprofile{CSWA2}{Low ionization}

\outflowprofile{CSWA164}{Low ionization}

\outflowprofile{CSWA40}{Low ionization}

\outflowprofile{CSWA19}{Low ionization}

\outflowprofile{CSWA103}{Low ionization}

\outflowprofile{cosmiceye}{Low ionization} 

\outflowprofile{horseshoe}{Low ionization}

\outflowprofile{J0004}{Low ionization}

\outflowprofile{J0108}{Low ionization}

\outflowprofile{J1429}{Low ionization}

\outflowprofile{J1458}{Low ionization}

\outflowprofile{J1527}{Low ionization}

\outflowprofile{RCSGA0327-G}{Low ionization}

\outflowprofile{AGEL014106}{Low ionization}

\outflowprofile{AGEL183520}{Low ionization} 

\outflowprofile{AGEL231935}{Low ionization} 

\outflowprofile{8oclock}{Low ionization}

\outflowprofile{clone}{Low ionization} 

\newpage

\begin{deluxetable*}{|chc|}[htb!]
    \tablecaption{Table of ISM lines used to construct the velocity profile for each of the targets.}
    \tablewidth{0.5\textwidth}
    \tabletypesize{\footnotesize}
    % Header %
    \tablehead{ 
     \colhead{objid} & \colhead{} & \colhead{Lines used}  
    }
    %data %
    \startdata
J0004 & 1.681 & Si II 1260, O I 1302, Si II 1304, C II  1334, Si II 1526, Fe II 1608, Al II 1670 \\
RCSGA0327-G & 1.704 & Si II 1260, O I 1302, Si II 1304, C II  1334, Si II 1526, Fe II 1608, Al II 1670 \\
J0108 & 1.910 & Si II 1260, O I 1302, Si II 1304, C II  1334, Si II 1526, Fe II 1608, Al II 1670 \\
CSWA103 & 1.960 & Fe II 1608, Al II 1670, Fe II 2344, Fe II 2374, Fe II 2382 \\
AGEL231935 & 1.993 & Fe II 2344, Fe II 2374, Fe II 2382, Fe II 2586, Fe II 2600 \\
clone & 2.003 & Al II 1670, Fe II 2344, Fe II 2382, Fe II 2586, Fe II 2600 \\
CSWA19 & 2.032 & C II  1334, Si II 1526, Al II 1670, Fe II 2344 \\
CSWA40 & 2.189 & O I 1302, Si II 1304, C II  1334, Fe II 1608, Al II 1670 \\
CSWA2 & 2.197 & Si II 1526, Fe II 1608, Al II 1670, Fe II 2344, Fe II 2374, Fe II 2382 \\
CSWA128 & 2.225 & O I 1302, Si II 1304, C II  1334, Si II 1526, Fe II 1608, Al II 1670 \\
horseshoe & 2.381 & Si II 1260, O I 1302, Si II 1304, C II  1334, Si II 1526, Al II 1670 \\
AGEL014106 & 2.437 & Si II 1526, Fe II 1608, Al II 1670, Fe II 2382 \\
CSWA164 & 2.512 & Si II 1260, O I 1302, Si II 1304, C II  1334, Si II 1526, Al II 1670 \\
8oclock & 2.735 & Si II 1260, O I 1302, Si II 1304, C II  1334, Si II 1526, Al II 1670 \\
J1527 & 2.762 & O I 1302, Si II 1304, C II  1334, Si II 1526, Fe II 1608, Al II 1670 \\
J1429 & 2.824 & Si II 1260, O I 1302, Si II 1304, C II  1334, Si II 1526, Al II 1670 \\
CSWA38 & 2.926 & Si II 1260, O I 1302, Si II 1304, C II  1334, Si II 1526, Al II 1670 \\
cosmiceye & 3.073 & Si II 1260, O I 1302, Si II 1304, C II  1334, Si II 1526, Fe II 1608, Al II 1670 \\
AGEL183520 & 3.388 & O I 1302, Si II 1304, C II  1334, Si II 1526, Al II 1670 \\
J1458 & 3.487 & Si II 1260, O I 1302, Si II 1304, C II  1334, Si II 1526, Al II 1670 \\
\enddata
\tablenotetext{}{} \label{tab:Lines-used}
\end{deluxetable*}

\begin{deluxetable*}{|cc|cc|cc|cc|}
    \tablecaption{Best-fit single Gaussian parameters to the covering fraction. The single Gaussian function is of the form $C_f(v) = A_{SG} \exp((v-v_{SG})^2/(2\sigma_{SG}^2))$. The $err$ subscript denotes 1$\sigma$ uncertainity in the measured quantities. Values for $v$ and $\sigma$ are in units of \kms.}\label{tab:single-gaussian-fit}
    
    \tabletypesize{\footnotesize}
    % Header %
    \tablehead{ 
    \colhead{objid} & \colhead{$z_s$} & \colhead{$A_{SG}$} & \colhead{$A_{SG,err}$} &  \colhead{$\sigma_{SG}$} & \colhead{$\sigma_{SG,err}$}  & \colhead{$v_{SG}$}  & \colhead{$v_{SG,err}$} 
    }
    %data %
    \startdata
J0004 & 1.681 & 0.665 & 0.004 & 171.351 & 1.149 & -155.597 & 1.351 \\
RCSGA0327-G & 1.704 & 0.778 & 0.008 & 181.368 & 2.578 & -189.593 & 2.232 \\
J0108 & 1.910 & 0.408 & 0.004 & 290.694 & 2.833 & -287.416 & 3.231 \\
CSWA103 & 1.960 & 0.513 & 0.005 & 197.387 & 2.858 & -232.264 & 2.316 \\
AGEL231935 & 1.993 & 0.378 & 0.005 & 147.741 & 2.174 & -122.190 & 2.331 \\
clone & 2.003 & 0.451 & 0.006 & 215.906 & 3.181 & -164.518 & 4.300 \\
CSWA19 & 2.032 & 0.280 & 0.006 & 166.937 & 4.188 & -172.176 & 6.582 \\
CSWA40 & 2.189 & 0.629 & 0.008 & 204.387 & 3.357 & -107.486 & 2.604 \\
CSWA2 & 2.197 & 0.758 & 0.011 & 213.617 & 2.966 & 42.658 & 3.255 \\
CSWA128 & 2.225 & 0.692 & 0.008 & 156.369 & 2.749 & -123.379 & 2.619 \\
horseshoe & 2.381 & 0.656 & 0.014 & 209.343 & 5.543 & -235.747 & 5.822 \\
AGEL014106 & 2.437 & 0.524 & 0.007 & 229.719 & 4.410 & -142.180 & 4.778 \\
CSWA164 & 2.512 & 0.385 & 0.004 & 233.208 & 2.957 & -77.331 & 3.290 \\
8oclock & 2.735 & 0.565 & 0.003 & 265.743 & 1.434 & -255.609 & 1.832 \\
J1527 & 2.762 & 0.335 & 0.011 & 149.403 & 6.262 & -129.960 & 5.025 \\
J1429 & 2.824 & 0.337 & 0.004 & 195.469 & 3.388 & -213.160 & 2.909 \\
CSWA38 & 2.926 & 0.809 & 0.006 & 170.661 & 1.449 & -205.611 & 1.931 \\
cosmiceye & 3.073 & 0.722 & 0.004 & 296.740 & 1.853 & 149.634 & 1.678 \\
AGEL183520 & 3.388 & 1.000 & 0.000 & 185.029 & 2.065 & 89.664 & 2.368 \\
J1458 & 3.487 & 0.759 & 0.005 & 310.094 & 2.696 & -185.610 & 2.854 \\
\enddata
\tablenotetext{}{} 
\end{deluxetable*}

\begin{deluxetable*}{|cc|cc|cc|cc|cc|cc|cc|}
    \tablecaption{Best-fit double Gaussian parameters to the covering fraction. The double Gaussian function is of the form $C_f(v) = A_0 \exp((v-v_0)^2/(2\sigma_0^2)) + A_1 \exp((v-v_1)^2/(2\sigma_1^2))$, where we adopt a convention that the `0' component is more blueshifted ($v_0 < v_1$). The $err$ subscript denotes 1$\sigma$ uncertainity in the measured quantities. Values for $v$ and $\sigma$ are in units of \kms.
    }\label{tab:double-gaussian-fit}
    \tabletypesize{\footnotesize}
    % Header %
    \tablehead{ 
\colhead{objid} & \colhead{$z_s$} & \colhead{$A_0$} & \colhead{$A_{0,err}$} &  \colhead{$A_1$} & \colhead{$A_{1,err}$} & \colhead{$\sigma_0$} & \colhead{$\sigma_{0,err}$}  & \colhead{$\sigma_1$} & \colhead{$\sigma_{1,err}$} & \colhead{$v_0$}  & \colhead{$v_{0,err}$} & \colhead{$v_1$} & \colhead{$v_{1,err}$} 
    }
    %data %
    \startdata
J0004 & 1.681 & 0.264 & 0.021 & 0.691 & 0.018 & 119.165 & 11.667 & 122.341 & 5.196 & -393.937 & 25.035 & -107.172 & 8.949 \\
RCSGA0327-G & 1.704 & 0.474 & 0.061 & 0.573 & 0.058 & 180.655 & 3.199 & 105.466 & 8.503 & -313.221 & 27.360 & -97.046 & 10.696 \\
J0108 & 1.910 & 0.219 & 0.019 & 0.424 & 0.008 & 68.983 & 11.687 & 238.097 & 7.482 & -689.599 & 10.675 & -239.882 & 7.211 \\
CSWA103 & 1.960 & 0.508 & 0.009 & 0.176 & 0.042 & 169.422 & 11.819 & 72.661 & 13.642 & -266.963 & 12.507 & -12.055 & 7.957 \\
AGEL231935 & 1.993 & 0.311 & 0.010 & 0.249 & 0.027 & 132.358 & 8.386 & 53.186 & 2.911 & -178.115 & 8.203 & -14.867 & 4.264 \\
clone & 2.003 & 0.414 & 0.021 & 0.212 & 0.031 & 196.160 & 6.591 & 65.790 & 12.433 & -212.046 & 15.460 & 27.799 & 13.959 \\
CSWA19 & 2.032 & 0.213 & 0.009 & 0.364 & 0.016 & 136.015 & 6.231 & 53.070 & 3.645 & -298.644 & 12.681 & -53.941 & 3.355 \\
CSWA40 & 2.189 & 0.173 & 0.019 & 0.648 & 0.017 & 162.913 & 11.204 & 182.493 & 7.671 & -635.098 & 34.824 & -93.399 & 7.399 \\
CSWA2 & 2.197 & 0.623 & 0.042 & 0.640 & 0.040 & 134.903 & 12.834 & 120.155 & 13.924 & -94.353 & 22.869 & 185.693 & 24.113 \\
CSWA128 & 2.225 & 0.239 & 0.023 & 0.703 & 0.058 & 119.469 & 28.562 & 116.438 & 8.922 & -358.799 & 59.712 & -88.857 & 15.418 \\
horseshoe & 2.381 & 0.488 & 0.076 & 0.402 & 0.346 & 204.224 & 22.332 & 94.989 & 41.874 & -301.625 & 98.518 & -132.461 & 20.543 \\
AGEL014106 & 2.437 & 0.100 & 0.000 & 0.525 & 0.035 & 150.737 & 34.704 & 201.872 & 15.100 & -503.679 & 138.772 & -119.511 & 17.981 \\
CSWA164 & 2.512 & 0.139 & 0.013 & 0.408 & 0.010 & 90.945 & 10.374 & 187.942 & 5.970 & -431.568 & 10.869 & -41.428 & 6.074 \\
8oclock & 2.735 & 0.577 & 0.004 & 0.216 & 0.009 & 238.401 & 1.853 & 50.000 & 0.000 & -279.742 & 2.223 & 137.199 & 2.870 \\
J1527 & 2.762 & 0.215 & 0.135 & 0.147 & 0.070 & 137.220 & 61.762 & 114.857 & 96.085 & -134.962 & 12.512 & -114.187 & 14.493 \\
J1429 & 2.824 & 0.144 & 0.029 & 0.311 & 0.055 & 189.953 & 18.825 & 137.336 & 10.027 & -461.169 & 108.144 & -154.013 & 15.660 \\
CSWA38 & 2.926 & 0.518 & 0.101 & 0.703 & 0.022 & 82.106 & 6.639 & 129.877 & 15.928 & -339.982 & 8.222 & -131.933 & 17.191 \\
cosmiceye & 3.073 & 0.572 & 0.005 & 0.779 & 0.009 & 197.324 & 3.721 & 114.665 & 1.317 & -64.461 & 3.993 & 331.493 & 1.776 \\
AGEL183520 & 3.388 & 0.718 & 0.091 & 0.971 & 0.036 & 92.928 & 9.688 & 121.912 & 11.488 & -57.855 & 14.147 & 171.525 & 16.030 \\
J1458 & 3.487 & 0.646 & 0.069 & 0.595 & 0.056 & 170.658 & 19.483 & 191.377 & 26.951 & -366.716 & 25.356 & 16.805 & 40.959 \\
\enddata
\tablenotetext{}{} 
\end{deluxetable*}

\begin{deluxetable*}{|cc|cc|ccc|cc|cc|cc|}
    \tablewidth{0.5\textwidth}
    \tabletypesize{\footnotesize }
    \tablecaption{Derived velocity parameters, obtained from single Gaussian (SG) and double Gaussian (DG) fits to the covering fraction profile of each galaxy. DG-V2 denotes values from a double Gaussian fit but considering only the absorption at velocities $v<0$ \kms, as described in the text. All velocity values are in units of \kms. }\label{tab:derived-velocity-vals}
    % Header %
    \tablehead{ 
    \colhead{objid} & \colhead{$z_s$} & \colhead{$\Delta v_{90}$} & \colhead{$\Delta v_{90}$} & \colhead{$v_{cent}$} & \colhead{$v_{cent}$} & \colhead{$v_{cent}$} & {$v_{75}$} & \colhead{$v_{75}$} & \colhead{$v_{95}$} & \colhead{$v_{95}$} & \colhead{$v_{99}$} & \colhead{$v_{99}$}\\ 
    \colhead{} & \colhead{} & \colhead{SG} & \colhead{DG} & \colhead{SG} & \colhead{DG} & \colhead{DG-V2} & {DG} & \colhead{DG-V2} & \colhead{DG} & \colhead{DG-V2} & \colhead{DG} & \colhead{DG-V2} % \\ 
    }
    %data %
    \startdata
J0004 & 1.681 & 564.0 & 578.0 & -156.0 & -163.0 & -196.0 & -303.0 & -331.0 & -502.0 & -513.0 & -605.0 & -613.5 \\
RCSGA0327-G & 1.704 & 597.0 & 616.0 & -190.0 & -198.0 & -225.0 & -350.0 & -369.0 & -562.0 & -572.5 & -697.0 & -705.0 \\
J0108 & 1.910 & 956.0 & 875.0 & -288.0 & -283.0 & -334.0 & -492.0 & -533.0 & -735.0 & -745.0 & -825.0 & -834.0 \\
CSWA103 & 1.960 & 649.0 & 589.5 & -233.0 & -236.0 & -263.0 & -363.0 & -380.0 & -537.0 & -547.0 & -656.0 & -663.0 \\
AGEL231935 & 1.993 & 486.0 & 440.0 & -123.0 & -126.0 & -162.0 & -236.0 & -256.0 & -377.0 & -390.0 & -472.0 & -482.0 \\
clone & 2.003 & 710.0 & 645.0 & -165.0 & -173.0 & -234.0 & -322.0 & -360.0 & -523.0 & -546.5 & -661.0 & -679.0 \\
CSWA19 & 2.032 & 549.0 & 499.0 & -173.0 & -170.0 & -198.0 & -326.0 & -337.0 & -484.0 & -490.0 & -585.0 & -590.0 \\
CSWA40 & 2.189 & 672.0 & 920.0 & -108.0 & -140.0 & -219.0 & -329.5 & -428.0 & -731.0 & -765.0 & -884.0 & -905.0 \\
CSWA2 & 2.197 & 703.0 & 616.0 & 42.0 & 41.0 & -131.0 & -109.0 & -212.0 & -278.0 & -337.0 & -382.5 & -427.0 \\
CSWA128 & 2.225 & 514.0 & 576.0 & -124.0 & -135.0 & -171.0 & -269.0 & -305.0 & -483.0 & -498.0 & -587.0 & -596.5 \\
horseshoe & 2.381 & 688.5 & 686.0 & -236.0 & -241.5 & -266.0 & -408.0 & -419.5 & -614.0 & -622.0 & -758.0 & -765.0 \\
AGEL014106 & 2.437 & 756.0 & 806.5 & -143.0 & -150.0 & -225.0 & -318.0 & -377.0 & -598.5 & -634.0 & -759.0 & -780.0 \\
CSWA164 & 2.512 & 767.0 & 738.0 & -78.0 & -80.0 & -191.0 & -249.0 & -345.0 & -481.0 & -512.5 & -581.0 & -601.5 \\
8oclock & 2.735 & 874.0 & 829.0 & -256.0 & -256.0 & -316.0 & -427.0 & -465.0 & -664.0 & -687.0 & -828.0 & -846.0 \\
J1527 & 2.762 & 491.0 & 551.0 & -130.0 & -128.0 & -165.0 & -233.0 & -262.0 & -404.0 & -422.0 & -523.0 & -537.0 \\
J1429 & 2.824 & 643.0 & 752.0 & -214.0 & -236.0 & -259.0 & -417.5 & -440.0 & -698.0 & -708.0 & -856.0 & -865.0 \\
CSWA38 & 2.926 & 562.0 & 492.0 & -206.0 & -202.0 & -228.0 & -315.0 & -328.0 & -432.0 & -437.0 & -501.0 & -505.0 \\
cosmiceye & 3.073 & 976.0 & 802.0 & 149.0 & 144.0 & -160.0 & -91.0 & -264.0 & -330.0 & -433.0 & -480.0 & -558.0 \\
AGEL183520 & 3.388 & 609.0 & 509.0 & 89.0 & 90.0 & -84.0 & -33.0 & -137.0 & -166.0 & -224.0 & -243.5 & -287.5 \\
J1458 & 3.487 & 1020.0 & 859.0 & -186.0 & -185.0 & -296.0 & -380.0 & -438.0 & -600.0 & -631.5 & -738.0 & -761.0 \\
\enddata
\tablenotetext{}{} 
\end{deluxetable*}

\end{document}